\RequirePackage{snapshot}

\documentclass[format=acmsmall, authorversion,screen=true]{acmart}
\setcopyright{rightsretained}
\acmPrice{}
\acmDOI{10.1145/3408979}
\acmYear{2020}
\copyrightyear{2020}
\acmSubmissionID{icfp20main-p7-p}
\acmJournal{PACMPL}
\acmVolume{4}
\acmNumber{ICFP}
\acmArticle{97}
\acmMonth{8}

\usepackage[
    type={CC},
    modifier={by},
    version={4.0},
	imagewidth={6em}
]{doclicense}

\usepackage{import}
\usepackage[utf8]{inputenc}
\usepackage{amsmath}
\usepackage{amssymb}
\usepackage{graphicx}
\usepackage{amsfonts}
\usepackage{amssymb}
\usepackage{cmll}
\usepackage{enumitem}
\usepackage{proof}
\usepackage{mathpartir}
\usepackage{stmaryrd}
\usepackage{tikz}
\usetikzlibrary
{
	calc ,
	matrix ,
	arrows ,
	angles ,
	quotes ,
	intersections ,
	positioning ,
	fit ,
	shapes.geometric ,
	shapes.misc ,
	decorations ,
	decorations.markings ,
	decorations.pathmorphing ,
	decorations.pathreplacing ,
	decorations.text 
}

\usepgflibrary
{
	arrows.meta
}

\usepackage{xcolor}

\tikzset
{
	diagram/.style =
	{
		line cap = butt ,
		line join = bevel ,
		arrows = -> ,
		> = angle 60 ,
		auto = left ,
		text height = 1.5ex , 
		text depth = 0.25ex , 
		align = center ,
	} ,
	code node/.append style =
	{
		every node/.append style =
		{
			execute at begin node = \begin{texttt} ,
			execute at end node = \end{texttt}
		}
	} ,
	code diagram/.style =
	{
		diagram ,
		code node
	} ,
	math node/.append style =
	{
		every node/.append style =
		{
			execute at begin node = \begin{math} ,
			execute at end node = \end{math}
		}
	} ,
	math diagram/.style =
	{
		diagram ,
		math node
	} ,
	string diagram/.style =
	{
		math diagram ,
		arrows = - ,
	} ,
	online/.style =
	{
		shape = rectangle ,
		rounded corners = 2mm ,
		inner sep = 1.5pt ,
		fill = white ,
		opacity = 1.0 ,
		anchor = center	
	} ,
	double arrow/.style =
	{
		double distance between line centers = 2pt ,
		-implies 
	} ,
	equals/.style =
	{
		double distance between line centers = 2pt ,
		-implies , 
		arrows = -
	} ,
	mapto/.append style =
	{
		arrows = |-> ,
	} ,
	paph/.append style =
	{
		decorate ,
		decoration = {snake , segment length = 5mm , amplitude = 0.5mm} ,
	} ,
	includel/.append style =
	{
		arrows = Hooks[right]-> ,
	} ,
	includer/.append style =
	{
		arrows = Hooks[left]-> ,
	} ,
	monomorphism/.append style =
	{
		arrows = >-> ,
	} ,
	epimorphism/.append style =
	{
		arrows = ->> ,
	} ,
	cofibration/.append style =
	{
		arrows = >-> ,
	} ,
	fibration/.append style =
	{
		arrows = ->> ,
	} ,
	equivalence/.append style =
	{
		decoration = {markings , mark = at position 1/2 with {\node[online , transform shape] {∼};}} ,
		postaction = {decorate} ,
	} ,
	proarrow/.append style =
	{
		decoration = {markings , mark = at position 1/2 with {\draw [solid , -] (0 , -0.6ex) to (0 , 0.6ex);}} ,
		postaction = {decorate} ,
	} ,
	string/.append style =
	{
		arrows = - ,
	} ,
	cofibration string/.append style =
	{
		decoration = {markings , mark = at position 1/4 with {\arrow{>}} , mark = at position 3/4 with {\arrow{>}}} ,
		postaction = {decorate} ,
	} ,
	fibration string/.append style =
	{
		decoration = {markings , mark = at position 4/8 with {\arrow{>}} , mark = at position 5/8 with {\arrow{>}}} ,
		postaction = {decorate} ,
	} ,
	overcross/.append style =
	{
		preaction = {draw = white , - , line width = 5pt}
	} ,
	label/.append style =
	{
		font = \scriptsize
	} ,
	incoming/.append style =
	{
		pos = 0 ,
		anchor = south
	} ,
	outgoing/.append style =
	{
		pos = 1 ,
		anchor = north
	} ,
	outcoming/.append style =
	{
		pos = 0 ,
		anchor = north
	} ,
	ingoing/.append style =
	{
		pos = 1 ,
		anchor = south
	} ,
	fromabove/.append style =
	{
		pos = 0 ,
		anchor = south
	} ,
	tobelow/.append style =
	{
		pos = 1 ,
		anchor = north
	} ,
	frombelow/.append style =
	{
		pos = 0 ,
		anchor = north
	} ,
	toabove/.append style =
	{
		pos = 1 ,
		anchor = south
	} ,
	fromleft/.append style =
	{
		pos = 0 ,
		anchor = east
	} ,
	toright/.append style =
	{
		pos = 1 ,
		anchor = west
	} ,
	fromright/.append style =
	{
		pos = 0 ,
		anchor = west
	} ,
	toleft/.append style =
	{
		pos = 1 ,
		anchor = east
	} ,
	forall/.append style =
	{
		line width = 0.5pt ,
	} ,
	exists/.append style =
	{
		densely dashed
	} ,
	structural/.append style =
	{
		opacity = 2/3 ,
	} ,
	0-cell/.style =
	{
		shape = rectangle ,
		rounded corners = 2mm ,
		inner sep = 2pt
	} ,
	1-cell/.style =
	{
		shape = rectangle ,
		rounded corners = 2mm ,
		inner sep = 1.5pt ,
		fill = white ,
		opacity = 1.0 ,
		anchor = center	
	} ,
	2-cell/.style =
	{
		draw = black!75 ,
		fill = white ,
		opacity = 0.9 ,
		inner sep = 0.75mm ,
		minimum size = 4.0mm
	} ,
	bead/.style =
	{
		2-cell ,
		shape = rectangle ,
		rounded corners = 2.0mm
	} ,
	diamond/.style =
	{
		2-cell ,
		shape = diamond
	} ,
	bra/.style =
	{
		2-cell ,
		shape = isosceles triangle ,
		isosceles triangle apex angle = 70 ,
		shape border rotate = -90 ,
		minimum size = 5mm
	} ,
	ket/.style =
	{
		2-cell ,
		shape = isosceles triangle ,
		isosceles triangle apex angle = 70 ,
		shape border rotate = 90 ,
		minimum size = 6mm
	} ,
	box/.style =
	{
		2-cell ,
		shape = rectangle
	} ,
	white dot/.style =
	{
		2-cell ,
		shape = circle ,
		minimum size = 1.5mm ,
		fill = white
	} ,
	black dot/.style =
	{
		2-cell ,
		shape = circle ,
		minimum size = 1.5mm ,
		fill = gray
	} ,
	dot/.style =
	{
		2-cell ,
		shape = circle ,
		inner sep = 0mm ,
		minimum size = 0.5mm ,
		fill = black
	} ,
	sheet/.style =
	{
		draw = black!75 ,
		fill = gray!10 ,
		opacity = 0.9 ,
		fill opacity = 0.5
	} ,
	torn/.append style =
	{
		decoration = {random steps , segment length = 2pt , amplitude = 1pt}
	} ,
	edge annotation/.append style =
	{
		anchor = west
	} ,
	callout/.append style =
	{
		minimum height = 2mm ,
		minimum width = 4mm ,
		draw ,
		draw opacity = 0.25
	} ,
	callout-pre/.append style =
	{
		callout ,
		inner sep = 1.2mm ,
		solid
	} ,
	callout-post/.append style =
	{
		callout ,
		inner sep = 0.8mm ,
		dashed
	} ,
	pullback/.pic =
	{
		\draw [semithick , arrows = -] (-0.1 , 0.1) -- (0.1 , 0.1) -- (0.1 , -0.1) ;
	} ,
	pushout/.pic =
	{
		\draw [semithick , arrows = -] (-0.1 , 0.1) -- (-0.1 , -0.1) -- (0.1 , -0.1) ;
	} ,
}

\pgfdeclaredecoration{simple line}{start}
{
  \state{start}[width = +0pt,
                next state=step]{
    \pgfpathmoveto{\pgfpoint{0pt}{0pt}}
  }
  \state{step}[auto end on length    = 3pt,
               auto corner on length = 3pt,               
               width=+1pt]
  {
    \pgfpathlineto{\pgfpoint{1pt}{0pt}}
  }
  \state{final}
  {}
}

\usepackage{thmtools}
\usepackage{versions}
\usepackage{listings}
\usepackage{microtype}

	\excludeversion{icfp2020}
	\includeversion{arxiv}

\title{Denotational Recurrence Extraction for Amortized Analysis}

\newcommand{\norm}[1]{\left\lVert#1\right\rVert}
\newcommand{\angles}[1]{\left\llangle #1 \right\rrangle}

\newcommand{\bdby}{\sqsubseteq}
\newcommand{\valbd}{\sqsubseteq_{\texttt{val}}}
\newcommand{\subbd}{\sqsubseteq_{\texttt{sub}}}

\newcommand{\savename}{\ensuremath{\texttt{save}}}
\newcommand{\xfername}{\ensuremath{\texttt{transfer}}}
\newcommand{\save}[3]{\ensuremath{\texttt{save}^{#1}_{#2} \; #3}}
\newcommand{\disc}[2]{\ensuremath{\texttt{spend}_{#1} \; #2}}
\newcommand{\waitname}[0]{\ensuremath{\texttt{create}}}
\newcommand{\discname}{\texttt{spend}}
\let\createname\waitname
\let\spendname\discname
\newcommand{\wait}[2]{\ensuremath{\texttt{create}_{#1} \; #2}}
\newcommand{\tsfer}[6]{\ensuremath{\texttt{transfer}_{#1} \, !^{#2}_{#3} \, #4 = #5 \; \texttt{to} \; #6}}
\newcommand{\ccase}[5]{\ensuremath{\texttt{case} \, (#1,\, #2. #3 \, , \, #4 . #5})}
\newcommand{\acase}[6]{\ensuremath{\texttt{case}_{#1} \, (#2,\, #3. #4 \, , \, #5 . #6})}
\newcommand{\inl}[1]{\ensuremath{\texttt{inl} \, #1}}
\newcommand{\inr}[1]{\ensuremath{\texttt{inr} \, #1}}
\newcommand{\elist}{\ensuremath{\texttt{[]}}}
\newcommand{\cons}[2]{\ensuremath{#1 \, :: \, #2}}
\newcommand{\listty}[1]{\ensuremath{\texttt{List}\left(#1\right)}}

\newcommand{\asplit}[5]{\ensuremath{\texttt{split}_{#1}(#2, \, #3.#4.#5)}}

\newcommand{\N}{\ensuremath{\mathbb{N}}}
\newcommand{\Z}{\ensuremath{\mathbb{Z}}}
\newcommand{\nrec}[3]{\ensuremath{\texttt{nrec}\left(#1,#2,#3\right)}}
\newcommand{\lrec}[3]{\ensuremath{\texttt{lrec}\left(#1,#2,#3\right)}}
\newcommand{\inj}{\overline}
\newcommand{\tick}[1]{\ensuremath{\texttt{tick} \; \, ; \; #1}}
\newcommand{\amp}{\ensuremath{\&}}
\newcommand{\amppair}[2]{\ensuremath{\langle #1,#2 \rangle}}
\newcommand{\scott}[1]{\ensuremath{\llbracket #1 \rrbracket}}
\newcommand{\snrec}{\ensuremath{\texttt{snrec}}}
\newcommand{\slrec}{\ensuremath{\texttt{slrec}}}
\newcommand{\scase}{\ensuremath{\texttt{scase}}}
\newcommand{\bbbc}{\mathbb{C}}
\newcommand{\tree}[1]{\ensuremath{\texttt{tree}\left(#1\right)}}
\newcommand{\trec}[6]{\ensuremath{\texttt{trec}\left(#1,#2,#3,#4,#5,#6\right)}}
\newcommand{\set}[1]{\ensuremath{\texttt{set}\left(#1\right)}}

\DeclareMathOperator{\Hom}{Hom}

\newcommand{\loli}{\multimap}
\newcommand{\tensor}{\otimes}

\newcommand{\const}[1]{\ensuremath{\text{const} \left(#1\right)}}

\newcommand{\pack}[3]{\ensuremath{\texttt{pack}_{#1 = #2} #3}}
\newcommand{\unpack}[4]{\ensuremath{\texttt{unpack } (#1,#2) = #3 \texttt{ in } #4}}
\newcommand{\toC}[1]{\ensuremath{\text{to}\mathbb{C}(#1)}}

\makeatletter
\DeclareFontFamily{OMX}{MnSymbolE}{}
\DeclareSymbolFont{MnLargeSymbols}{OMX}{MnSymbolE}{m}{n}
\SetSymbolFont{MnLargeSymbols}{bold}{OMX}{MnSymbolE}{b}{n}
\DeclareFontShape{OMX}{MnSymbolE}{m}{n}{
    <-6>  MnSymbolE5
   <6-7>  MnSymbolE6
   <7-8>  MnSymbolE7
   <8-9>  MnSymbolE8
   <9-10> MnSymbolE9
  <10-12> MnSymbolE10
  <12->   MnSymbolE12
}{}
\DeclareFontShape{OMX}{MnSymbolE}{b}{n}{
    <-6>  MnSymbolE-Bold5
   <6-7>  MnSymbolE-Bold6
   <7-8>  MnSymbolE-Bold7
   <8-9>  MnSymbolE-Bold8
   <9-10> MnSymbolE-Bold9
  <10-12> MnSymbolE-Bold10
  <12->   MnSymbolE-Bold12
}{}

\let\llangle\@undefined
\let\rrangle\@undefined
\DeclareMathDelimiter{\llangle}{\mathopen}%
                     {MnLargeSymbols}{'164}{MnLargeSymbols}{'164}
\DeclareMathDelimiter{\rrangle}{\mathclose}%
                     {MnLargeSymbols}{'171}{MnLargeSymbols}{'171}
\makeatother

\newcommand{\codeInc}{\texttt{inc}}
\newcommand{\codeSet}{\texttt{set}}
\newcommand{\codebitlist}{\texttt{bit\ list}}
\newcommand{\codenat}{\texttt{nat}}

\author{Joseph W. Cutler}
\affiliation{%
  \institution{Wesleyan University}
  \city{Middletown}
  \state{Connecticut}
}
\email{jwcutler@wesleyan.edu}

\author{Daniel R. Licata}
\affiliation{%
  \institution{Wesleyan University}
  \city{Middletown}
  \state{Connecticut}
}
\email{dlicata@wesleyan.edu}

\author{Norman Danner}
\orcid{0000-0003-1119-4982}
\affiliation{%
  \institution{Wesleyan University}
  \city{Middletown}
  \state{Connecticut}
}
\email{ndanner@wesleyan.edu}

\renewcommand{\shortauthors}{Cutler et al.}

\begin{abstract}
A typical way of analyzing the time complexity of functional programs is to
extract a recurrence expressing the running time of the program in terms of the
size of its input, and then to solve the recurrence to obtain a big-O bound.
For recurrence extraction to be compositional, it is also necessary to extract
recurrences for the size of outputs of helper functions.  Previous work has
developed techniques for using logical relations to state a formal correctness
theorem for a general recurrence extraction translation: a program is bounded
by a recurrence when the operational cost is bounded by the extracted cost, and
the output value is bounded, according to a value bounding relation defined by
induction on types, by the extracted size.  This previous work supports
higher-order functions by viewing recurrences as programs in a lambda-calculus,
or as mathematical entities in a denotational semantics thereof.  In this
paper, we extend these techniques to support amortized analysis, where costs
are rearranged from one portion of a program to another to achieve more precise
bounds.  We give an intermediate language in which programs can be annotated
according to the banker's method of amortized analysis; this language has an
affine type system to ensure credits are not spent more than once.  We give a
recurrence extraction translation of this language into a recurrence language,
a simply-typed lambda-calculus with a cost type, and state and prove a bounding
logical relation expressing the correctness of this translation.  The
recurrence language has a denotational semantics in preorders, and we use this
semantics to solve recurrences, e.g analyzing binary counters and splay trees.
\end{abstract}

\keywords{recurrence extraction, resource analysis, amortized analysis, cost
semantics, higher order recurrences, denotational semantics}

\begin{document}
\maketitle

\section{Introduction}\label{sec:intro}

A common technique for analyzing the asymptotic resource
complexity of functional programs is the
\emph{extract-and-solve} method, in which one extracts a recurrence
expressing an upper bound on the cost of the program in terms of the size of
its input, and then solves the recurrence to obtain a big-$O$ bound.
Typically, the connection between the original program and the extracted
recurrence is left informal, relying on an intuitive understanding that the
extracted recurrence correctly models the program.  Previous
work~\cite{danner-et-al:plpv13,danner-et-al:icfp15,hudson,kavvos-et-al:popl20,
danner-licata:jfp-in-prep} has begun to explore more formal techniques for
relating programs and extracted recurrences.  The process of extracting a
recurrence consists of two phases.  The first is a monadic translation into
the writer monad~$\bbbc\times\cdot$, translating a program to also
``output'' its cost along with its value.  We call the result a
\emph{syntactic recurrence}, and at function type, the result is essentially
a function that maps a value to a pair consisting of the cost of evaluating
that function along with its result.  At higher type, the syntactic
recurrence maps a recurrence for the argument to a recurrence for the
result.  A \emph{bounding logical relation} relates programs to syntactic
recurrences, and the fundamental \emph{bounding theorem} states that a
program and its syntactic recurrence are related, which in particular
implies that its actual runtime cost is bounded by the extracted prediction.
Since inductive values are translated to (essentially) themselves, this
phase does not abstract values to sizes; in effect, the syntactic recurrence
describes the cost of the program in terms of its actual arguments.  The
second phase performs this size abstraction by interpreting (the language
of) syntactic recurrences in a denotational model.  The interpretation of
each type is intended to be a domain of sizes for values of that type, and
different models can implement different notions of size.  For example, a
list value (i.e., the list type and constructors) may be interpreted by its
length in one model, or even more exotic notions of size, such as the number
of pairwise inversions (as required for an analysis of insertion sort) for a
list of numbers.  Thus the interpretation of the syntactic recurrence
extracted from a source program (what we might call the \emph{semantic
recurrence}) is a function that maps sizes (of source-program values) to a
bound on the cost of that program on those values.  
It is these semantic recurrences
that match the recurrences that arise from the typical ``extract-and-solve''
approach to analyzing program cost.  Our previous work develops this
methodology for functional programs with numbers and
lists~\cite{danner-et-al:plpv13}, inductive types with structural
recursion~\cite{danner-et-al:icfp15}, general
recursion~\cite{kavvos-et-al:popl20}, and
let-polymorphism~\cite{danner-licata:jfp-in-prep}.

As an example that demonstrates both the approach and a weakness of the
underlying technique for cost analysis that it formalizes, let us consider
the binary increment function, a standard motivating example for amortized
analysis:
\begin{small}
\[
\begin{array}[t]{lcl}
\codeInc &:& \codebitlist \to \codebitlist \\
\codeInc\,[\,] &=& [1] \\
\codeInc\,(0 :: bs) &=& 1 :: bs \\
\codeInc\,(1 :: bs) &=& 0 :: \codeInc\,bs
\end{array}
\qquad
\begin{array}[t]{lcl}
\codeSet &:& \codenat \to \codebitlist \\
\codeSet\,0 &=& [\,] \\
\codeSet\,(S\,n) &=& \codeInc(\codeSet\,n)
\end{array}
\]
\end{small}

\noindent
The value part of a monadic translation of a function into~$\bbbc\times\cdot$
is a function into a pair, but here we
sugar that into a pair of functions, which may be mutually recursive.  We
denote the cost and value components by $(\cdot)_c$ and $(\cdot)_p$,
respectively (this notation is explained in
Section~\ref{sec:monadic-translation}), and charge one unit of cost for each
$::$ operation:
\begin{small}
\[
\begin{array}[t]{lcl}
\codeInc_c &:& \codebitlist \to \bbbc \\
\codeInc_c\,[] &=& 1 \\
\codeInc_c\,(0 :: bs) &=& 1 \\
\codeInc_c\,(1 :: bs) &=& 1 + \codeInc_c\,bs
\\ \\
\codeSet_c &:& \codenat \to \bbbc \\
\codeSet_c\,0 &=& 0 \\
\codeSet_c\,(S\,n) &=& \codeSet_c(n) + \codeInc_c(\codeSet_p\,n)
\end{array}
\qquad
\begin{array}[t]{lcl}
\codeInc_p &:& \codebitlist \to \codebitlist \\
\codeInc_p\,[] &=& [1] \\
\codeInc_p\,(0 :: bs) &=& 1 :: bs \\
\codeInc_p\,(1 :: bs) &=& 0 :: \codeInc_p\,bs
\\ \\
\codeSet_p &:& \codenat \to \codebitlist \\
\codeSet_p\,0 &=& [] \\
\codeSet_p\,(S\,n) &=& \codeInc_p(\codeSet_p\,n)
\end{array}
\]
\end{small}

We obtain the usual recurrences that we expect when we interpret these
syntactic recurrences in an appropriate denotational semantics.  We
interpret $\codebitlist$ and $\codenat$ by~$\N$, the natural numbers, and
interpret the constructors so that a $\codebitlist$ is interpreted by its
length and a $\codenat$ by its value.  Doing so, we obtain semantic
recurrences for the the cost and size of~$\codeInc$:
\[
\begin{aligned}
T_{\codeInc}(0) &= 1 \\
T_{\codeInc}(n+1) &= \max\{1, 1 + T_{\codeInc}(n)\}
\end{aligned}
\qquad
\begin{aligned}
S_{\codeInc}(0) &= 1 \\
S_{\codeInc}(n+1) &= \max\{1 + n, 1 + S_{\codeInc}(n)\}
\end{aligned}
\]
The usual techniques (in the semantics) then allow us to conclude that
$T_{\codeInc}(n) \leq n + 1$ and $S_{\codeInc}(n) \leq n + 1$, which are correct
and tight bounds on the cost and size of the $\codeInc$ function.  The
semantic recurrences for $\codeSet$ are
\[
\begin{aligned}
T_{\codeSet}(0) &= 0 \\
T_{\codeSet}(n+1) &= T_{\codeSet}(n) + T_{\codeInc}(S_{\codeSet}(n)) \\
                  &\leq T_{\codeSet}(n) + S_{\codeSet}(n) + 1
\end{aligned}
\qquad
\begin{aligned}
S_{\codeSet}(0) &= 0 \\
S_{\codeSet}(n+1) &= S_{\codeInc}(S_{\codeSet}(n)) \\
                  &\leq S_{\codeSet}(n) + 1
\end{aligned}
\]
and so we conclude that $S_{\codeSet}(n)\leq n$ and hence
$T_{\codeSet}(n)\in O(n^2)$, both of which are correct, but not tight,
bounds.  

On the one hand, through syntactic recurrence extraction, the bounding
theorem, and soundness of the semantics, we have a formal connection between
the original programs and the semantic recurrences that bound their cost and
size.  On the other, this example demonstrates a well-understood weakness in
the informal technique:  while the cost of a composition of functions is
bounded by the composition of their costs, the bound is not necessarily
tight.  The tight bound is usually established with some form of amortized
analysis, and \emph{the goal of this paper is to provide a formalization of
the banker's method for amortized analysis comparable to the formalization
of \cite{danner-et-al:plpv13,danner-et-al:icfp15,hudson} for non-amortized
analysis.}

The \emph{banker's method}
for amortized analysis~\cite{tarjan:amortized-complexity}
permits one to ``prepay'' time
cost to generate ``credits'' that are ``spent'' later to reduce time
cost, rearranging the accounting of costs from one portion of a program
to another (in particular, generating a credit costs 1 unit of time,
while spending a credit reduces the cost by 1 unit of time).  In this
example, we maintain the invariant that one credit is attached to every $1$ bit in
the counter representation.  The \emph{amortized cost} of flipping a bit
from $0$ to $1$ is then $2$ units of time---one for the actual bit flip
plus one to generate the credit. However, the amortized cost of flipping
a bit from $1$ to $0$ is $0$ units of time---the bit flip takes one unit
of time, but that is paid for by the credit.  Using these new amortized
costs, we can see that $T_{\texttt{inc}}(n)$ is $O(1)$ amortized: in the
case where the first bit is $0$, we flip it to $1$, which costs $2$
units of time, and stop. In the case where the first bit is $1$, we flip
it \emph{for free} to $0$, and then make a recursive call, which
inductively is bounded by 2. So $T_{\texttt{inc}}(n) = 2$, which means
that $T_{\texttt{set}}(n) = 2n$, amortized. Since a single run of
$\texttt{set}$ starts with no credits, its actual cost will be bounded
by the amortized cost $2n$: all of the credits spent during the call to
$\texttt{set}$, which subtract from the cost, must have been created
earlier, incurring a cost which balances out the gain garnered from
spending it.


Formalizing recurrence extraction for the banker's method for amortized
analysis requires us to move
from a relatively standard source language based on the simply-typed
$\lambda$-calculus with inductive datatypes to a more specialized one.
We do not expect amortization policies (e.g.\ generate a credit when
flipping a bit from 0 to 1, to be spent when flipping a bit from 1 to 0)
to be automatically inferable in the general case---these policies are the part
of an amortized analysis that requires the most cleverness.  To notate
these policies, we use an \emph{intermediate language} $\lambda^A$
(Section~\ref{sec:la}), which has ``effectful'' operations for
generating and spending credits ($\waitname$ and $\discname$), as well
as a modal type operator $!_\ell$ for associating credits with values
(e.g.\ storing a credit with each 1 in a bit list).  The type~$!_\ell A$
classifies a value of type $A$ that has $\ell$ credits associated with
it.  To correctly manage credits, this intermediate language is based on
a form of linear logic, which prevents spending the same credit more
than once; in particular, $\lambda^A$ is an affine lambda calculus with
all of the standard connectives $\otimes, \oplus, \&, \multimap, !$ plus
multiplicities $!^k A$ (where $k$ is a positive number) for tracking
multiple-use values.  The type structure of the intermediate language is
inspired by the credits (written as $\Diamond$) of
\cite{hofmann02diamonds,hofmann03diamonds-journal}, $n$-linear types
(e.g. \cite{girard-et-al:tcs92:bll,reed:names-useless,mcbride:plenty-o-nuttin,atkey:lics18}),
and the uses of credits and linear logic in in automatic amortized
resource analysis (AARA)
(e.g. \cite{hofmannjost03aara,hoffmann-et-al:toplas12:multivariate-amortized,knoth+19resourceguided}).

The target of the monadic translation is the \emph{recurrence
language}~$\lambda^{\bbbc}$, which,
following~\cite{danner-et-al:icfp15,hudson}, is a
standard simply-typed $\lambda$-calculus with a base type for costs
(linearity is not needed at this stage). It is equipped with an
inequality judgment $E \le_T E'$ that can be used to express upper
bounds.  The translation we define here extracts a recurrence for the
\emph{amortized} cost of the program (where the costs have been
``rearranged''), by translating the credit generation and spending
operations in $\lambda^A$ to modifications of the cost.  We define a
bounding relation (a cross-language logical relation) for the amortized
case, and prove that a term is related to its extraction.  As a
corollary, we obtain that the amortized cost of running a program from
$\lambda^A$ is bounded by the cost component of its translation into
$\lambda^{\bbbc}$; for programs that use no external credits, this gives
a bound on its actual cost as well.  The recurrence language, recurrence
extraction and bounding theorem are described in Section~\ref{sec:cl}.
Next, we use a denotational semantics of the recurrence language in
preorders, similar to~\cite{danner-et-al:icfp15}, to justify the
consistency of the recurrence language $\le$ judgment, and to simplify
and solve extracted recurrences (Section~\ref{sec:preorder}).

The version of $\lambda^A$ and the recurrence extraction presented
through Section~\ref{sec:preorder} allows a statically fixed number of
credits to be stored with each element of a data structure (e.g. 1
credit on element of a list, so $n$ credits overall).  For some
analyses, it is necessary to choose the number of credits stored with an
element dynamically.  For example, when analyzing
splay trees \cite{sleator-tarjan-85}, the number of
credits stored at each node in the tree is a function of the size of the
subtree rooted at that node, which varies for different tree nodes.  To
support such analyses, we extend $\lambda^A$
with existential quantifiers over credit variables
in Section~\ref{sec:ex}, and use them to code
a portion of \citet{okasaki:purely-functional-data-structures}'s
analysis of splay trees in our system.  

\begin{figure}[t]
  \vspace{-.25in}
  \tikzset
{
		line cap = butt ,
		line join = bevel ,
		arrows = -> ,
		> = angle 60 ,
		auto = left ,
		text depth = 0.25ex , 
		align = center ,
}

\begin{tikzpicture}
	\node (source) at (0,0) {Source \\ language};
	\node [draw] at ($(source) + (3.5,0)$ ) (affine) {Intermediate \\ language $\lambda^A$};
	\node [draw] at ($(affine) + (3.5,0)$ ) (syntactic) {Recurrence \\ language $\lambda^\bbbc$};
	\node [draw] at ($(syntactic) + (3.5,0)$) (semantic) {Semantic \\ recurrences} ;
	
	\draw[dotted] (source) to node [auto] {Annotate} (affine);
	\draw (affine) to node [auto] {$\norm{-}$} (syntactic);
	\draw (syntactic) to node [auto] {$\scott{-}$} (semantic);
	\draw[dotted, loop above] (semantic) to node [auto] {Solve} (semantic);
	
\end{tikzpicture}
  \caption{Recurrence Extraction Pipeline}
  \label{fig:pipeline}
\end{figure}
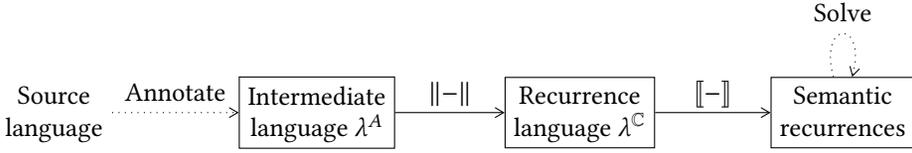

The process of extracting and solving a recurrence in diagrammed in
Figure~\ref{fig:pipeline}.
While automation of the annotation and solving steps
is a worthwhile goal (something we discuss in
Section~\ref{sec:future-work}),
our
main motivation in this paper is to formally justify the
extract-and-solve method for amortized analysis, a technique that we teach and that is
typically used by practitioners.  Connecting the extracted recurrence in
terms of user-defined notions of size to the operational cost is the
least justified step in this process, and so a formal account of it has
important foundational value.  It could likewise have important
practical value: because students and practitioners are trained in the
use of cost recurrences, reverse-engineering a recurrence that yields a
worse-than-expected cost bound to the (mis)implementation may require
a lower cognitive load than doing the same with more
sophisticated techniques.  Moreover, though this technique is less
automated than others, it can handle at least some examples that
existing techniques cannot---to our knowledge, splay trees cannot be
analyzed by the existing automatic techniques.
We give a detailed comparison with related work in Section~\ref{sec:related-work}.



\section{Intermediate Language \texorpdfstring{$\lambda^A$}{}}\label{sec:la}

In this section we discuss the static and operational semantics of
$\lambda^A$, which is an \emph{affine} lambda calculus---it permits
weakening (unused variables) but not contraction (duplication of
variables).  It includes some standard connectives of linear logic, such
as positive/eager/multiplicative products ($\otimes$ and $1$),
sums/coproducts ($\oplus$), and functions ($\loli$), as well as
negative/lazy/additive products ($\amp$).  The language has two basic
datatypes, natural numbers ($\N$) and (eager) lists ($\listty A$), both
with structural recursion (though we expect these techniques to extend
to all strictly positive inductive
types~\cite{danner-et-al:icfp15,danner-licata:jfp-in-prep}).

In addition to these, $\lambda^A$ contains some constructs specific to
its role as an intermediate language for expressing amortized analyses.
First, instead of fixing the operational costs of $\lambda^A$'s programs
themselves, we include a \texttt{tick} operation which costs 1 unit of
time, and assume that the translation of a program into $\lambda^A$ has
annotated the program with sufficient ticks to model the desired
operational cost~\cite{danielsson:popl08} (for example, we can
charge only for bit flips in the above binary counter
program).

Second, we have operations \waitname\ and \discname\ for creating and spending
credits, which respectively increase and decrease the
\emph{amortized} cost of the program \textit{without changing} the true
operational cost.

Third, we have a type constructor $!_\ell A$, where a value of this type
is a value of type $A$ with $\ell$ credits attached; its introduction
and elimination rules allow for the movement of credits around a
program.  The combination of of \discname\/ and the $!_\ell$ modality
motivates our affine type system: because spending credits decreases the
amortized cost of a program, we must ensure that a credit is spent only
once, so credits should not be duplicated; because credits can be stored
in values, values cannot in general be duplicated as well.  However,
$\lambda^A$ does allow credit weakening---choosing not to spend
available credits---because this increases the amortized cost (relative
to spending the credits), and we are interested in upper bounds on
running time.  While the basic affine type system allows a variable to
be used only once, to simplify the expression of programs that use a
variable a fixed number of times, we use $n$-linear types (see e.g.
\cite{girard-et-al:tcs92:bll,reed:names-useless,mcbride:plenty-o-nuttin,atkey:lics18}),
where variables are annotated with a multiplicity $k$, and can be used
at most $k$ times.\footnote{While Girard's notation for multiplicities
  is $!_k A$~\cite{girard-et-al:tcs92:bll}, we write superscripts
  following~\citet{atkey:lics18}, and write subscripts for the
  credit-storing modality, which is used more frequently in our system.}
This is internalized by a modality $!^k A$, which represents an $A$ that
can be used at most $k$ times.  We additionally allow $k$ to be
$\infty$, in which case $!^\infty A$ is the usual exponential of linear
logic, allowing unrestricted use.  Using this modality, standard
functional programs can be coded in $\lambda^A$, but our current
recurrence extraction does not handle the $!^\infty$ fragment very well,
as explained below---at present, we use $!^\infty$ mainly as a technical
device for typing recursors.  It is technically convenient to combine
the two modalities into one type former $!^k_\ell A$, which represents
an $A$ that can be used $k$ times, which also has $\ell$ credits
attached (total, not $\ell$ credits with each use).  Because $k$ is a
coefficient but $\ell$ is an additive constant, the individual
modalities are recovered as $!^k A := !^k_0 A$ and $!_\ell A := !^1_\ell
A$.  In pure affine logic, one can think of $!^k_\ell A$ as $X \otimes
\ldots \otimes X \otimes A \otimes \ldots \otimes A$ with $\ell$ $X$s
and $k$ $A$'s (in the case where $k$ and $\ell$ are finite), for an
atomic proposition $X$ representing a single credit.  However, our
judgmental presentation is easier to work with for our bounding relation
and theorem below, and the $n$-linear modality $!^k A$ ensures that
additional invariant that it is the \emph{same} value that can be used
$k$ times, i.e. it only allows the diagonal of $A \otimes \ldots \otimes
A$.

\begin{figure}
  \input{figs/la-bnf}
  \vspace{-0.2in}
  \caption{$\lambda^A$ Grammar}
  \label{fig:la-bnf}
\end{figure}

\subsection{Type System}

In Fig.~\ref{fig:la-ty-rules} we define
a typing judgment of the form
$\Gamma \vdash_f M : A$, where $\Gamma$ is a standard context $x_1:A_1,
x_2:A_2, \ldots, x_n : A_n$ and $f$ is a \textit{resource} term of the
form $a_1 x_1 + a_2 x_2 + \ldots + a_n x_n + \ell$, where
$x_1,\ldots,x_n$ are the variables in $\Gamma$ and $a_i$ and $\ell$ are
natural numbers or $\infty$.  The resource term $f$ can be
thought of as annotating each variable $x_i$ with the number of times
$a_i$ that it is allowed to occur, and additionally annotating the
judgment with a nonnegative ``bank'' $\ell$ of available credits
that can be used.  For example, the judgment $x : A, y : B, z : C
\vdash_{3x+2y+0z+2} M : D$, means that $M$ is a term of type $D$, which
may use $x$ at most $3$ times, $y$ at most twice, $z$ not at all, and
has access to $2$ credits.  We consider these resource terms up to the
usual arithmetic identities (associativity, unit, commutativity,
distributivity, $0 f = 0$, $\infty k = \infty$ otherwise, etc.).  In the
admissible substitution rule, we write $g[f/x]$ to denote the result of
normalizing the textual substitution of $f$ for $x$ in $g$ according to
these identities; e.g. $(3x+2y+2)[10a+11b+3/x] = 30a+33b+2y+11$. 
Our judgmental presentation of $n$-linear types differs from some 
existing ones-- the reader more familiar with Girard's BLL~\cite{girard-et-al:tcs92:bll}
may read $\Gamma \vdash_f M : A$ as analogous to $!_{\vec{f}} \Gamma \vdash M : A$
-- but this type system was derived as an instance of a general framework for modal
types~\cite{lsr}, which, for our purposes, simplifies the presentation of standard 
metatheorems like substitution. Note that
the resource terms $f$ play a different role than the resource
polynomials in Bounded Linear Logic and AARA~\cite{girard-et-al:tcs92:bll,hoffmann-et-al:toplas12:multivariate-amortized}, 
which provide a mechanism for measuring the size and credit allocation in a
data structure.  The resource terms are also affine in the sense of a
polynomial---the exponent of every variable is 1, except for the constant term
$\ell$---but we will avoid this meaning of affine to avoid
confusion with ``affine logic'' (allowing weakening but not contraction).

\begin{figure}
  \input{figs/la-ty-rules}
  \vspace{-0.2in}
  \caption{$\lambda^A$ Typing Rules}
  \label{fig:la-ty-rules}
\end{figure}

\subsubsection{Structural Rules.}

The rules make three structural principles admissible:

\begin{restatable}[Admissible structural rules]{theorem}{lastructural} \label{thm:la-structural}\hfill
  \begin{itemize}
\item Resource Weakening: Write $g \ge f$ for the coefficient-wise
  partial order on resource terms ($a_1 x_1 + a_2 x_2 + \ldots + \ell
  \ge$ $b_1 x_1 + b_2 x_2 + \ldots + \ell'$ iff $a_i \ge b_i$ for all
  $i$ and $\ell \ge \ell'$).  Then if $\; \Gamma \vdash_f M : A$ and $g
  \geq f$ then $\Gamma \vdash_g M : A$.

\item Variable Weakening:
If $\Gamma \vdash_f M : A$ and $y$ does not occur in $\Gamma$, then $\Gamma,y:B \vdash_{f+0y} M : A$.
  
\item
    Substitution: 
If $\Gamma \vdash_f M : A$ and $\Gamma, x : A \vdash_g N : B$, then
$\Gamma \vdash_{g[f/x]} N[M/x] : B$

  \end{itemize}
\end{restatable}
\begin{proof}
By induction on derivations.
\end{proof}

First, we can weaken the resource subscript, allowing more uses of a
variable or more credits in the bank (e.g.\ if $\cdot \vdash_3 M : A$,
then $\cdot \vdash_5 M : A$).  Second, we can weaken a context
to include an unused variable (we write $f+0y$ for emphasis, but by
equating resource terms up to arithmetic identities, this is just $f$).
Third, we can substitute one term into another, performing the
corresponding substitution on resource terms.  The idea is that, if $N$
uses a variable $x$ say $3$ times, then it requires 3 times the
resources needed to make $M$ to duplicate $M$ three times; this
multiplication occurs when substituting $f$ for the occurrence of $x$ in
$g$.

\subsubsection{Multiplicative/Additive Rules in $n$-linear Style.}
In the $n$-linear types style of presentation, rules of linear logic
that traditionally split the context (e.g. $\otimes$ introduction,
$\loli$ elimination) sum the resources used in each premise, but keep
the same underlying variable context $\Gamma$ in all premises.  For
example, in a positive pair $(M,N) : A \otimes B$, if $M$ is allowed to
use $x$ 3 times and $N$ is allowed to use $x$ 4 times, then the whole
pair must be allowed to use $x$ 7 times.  As a special case, if a
variable is not allowed to occur in, e.g., $N$, it can be marked with a
coefficient of 0.  On the other hand, rules for additives (e.g. pairing
for $A \& B$) use the same resource term in multiple premises.  While
the elimination rule for $\oplus$ is additive in sequent calculus style,
in natural deduction there is some summing because it builds in a cut
for the term being case-analyzed.

\subsubsection{Ticks, and Creating/Spending Credits.}\label{ssec:wdt}

The $\texttt{tick} \; ; \; M$ construct is used to mark program points that
are intended to incur one unit of time cost (e.g.\ bit flips in the binary
counter example); it uses the same resources as $M$.  

$\waitname$ is the means to create credits, where $\waitname_\ell$
gives $M$ access to $\ell$ extra credits to use, along with whatever
resources are present in the ambient context; formally, this is
represented by adding to the ``bank'' in the premise of the typing rule
for $M$.  In the operational semantics and recurrence extraction below,
\waitname\/ adds $\ell$ steps to the amortized cost of $M$---it is used to
``prepay'' for later costs.  

$\discname$ is the means to spend credits, where $\discname_\ell$ spends
$\ell$ credits; because credits can only be spent once, these $\ell$
credits in the conclusion of the typing rule are not also available in
the premise for $M$.  In the operational semantics/recurrence
extraction, \discname\/ subtracts $\ell$ steps from the amortized cost of
$M$---it is used to take advantage of prepaid steps.  Note that
\discname\/ satisfies the same typing judgments as an instance of
resource weakening (because $f + \ell \ge f$); the ``silent'' weakening
does not change the amortized cost, but instead is a case where our recurrence extraction might
obtain a non-tight upper-bound.

\subsubsection{$!^k_\ell$ Modality.}
Instead of having two separate modalities, one for $n$-use types and the
other for types storing credits, we combine them into a single modality
$!^k_\ell A$. A value of type $!^k_\ell A$ is a $k$-use $A$ with $\ell$
credits attached (not $k \cdot \ell$ credits, which is what one would
expect if each use had $\ell$ credits attached---though that could be
modeled by the type $!^k_0 (!^1_\ell A)$).  While we write $a$ and
$\ell$ for nonnegative numbers or $\infty$, we restrict $k$ to range
over a \emph{positive} number or $\infty$ -- i.e. we do not allow a
``zero-use'' modality $!^0_\ell A$, which would complicate the
erasure of $\lambda^A$ to regular simply typed lambda calculus.

The introduction rule for $!^k_\ell$ says that if we can prove $M$ has
type~$A$ with~$f$ resources, then a version of~$M$ that can be used $k$
times requires $kf$ resources.  If in addition, $\ell$ credits are to be
attached, then $kf+\ell$ resources are required.  Intuitively, one can
think of $\save k \ell M$ as the act of running $M$ once to obtain its
value, but repeating whatever requirement it imposes on the bank $k$
times, which justifies making $k$ uses of its value, and then attaching
$\ell$ credits to this value.  In order to make resource weakening
admissible in general, it is necessary to build weakening into this
rule (the second premise).

The elimination rule for the modality allows for the credit stored
on a term to be released into the ambient context of another in order to
be redistributed or spent. We first present a simplified version, and
then explain the general version. Given $\Gamma \vdash_f M : !^k_\ell
A$, we essentially have $k$ copies of an $A$, along with $\ell$ extra
credits. Given a term $N$ which can use $k$ copies of an $A$ and $\ell$
credits, $\Gamma,y : A \vdash_{ky + \ell} N : C$, we can form the term
$\Gamma \vdash_{f} \texttt{transfer} \, !^k_\ell y = M \; \texttt{to} \;
N : C$, which, intuitively, deconstructs $M$ into its $k$-usable value
and $\ell$ credits, and moves them to $N$, where they can be used. On
top of this version, we make two modifications. Firstly, $N$ should have
access to resources other than just what's provided to it by $M$-- so we
add a resource term $g$ available in $N$ (and
therefore required to type the
$\texttt{transfer}$). Secondly, it may be necessary at the site of the
transfer to further duplicate the $M : !^k_\ell A$ --- this is required to
prove a fusion law below, for example.  To support this,
we parameterize the $\texttt{transfer}$ term
by another number, $k'$, arriving at the version of the rule
presented in Figure~\ref{fig:la-ty-rules}, which should be thought of as
eliminating $k'$ copies of a $!^k_\ell A$ at once.
The rules for other positive types ($\oplus,\otimes$) similarly permit elimination of multiple copies at
once.

The $!$ modality satisfies the following interactions with other logical
connectives, where we write $A \dashv \vdash B$ to mean
interprovability/functions in both directions:

\begin{restatable}[Fusion Laws]{theorem}{fusion}\hfill
\label{thm:fusion}
\begin{enumerate}
  \item $!^{k_1k_2}_{\ell_1 + k_1 \cdot \ell_2} A \dashv \vdash \, !^{k_1}_{\ell_1} !^{k_2}_{\ell_2} A$
  \item $!^k_{\ell_1 + \ell_2} (A \otimes B) \dashv \vdash \, !^k_{\ell_1} A \otimes !^k_{\ell_1} B$
  \item $!^k_\ell (A \oplus B) \dashv \vdash \, !^k_\ell A \oplus !^k_\ell B$
\end{enumerate}
\end{restatable}

\subsubsection{Natural Number Recursor} \label{sec:ns-rules}

For natural numbers, while the rules for zero and successor are standard,
the recursor takes a bit of explanation.  We think of
the recursor \texttt{nrec} as a function constant of type
$\N \loli (1 \loli C) \loli !^\infty_0(\N \times (1 \loli C) \loli C) 
 \loli C$.
The base case is ``thunked'' because we think of $\loli$ as
a call-by-value function type, but the base case should not be evaluated until the recurrence
argument is~$0$.  The ordinary type for the step function (inductive case)
would be $(\N \times C \loli C)$, but we also suspend the recursive
call, to allow for a simple case analysis that chooses not to use the
recursive call.  
The $!^\infty_0$ modality surrounding the step function is needed to
ensure that the step function itself does not use any ambient credits,
which is necessary because the step function is applied repeatedly by
the recursor ($n$ times if the value of $M$ is $n$).  Without this
restriction, one could, for example, iterate a step function that spends
$k$ credits to subtract $Mk$ credits from the amortized cost, while only
having $k$ credits in the bank to spend.  For example, without the use of $!^\infty_0$, the term
$\cdot\vdash_1\nrec 7 {\lambda\_.0}{\lambda\_.\disc 1 0} : \N$ typechecks
with only one credit in the ambient bank, but 
intuitively subtracts 7 from the amortized cost, rather than just the
1 credit that was allowed.  
We solve this problem using the type $!^\infty_0 A$ (where $A$ is the ordinary
type of the step function $\N \otimes (1 \loli C) \loli C$),
which represents an infinitely duplicable $A$ that stores no additional credits.
Being infinitely duplicable is an over-approximation, because
the step function really only needs to be run $M$ times, but
being more precise would require reasoning about such values in the
type system.

In the common case, the step function will use other
infinite-use variables but no credits from the bank.  A typical 
typing derivation for this case, where $H$ is the type of a helper
function and $A$ is the type of the step function, would be
\[
\infer{f : H \vdash_{\infty (\infty f) = \infty f} \save{\infty}{0}{N_2'} : !^\infty_0 A}
      {f : H \vdash_{\infty f} N_2' : A}
\]
Using this as the third premise of the typing rule of \texttt{nrec}, we
see that such an \texttt{nrec} itself requires only the credits demanded
by the number argument ($M$) and base case ($N_1$), assuming $f$ is
substituted by a helper function that uses no credits.

The way in which the $!^\infty$ modality ``prevents'' the use of credits
from the bank is somewhat subtle: a step function \emph{can} use credits
from the bank, but this will require the bank to be infinite in the
conclusion.  This is because the introduction rule for $!^\infty_0$ 
inflates any finite resources to $\infty$ in the conclusion:
\[
\infer{f : H \vdash_{\infty(2f + 3) = \infty f + \infty} \save{\infty}{0}{N_2'} : !^\infty_0 A}
      {f : H \vdash_{2x + 3} N_2' : A}
\]
Thus, the step function is only permitted to use credits from the bank
when the bank has $\infty$ credits in the conclusion, while we are
generally interested in programs that use finitely many credits.

\subsubsection{List Recursor}
The list recursor $\lrec {M} {N_1} {N_2}$ has the same ``credit
capture'' problem as the recursor on naturals, which we solve using
$!^\infty_0$.  The list recursor has another challenge, though,
because unlike a natural number, the values of the list can themselves
store credits.  Because of this, to prevent credits from being
duplicated, in the cons case, the recursor may use \emph{either} the
tail of the list or the recursive result, but not both.  We code this
using an internal choice/negative product $\&$.  The negative product
will itself be treated as a lazy type constructor, where an $A \with B$
pair is a value even when the $A$ and $B$ are not, so we do not need
to further thunk the recursive result $C$ here.

\subsection{Operational Semantics for \texorpdfstring{$\lambda^A$}{the
intermediate language}} \label{ssec:la-sem}

We present a call-by-value big-step operational semantics for
$\lambda^A$ in Figure~\ref{fig:la-sem-rules}, whose primary judgment
form is $M \downarrow^{(n,r)} v$, which means that $M$ evaluates to the
value $v$ with cost $(n,r)$.  The first component of the cost, $n$ (a
non-negative number) indicates the \textit{real cost} of evaluating $M$,
in this case the number of $\texttt{ticks}$ performed while evaluating
$M$.  The second component, $r$ (which can be any integer), tracks
$\waitname$s and $\discname$s --- the (possibly negative) sum total of
credits created and spent while evaluating $M$, where creating is
positive and spending is negative.  The \emph{amortized cost} of
evaluating $M$ is $n + r$: the number of ``actual" steps taken, plus the
number of credits created, minus the number spent.

One reason we separate $n$ and $r$ in the judgment form is that there is
a straightforward \emph{erasure} of $\lambda^A$ to ordinary simply typed
$\lambda$-calculus (STLC with a \texttt{tick} operation), in which
evaluating the STLC program has cost (number of ticks) $n$.  Briefly,
this translation translates $!^k_\ell A$ to $A$, translates all of the
linear connectives to their unrestricted counterparts, drops all
\waitname, \discname, \texttt{save} term constructors, and translates
\texttt{transfer} to a \texttt{let}.  The definition of $n$ in each of
our inference rules for $M \downarrow^{(n,r)} v$ is the same as the
usual cost for STLC with a tick operation, so this erasure preserves
cost.  Because of this erasure, the $n$ in $M \downarrow^{(n,r)} v$ is a
meaningful cost to bound. Further, the distinction between $n$ and $r$
is why we have separate terms $\waitname$ and $\texttt{tick}$:
$\texttt{tick}$ increases the operational cost which should be preserved
under erasure, while $\texttt{create}$ increase the amortized cost only.

\begin{figure}
  \input{figs/la-sem-rules}
  \caption{$\lambda^A$ Operational Semantics}
  \label{fig:la-sem-rules}
\end{figure}

As discussed in
Section~\ref{ssec:wdt}, $\wait \ell M$ creates $\ell$ credits for $M$ to
use for the price of $\ell$ units of time cost, whereas
\discname\/ subtracts from the amortized cost of an expression ---
a speedup which is paid for by the $\ell$ credits which the body is no
longer allowed to use.  Both are reflected by corresponding changes
to~$r$.


The operational intuition for $\save{k}{\ell}{M} : \: !^k_\ell A$ is that
it runs $M$ once, but repeats whatever effect this had on the credit
bank $k$ times, which justifies using the credits in the value of $M$
$k$ times.  (The erasure to STLC discussed above runs $M$ only once, not
$k$ times---which would be challenging when $k$ is $\infty$.)  Formally,
this means that the $n$ in the conclusion is just the $n$ in the
premise, but the $r$ is multiplied by $k$.  Running
$\texttt{save}^k_\ell$ does \textit{not} add $\ell$ to the $r$ component
because \texttt{save} does not create credits (adding to the amortized
cost), but only attaches some already existing credits to the value $v$.
Recall that \texttt{transfer} detaches the credits from a $!^k_\ell$
value, and allows for them, along with the $k$ copies of the value, to
be used in another term. The evaluation rule says that, in order to
evaluate $\tsfer {k'} k \ell x M N$, we first evaluate $M$ to a
\texttt{save} value, and then evaluate the substitution instance
$N[v_1/x]$. The $k'$ in \texttt{transfer} means to repeat the evaluation
of $M$ $k'$ times, allowing $k \cdot k'$ uses in the body of $N$, so
this (similarly to \texttt{save}) repeats the credit effects $r_1$ of
$M$ $k'$ times in the conclusion.  The other positive elimination forms are similar.

\subsection{Syntactic Properties}


In the operational semantics judgment $M \downarrow^{(n,r)} v$, we
think of $n + r$ (the actual cost $n$ plus the credit difference $r$) as
the amortized cost of the program.  A key property of amortized analysis
is that the amortized cost is an upper bound on the true cost, which
means in this case that $n + r \ge n$, so we would like $r \ge 0$.
While $r$ is in general allowed to be a negative number, it is
controlled by the credits $a$ of the typing judgment $\cdot \vdash_a M
: A$, intuitively because it is only \discname\/ operations that
subtract from $r$, and \discname\/ operations are only allowed when the
type system deems there to be sufficient credits available.  Thus, we
will be able to prove that $r \ge 0$ for well-typed terms.  To do so,
we strengthen the induction hypotheses to prove that $\cdot \vdash_a M :
A$ and $M \downarrow^{(n,r)} v$ imply $a + r \geq 0$, which gives $r
\geq 0$ for closed programs that use no external credits (so $a = 0$),
which is what a ``main'' function is expected to be (e.g. \texttt{set}
in the binary counter example).  It is technically convenient to combine
this with a preservation result, stating that the credits of $v$ is in
fact $a + r$ (the resource term in a typing judgment must be
non-negative, so $a + r \geq 0$ is in fact a prerequisite for even
asserting that $\cdot \vdash_{a+r} v : A$).  The proofs of the following are
relatively straightforward and may be found
\begin{icfp2020}in the full version of this
paper~\citep{cutler-et-al:icfp2020-full}.\end{icfp2020}
\begin{arxiv}in Appendix~\ref{sec:appendix}.\end{arxiv}

\begin{restatable}[Preservation Bound]{theorem}{pres}
\label{thm:pres}
If $\cdot \vdash_a M : A$ and $M \downarrow^{(n,r)} v$, then $a + r \geq 0$ and $\cdot \vdash_{a + r} v : A$. 
\end{restatable}

We also have that values evaluate in 0 steps:
\begin{restatable}[]{theorem}{valevalzero}\label{thm:val-eval-none}
If $v$ is a value, and $v \downarrow^{(n,r)} v$, then $n = r = 0$.
\end{restatable}

and that values of type $\N$ contain no credits:
\begin{restatable}[Resource strengthening for $\N$]{theorem}{natstren}\label{thm:nat-strengthening}
If $\cdot \vdash_a v : \N$, then $\cdot \vdash_0 v : \N$
\end{restatable}

\subsection{Binary Counter Annotation}

\begin{figure}
  \input{figs/bc-term}
  \vspace{-0.15in}
  \caption{Binary Counter Terms in $\lambda^A$}
  \label{fig:bc-term}
\end{figure}

As an example, we translate the binary counter program from
Section~\ref{sec:intro} to $\lambda^A$, decorating the program with
\createname, \spendname, \savename, and \xfername\/ in order to emulate the
analysis described in Section~\ref{sec:intro}.  Since the analysis
stores credits on 1 bits, the type of bits is $\texttt{bit} = 1 \oplus
!^1_1 1$; a value $\inl {(\,)}$ represents a $0$ bit, and a value $\inr
{(\save 1 1 {(\, )})}$ represents a $1$~bit, with a credit attached. A
binary number is represented as a list of bits, $\listty
{\texttt{bit}}$.
The cost of interest is the number of bit flips, so 
we insert $\texttt{tick}$s everywhere a bit is flipped from
$0$ to $1$ or vice versa. Next, to handle the credits, we
$\createname$ and subsequently $\texttt{save}$ a credit when we
flip a bit from $0$ to $1$, and $\xfername$ then $\spendname$
when flipping bits from $0$ to $1$.
This annotation is shown in Figure~\ref{fig:bc-term} -- for simplicity, we use \texttt{inc} as a meta-level name for the term
implementing the function, so its occurrence in \texttt{set} really means
a copy of that entire term (to do this at the object level, we could
alternatively think of a top-level definition of \texttt{inc} as binding
an infinite-use variable).


\section{Recurrence Language \texorpdfstring{$\lambda^\bbbc$}{}, Amortized Recurrence Extraction, and Bounding Theorem} \label{sec:cl}

Next, we define a translation from $\lambda^A$ into a \emph{recurrence
  language} $\lambda^{\bbbc}$. Unlike $\lambda^A$, $\lambda^{\bbbc}$ has
a fully structural (weakening and contraction) type system, and no
special constructs for amortized analysis (it is mostly unchanged from
\cite{danner-et-al:icfp15,hudson}). Further, because we view $\lambda^\bbbc$
as a syntatx for mathematical expressions, it is designed as a call-by-name language--
this is in contrast to $\lambda^A$, which is by-value.
The recurrence translation takes a function in
$\lambda^A$ to a function that outputs the original function's cost in
$\lambda^\bbbc$, using a cost type $\bbbc$ (which we will often
take to be integers).  Formally, $\bbbc$ can be any commutative ring
with an $\infty$ element, the typical example being the (``tropical'')
max-plus ring on the integers, i.e. integers with addition and binary
maxes.  Some of the typing rules for $\lambda^\bbbc$ are presented in
Figure~\ref{fig:lc-rules}.

Relative to our previous work, the main conceptual change for supporting
amortized analysis is that, instead of extracting recurrences for the
true cost of a program ($n$ in $M \downarrow^{(n,r)} v$), we extract
recurrences that given an upper bound on the program's amortized cost $n
+ r$, which is itself a bound on the true cost for programs which begin
with an empty bank of credits.

\begin{figure}[h]
  \input{figs/lc-rules}
  \vspace{-0.25in}
  \caption{Recurrence Language $\lambda^\bbbc$ Definition}
  \label{fig:lc-rules}
\end{figure}

\subsection{Monadic Translation from \texorpdfstring{$\lambda^A$}{intermediate
language} to \texorpdfstring{$\lambda^\bbbc$}{recurrence language}}\label{ssec:mt}
\label{sec:monadic-translation}

Following~\cite{danner-et-al:plpv13,danner-et-al:icfp15}, a function $A
\loli B$ in $\lambda^A$ will be translated to a function \mbox{$\angles{A}
\to \bbbc \times \angles{B}$}, where for a $\lambda^A$ type $A$, a value of
$\lambda^\bbbc$ type $\angles{A}$ represents the size of a value in
$\lambda^A$.  Intuitively, this means that a function in $\lambda^A$ is
translated to a $\lambda^{\bbbc}$ function that, in terms of the size of the
input, gives the cost of running the function on that argument and the size
of the output.  Generalized to higher-type, ``size'' is properly viewed as
``use-cost;'' it is a property that tells us how the value affects the cost
of a computation that uses it.  In an unfortunate terminological clash,
prior work~\cite{danner-royer:ats-lmcs} refers to this concept as
\emph{potential} (as in ``potential cost'' or ``future cost''), with no
intentional connotation of potential functions from the physicist's method
of amortized analysis.  In order to keep this work consistent with the
sequence of papers it follows, and since $\lambda^A$ is based on the
banker's method, we will only use ``potential" to refer to the use-cost of a
value, and so call $\angles{A}$ the \emph{potential type} for~$A$ and a
value of type~$\angles A$ a \emph{potential}.  The size of the output is
needed for the translation to be compositional: the recurrence extracted for
a term should be composed of the recurrences extracted for its subterms, but
the cost of e.g.\ a function application depends on the size of the argument
itself, not just its cost.  A recurrence extraction of this form can be
packaged as a monadic translation into the writer monad $\bbbc \times A$.  

As discussed in Section~\ref{sec:intro}, the proper notion of size for a
specific datatype may vary from analysis to analysis. To this end, we
follow~\cite{danner-et-al:icfp15} in deferring the
abstraction of values as sizes to denotational semantics of
$\lambda^\bbbc$ defined in Section~\ref{sec:preorder}, which allows the
same recurrence extraction and bounding theorem to be reused for
multiple models with different notions of size.

We call the pair of a cost and a potential a \textit{complexity}.  The
translation consists of three separate functions, the definitions of
which are shown in Figure~\ref{fig:rec-extr}. Firstly, $\angles{\cdot}$
takes a type $A$ in $\lambda^A$ and maps it to the type $\angles{A}$
whose elements are the potentials of type $A$. We extend this to contexts pointwise:
$\angles{\Gamma,x : A} = \angles{\Gamma},x:\angles{A}$.
The second is $\norm{A}
:= \bbbc \times \angles{A}$, which takes a type $A$ to the corresponding
type of complexities. Finally, we overload $\norm{\cdot}$ to denote the
recurrence extraction function from terms of $\lambda^A$ to terms in
$\lambda^\bbbc$.  For convenience, when $E : \bbbc \times T$, we often
write $\pi_1 E$ as $E_c$ (cost) and $\pi_2 E$ as $E_p$ (potential).
\footnote{We regard the subscript notation as binding tighter than
ordinary projection: i.e. $\pi_1E_p = \pi_1(E_p)$.
}
We
also use special notation for adding a cost to a complexity, writing
$E+_c E'$ for $(E + E'_c,E'_p)$ when $E : \bbbc$ and $E' : \bbbc \times
T$.

\begin{figure}
  \input{figs/rec-extr}
  \caption{Recurrence Extraction}
  \label{fig:rec-extr}
\end{figure}

Overall, the idea is that a term is translated to a function from
potentials of its context to complexities of its type:
\begin{restatable}[Extraction Preserves Types]{theorem}{extrsound}\label{thm:extr-sound}
If $\Gamma \vdash_a M : A$ then $\angles{\Gamma} \vdash \norm M : \norm A$
\end{restatable}

We comment on some of the less obvious aspects of this translation:

\begin{itemize}
  \item $!^k_\ell A$: The type translation erases the $!^k_\ell$ modality. 
  
  \item $A \amp B$: Since the negative product in $\lambda^A$ is lazy, a
    value of type $A \amp B$ is a pair of un-evaluated terms. Thus, the
    potential of a term of type $A \amp B$ must include the cost of
    evaluating each term, since that will factor into the cost of
    using such a value.
  
  \item $\texttt{tick}$: Since $\tick M$ evaluates with (true cost and)
    amortized cost $1$ higher than $M$'s, the cost component of
    $\norm{\tick M}$ is $1 + \norm{M}_c$.

  \item $\texttt{save}^k_\ell$: The extracted amortized cost of $\save k
    \ell M$ is $k$ times the extracted cost of $M$, with the potential
    remaining the same.  This is in principle a non-exact bound, because we
    are conceptually multiplying the operational amortized cost of $M
    \downarrow^{(n,r)} v$, which is $n + r$, by $k$, whereas the
    operational semantics gives the more precise $n + k r$.  We view
    this as a consequence of the fact that amortized analyses extract
    recurrences for the amortized cost $n+r$, rather than $n$ and $r$
    separately. However, this inflation is not a
    problem for our uses of $!^\infty$ in typing recursors because the
    branches of the recursor are usually values, which have 0 cost, and
    $\infty \times 0 = 0$. In future work, we might consider a recurrence
    translation into the $\bbbc \times \bbbc \times A$ monad, with
    separate extractions of $n$ and $r$, if more precision is needed.
    This would allow for $\lambda^A$ to be used in the place of the
    (linear fragment) of the source language in previous 
    work~\cite{danner-et-al:icfp15}. Embedding that language into the $!^\infty$
    fragment of $\lambda^A$ and then extracting recurrences into 
    $\bbbc \times \bbbc \times A$ would yield the same results as
    applying the non-amortized recurrence extraction. We emphasize
    that the loss of precision from not making this change has no bearing
    on \textit{amortized} algorithm analyses, it would only
    allow for \textit{non-amortized} analyses to also be performed
    with $\lambda^A$-- but such analyses are already handled by prior work \cite{danner-et-al:icfp15,kavvos-et-al:popl20}
    
  \item $\texttt{transfer}$: A similar imprecision arises with respect
    to the multiplicity $k'$ here, but otherwise $\texttt{transfer}$ is
    translated like a \texttt{let}.  

%
%
  \item $\texttt{nrec}$: As in the operational semantics, because we
    think of the recursor as a call-by-value function constant, some
    cost is in principle incurred for evaluating the branches to
    function values, though the branches are usually values in practice.

  \item \sloppypar $\texttt{lrec}$: The type of the step function in a
    list recursor is $!^\infty_0(A \otimes (\listty A \amp C) \loli C)$,
    and the potential translation of this type is
    \mbox{$\angles{A} \times \left(\left(\bbbc \times \listty {\angles
      A}\right) \times \left(\bbbc \times \angles C\right)\right) \to
    \bbbc \times \angles C$}. However, this does not match the
    required type of the step function of the list recursor in
    $\lambda^\bbbc$, which must be $T_1 \times (\listty {T_1} \times
    T_2) \to T_2$.  Taking $T_1 = \angles{A}$ and $T_2 = \bbbc \times
    \angles C$, the translation of the step function additionally
    requires a $\bbbc$ input representing the cost of the tail of the
    list.  However, lists are eager, so the step function is always applied
    to a value, so we can supply $0$ cost here.
\end{itemize}

\subsection{Recurrence Language Inequality Judgment}\label{sec:so}

$\lambda^\bbbc$ has a syntactic inequality judgment $\Gamma
\vdash E_1 \leq_T E_2$ (Figure~\ref{fig:syn-ord}), which intuitively means
that the recurrence $E_1$ is bounded above by $E_2$.  For now, we
include only those inequalities that are necessary to prove the
bounding theorem; this allows for the most models of the
recurrence language, and additional axioms valid in particular models
can be added in order to simplify recurrences syntactically.  The
necessary axioms are congruence in the principal positions of
elimination forms, as well as the fact that $\beta$-reducts are bounded
above by their redexes.  We often omit the context and type subscript
from $\Gamma \vdash E_1 \leq_T E_2$, writing $E_1 \leq_T E_2$ or $E_1
\leq E_2$, though formally it is a relation on well-typed terms in
context. This relation is primarily a technical device to provide closure
properties for the bounding relation. Because of this, we omit a more lengthy
discussion of the relation here, and refer the reader to the prior work
\cite{danner-et-al:icfp15} which introduces this type of relation.

\begin{figure}
  \input{figs/syn-ord}
  \vspace{-0.25in}
  \caption{Syntactic Ordering on $\lambda^\bbbc$}
  \label{fig:syn-ord}
\end{figure}

\subsection{Bounding Relation and Its Closure Properties}

The correctness of the recurrence extraction is stated in terms of a
logical relation between terms in $\lambda^A$ and terms in
$\lambda^\bbbc$. The intended meaning is that the $\lambda^\bbbc$
recurrence term is an upper bound on the $\lambda^A$ term's cost and
potential.

\begin{definition}[Bounding Relation] \label{def:bounding}
When $\cdot \vdash_a M : A$ and $\cdot \vdash E : \norm{A}$, then $M \bdby^{A,a} E$ if and only if, when $M \downarrow^{(n,r)} v$,
\begin{itemize}
  \item $n \leq E_c - r$
  \item $v \valbd^{A,a+r} E_p$
\end{itemize}
When $\cdot \vdash_a v : A$ and $\cdot \vdash E : \angles A$, we define $v
\valbd^{A,a} E$ by induction on $A$.
\begin{itemize}
    \item $\save k \ell v \valbd^{!^k_\ell A,c} E$ if there exists $d \geq 0$ so that $kd + \ell \leq c$, and $v \valbd^{A,d} E$
    \item $\lambda x.M \valbd^{A \loli B, c} E$ if whenever $v \valbd^{A,d} E'$, we have that $M[v/x] \bdby^{B,c+d} E \; E'$
    \item $(v_1,v_2) \valbd^{A_1 \otimes A_2,a} E$ if there are $a_1,a_2$ such that $a_1+a_2 = a$ and $v_i \valbd^{A_i,a_i} \pi_i E$ for $i \in \{1,2\}$
    \item $[] \valbd^{\listty A,a} E$ iff $[] \leq_{\listty {\angles A}} E$
    \item $\cons {v_1} {v_2} \valbd^{\listty A,a} E$ iff there are $E_1,E_2$ with $\cons {E_1} {E_2} \leq_{\listty {\angles A}} E$, and there are $a_1,a_2$ such that $a_1 + a_2 = a$ such that $v_1 \valbd^{A,a_1} E_1$ and $v_2 \valbd^{\listty A, a_2} E_2$.
    \item $0 \valbd^{\N,a} E$ iff $0 \leq E$
    \item $S(v) \valbd^{\N,a} E$ iff there is some $E'$ such that $S(E') \leq_\N E$, and $v \valbd^{\N,a} E'$
    \item $\inl v \valbd^{A \oplus B,a} E$ if there exists $E'$ such that $\inl E' \leq_{\angles A} E$ and $v \valbd^{A,a} E'$.
    \item $\inr v \valbd^{A \oplus B,a} E$ if there exists $E'$ such that $\inr E' \leq_{\angles B} E$ and $v \valbd^{B,a} E'$.
    \item $() \valbd^{1,a} E$ if $() \leq_1 E$.
    \item $\amppair M N \valbd^{A \amp B,a} E$ if $M \bdby^{A,a} \pi_1 E$, and $N \bdby^{B,a} \pi_2 E$.
\end{itemize}
We extend the value bounding relation to substitutions pointwise: $\theta \subbd^{\Gamma,\sigma} \Theta$ if for all $x : A \in \Gamma$, $\theta(x) \valbd^{A,\sigma(x)} \Theta(x)$. Finally, we define the bounding relation for open terms: when $\Gamma \vdash_f M : A$, we say that $M \bdby E$ if for all $\theta \subbd^{\Gamma,\sigma} \Theta$, we have $M[\theta] \bdby^{A,f[\sigma]} E[\Theta]$.
\end{definition}

The \emph{term/expression bounding relation} $M \bdby^{A,a} E$ says
first that the cost component of $E$ is an upper bound on the amortized
cost of $M$, which is $n + r \leq E_c$ (since we will eventually be
interested in bounding the actual cost of evaluating $M$, we write this
as $n \leq E_c - r$).  Additionally, expression bounding says that the
potential component of $E$ is an ``upper bound'' on the value that $M$
evaluates to; this is expressed via a mutually-defined type-varying
\emph{value bounding relation} $M \valbd^{A,a} E$.  The value bounding
relation is defined first by induction on the type $A$, and the cases
for natural numbers and lists have a local induction on the number/list
value as well.\footnote{In general, it is necessary to define the
  relations for inductive types inductively~\cite{danner-et-al:icfp15},
  but the values of $\N$ and $\listty{A}$ are simple enough that
  induction on values suffices here.}  We write the credit bank $a$ as a
parameter of the bounding relations, but it is a presupposition that
this number is the same one that was used to type check $\cdot \vdash_a
\{M,v\} : A$ (because the bounding relation is on closed terms, the resource
subscript is just a single number $a$).  


We extend the bounding relation to open terms by considering all closing
substitutions: a term $\Gamma \vdash_f M : A$ is bounded by $E$ if for
every substitution $\theta$ which is bounded pointwise by $\Theta$ with
some credit function $\sigma$, then the closed term $M[\theta]$ is
bounded by $E[\Theta]$ with $f[\sigma]$ credits.  In this definition,
$\sigma$ gives a number of credits $a_i$ for each variable $x_i$,
because $\theta$ is a substitution of closed terms for variables $(\cdot
\vdash_{a_1} v_1 : A_1) / x_1, (\cdot \vdash_{a_2} v_2 : A_2) / x_2,
\ldots$.

\subsection{Bounding Theorem}

As usual for a logical relation, we first require some lemmas about the
bounding relation, before a main loop proving the fundamental theorem
that terms are related to their extractions.  The proofs of the following
theorems can be found
\begin{icfp2020}in the full version of this
paper~\citep{cutler-et-al:icfp2020-full}.\end{icfp2020}
\begin{arxiv}in Appendix~\ref{sec:appendix}.\end{arxiv}

First, we have an analogue of Theorem~\ref{thm:nat-strengthening}:

\begin{theorem}[$\N$-strengthening]
For all $\cdot \vdash_a v : \N$, if $v \valbd^{\N,a} E$, then $v \valbd^{\N,0} E$.
\end{theorem}

Second, we can weaken a bound by recurrence language inequality:

\begin{restatable}[Weakening]{theorem}{weakening} \hfill
\label{thm:weakening}
\begin{enumerate}
    \item If $M \bdby^{A,a} E$, and $E \leq_{\norm{A}} E'$, then $M \bdby^{A,a} E'$
    \item If $v \valbd^{A,a} E$, and $E \leq_{\llangle A \rrangle} E'$, then $v \valbd^{A,a} E'$
\end{enumerate}
\end{restatable}

Next, we have an analogue of resource weakening in
Theorem~\ref{thm:la-structural}:

\begin{restatable}[Credit Weakening]{theorem}{credwkn}
If $a_1 \leq a_2$, then:
\begin{enumerate}
  \item[(1)] If $M \bdby^{A,a_1} E$, then $M \bdby^{A,a_2} E$
  \item[(2)] If $v \valbd^{A,a_1} E$, then $v \valbd^{A,a_2} E$ 
\end{enumerate}
\end{restatable}

Next, we have inductive lemmas that will be used in the recursor cases
of the fundamental theorem:

\begin{restatable}[$\N$-Recursor]{theorem}{nreclemma}
\label{thm:nrec-lemma}
If $\lambda x.N_1' \valbd^{1 \loli C,c_3} E_1$, $\lambda x.N_2' \valbd^{\N
\otimes (1 \loli C) \loli C,d} E_2$ with $d \geq 0$, then $\forall n \geq 0$, if $\inj n \valbd^{\N,0} E$, then $\nrec {\inj n} {\lambda x.N_1'} {\save \infty 0 (\lambda x.N_2')} \bdby^{C,c_3 + \infty \cdot d} \nrec E {E_1} {\lambda p. E_2 \; (\pi_1 p, \lambda z.\pi_2 p)}$
\end{restatable}

\begin{restatable}[$\listty A$-Recursor]{theorem}{lreclemma}
\label{thm:lrec-lemma}
If $\lambda x.N_1' \valbd^{1 \loli C,c_1} E_1$ and $\lambda x.N_2'
\valbd^{A \otimes (\listty A \amp C) \loli C,c_2} E_2$, then for all
values $\cdot \vdash_d v : \listty A$ such that $v \valbd^{\listty A,d}
E$, we have that\\
$\lrec v {\lambda x.N_1'} {\save \infty 0 {(\lambda x.N_2')}} \bdby^{C,c_1+d + \infty \cdot c_2} \lrec E {E_1} {\lambda x. E_2 (\pi_1 x,((0,\pi_1 \pi_2 x),\pi_2\pi_2 x)) }$
\end{restatable}

Using these, we prove the main result:

\begin{restatable}[Bounding Theorem]{theorem}{bounding}
\label{thm:bounding}
If $\Gamma \vdash_f M : A$, then $M \bdby^A \norm{M}$
\end{restatable}

Finally, for terms that use no external credits, the true cost is
bounded by the extracted recurrence: 

\begin{corollary}[True cost bounding] \label{cor:true-cost}
If $\cdot \vdash_0 M : A$ and $M \downarrow^{(n,r)} v$ then $n \le
\norm{M}_c$.
\end{corollary}
\begin{proof}
By Theorem~\ref{thm:bounding}, we have $n \le \norm{M}_c - r$, but by
preservation~(Theorem~\ref{thm:pres}), we have that $0 + r \ge 0$, so
$n \le \norm{M}_c$.  
\end{proof}

\subsection{Binary Counter Recurrences}

\begin{figure}
  \input{figs/bc-rec}
  \vspace{-0.25in}
  \caption{Binary Counter Recurrences in $\lambda^\bbbc$}
  \label{fig:bc-rec}
\end{figure}

As an example, the binary counter program in $\lambda^A$
(Figure~\ref{fig:bc-term}) is translated by the recurrence extraction
translation to the terms in Figure~\ref{fig:bc-rec}.
Next, we will use a denotational semantics of the recurrence language to
simplify these recurrences to the desired closed form.

\section{Recurrence Language Semantics} \label{sec:preorder}

The final step of our technique is to simplify recurrences to closed
forms.  This can be done semantically, in a denotational model of the
recurrence languages, or syntactically, by adding axioms to the
inequality judgment $\Gamma \vdash E \le_T E'$ corresponding to
properties true in a particular model.  Here, we will work in a
denotational model of $\lambda^\bbbc$ in preorders, which mostly follows
previous work~\cite{danner-et-al:plpv13,danner-et-al:icfp15,hudson}.

\subsection{Semantic Interpretation}

We describe the semantic interpretation of $\lambda^\bbbc$ in preorders
here, and highlight the differences from \cite{hudson}, which gives a
similar presentation with mechanized proofs.

The semantics of types and terms is given in
Figure~\ref{fig:sem-interp}, omitting function and product types, which are interpreted using the standard cartesian product and exponential objects of preorders.  For each type $A$ of $\lambda^\bbbc$, we
associate a partially ordered set $\scott{A}$ equipped with a top
element ($\infty$) and binary maximums ($\vee$) for which the top
element is an annihilator.
We write $1$ for the one-element poset, and $\N \cup \infty$ for the
natural numbers with an infinite element added, with the usual $0 \le 1
\le 2 \le \ldots \le \infty$ total order, and $\mathbb{Z} \cup \infty$
for the integers with an infinite element added, with the usual total
order.  We write $P \times Q$ for the cartesian product of posets with
the pointwise order, and $Q^P$ for the poset of monotone functions from
$P$ to $Q$, ordered pointwise; these have binary maxes and top elements
given pointwise.  We write $P + Q /\mathord\sim$ for the ``coalesced'' sum,
which first takes the disjoint union of $P$ and $Q$, with only
$\texttt{inl}(x) \le \texttt{inl}(y)$ if $x \le_P y$ and similarly for
\texttt{inr}, and then equates $\texttt{inl}(\infty_P)$ and
$\texttt{inr}(\infty_Q)$ to create a top element $\infty_{P+Q/\mathord\sim}$;
binary maxes are defined using maxes in $P$ and $Q$ for two elements
whose injections match, and to be $\infty$ otherwise.  The translation
on types is extended to contexts: $\scott{\cdot} = 1$,
$\scott{\Gamma,x:A} = \scott{\Gamma} \times \scott{A}$. Finally, we
interpret terms of $\lambda^\bbbc$ as \textit{monotone} (but not
necessarily infinity- or max-preserving) maps\footnote{ We write the
  composition of maps $f : A \to B$ and $g : B \to C$ in diagrammatic
  order, $f ; g : A \to C$.  } from the interpretation of their contexts
into the interpretation of their types. These maps are morphisms in the category
\textbf{Poset} of partially ordered sets and monotone maps, and
so we write them as elements of $\Hom_{\textbf{Poset}}(A,B)$, the
set of monotone maps between posets $A$ and $B$.

\begin{figure}
  \input{figs/sem-interp}
  \caption{Semantic Interpretation Definition}
  \label{fig:sem-interp}
\end{figure}

In Figure~\ref{fig:sem-interp}, we show some representative cases of the
interpretation of terms for sums, natural numbers and lists.  For costs,
the interpretation of cost constants and addition uses the elements and
addition of $\mathbb{Z} \cup \infty$.
In this model, we interpret both natural numbers and
lists as $\N \cup \infty$; for lists, this interprets a list as its
length.  $\N \cup \infty$ has a 0 element and a monotone successor
function $S$, where $S(\infty) = \infty$; these are used to interpret
0/the empty list and successor/cons.
The elimination forms for positives are more complex, and use some
auxiliary monotone functions (which are the morphisms in the category of
posets):
\begin{restatable}{theorem}{auxsemlemma}\label{thm:aux-sem-lemma}
For any posets $A,B,C,G$ with $\infty$ and $\vee$,
\begin{enumerate}
  \item $\texttt{snrec} \in \Hom_{Poset}\left({\left(C^1\right)}^G\times {\left(C^{\N\times C}\right)}^G,C^{G\times\N}\right)$
  \item $\texttt{slrec} \in  \Hom_{Poset}\left({\left(C^1\right)^G} \times {\left(C^{A \times (\N \times C)}\right)^G},C^{G \times \N}\right)$
  \item $\texttt{scase} \in \Hom_{Poset}\left(C^{G \times A}\times C^{G \times B},C^{G\times(A + B)}\right)$
\end{enumerate}
\end{restatable}

The definition of \texttt{scase} is required to respect the quotienting
$\texttt{inl}(\infty) = \texttt{inr}(\infty)$; by maxing each branch the
image of $\infty$ from the other branch, we obtain $f(\gamma,\infty)
\vee g(\gamma,\infty)$ as the image of both of those.  The definition of
\texttt{snrec} is required to be monotone in the $0 \le 1 \le \ldots \le
\infty$ ordering; taking the maximum of the base case and the inductive
step achieves this, because it forces the image of 1 to dominate the
image of 0.  The definition of \texttt{slrec} is similar; the new
question that arises is that, because we have abstracted lists as their
lengths, forgetting the elements, we do not have a value for the head of
the list to supply to $g$ (which, when we use this operation, will be
the translation of the cons branch given to the $\lambda^\bbbc$
recursor).  Here, we always supply $\infty$ as the head list element,
which is sufficient when the analysis really does not require any
information about the elements of the list (otherwise, one can make a
model where lists are interpreted more precisely than as their
lengths~\cite{danner-et-al:icfp15,danner-licata:jfp-in-prep}).

The interpretation satisfies standard soundness theorems, the 
proofs of which 
\begin{icfp2020}can be found in the full version of this
paper~\citep{cutler-et-al:icfp2020-full}.\end{icfp2020}
\begin{arxiv}are in Appendix~\ref{sec:appendix}.\end{arxiv}


\begin{restatable}[Compositionality]{theorem}{semsubst}
\label{thm:sem-subst}
If $\Gamma, x : T_1 \vdash E : T_2$, and $\Gamma \vdash E' : T_1$, then $\scott{\Gamma \vdash E[E'/x] : T_2} =  \left(1_{\scott{\Gamma}},\scott{\Gamma \vdash E' : T_1}\right) ; \scott{\Gamma, x:T_1 \vdash E : T_2}$
\end{restatable}

\begin{restatable}[Soundness (Terms)]{theorem}{interpsound}
\label{thm:term-soundness}
If $\Gamma \vdash E : T$, then $\scott{\Gamma \vdash E : T} \in \Hom\left(\scott{\Gamma},\scott{T}\right)$
\end{restatable}

\begin{restatable}[Soundness (Inequality)]{theorem}{preordsound}
\label{thm:soundness-inequality}
If $\Gamma \vdash E \leq E'$, then for all $\gamma \in \scott{\Gamma}$, $\scott{\Gamma \vdash E : T}(\gamma) \leq \scott{\Gamma \vdash E' : T}(\gamma)$
\end{restatable}

\subsection{Binary Counter Conclusion}

We interpret the binary counter recurrences from Figure~\ref{fig:bc-rec}
in preorders by unfolding the definitions in
Figure~\ref{fig:sem-interp}; the result is shown in
Figure~\ref{fig:bc-poset}.  For the function \texttt{inc}, this yields a
monotone map $\scott{\norm{\texttt{inc}}_p} \in \Hom(1,\N \to \Z \times
\N)$, which is (essentially) a function from an input list size to the
cost of evaluation and the length of the output.  For the function
\texttt{set}, this yields a monotone map $\scott{\norm{\texttt{set}}}
\in \Hom(1,\Z\times(\N \to \Z \times \N))$, which is a pair of a cost
(the cost of evaluating the function definition --- $0$ since
$\texttt{set}$ is a value) and a function from input size to the cost of
evaluation and the length of the output.

\begin{figure}
  \input{figs/bc-poset}
  \vspace{-0.2in}
  \caption{Binary Counter Recurrences Interpreted}
  \label{fig:bc-poset}
\end{figure}

We have boxed the parts of the term that are related to computing the
cost.  The boxed portions of \texttt{inc} express that its amortized
cost is 2 on the empty list (to create a 1 bit with a credit), is 2 when
the bit is 0, and is exactly the same number of steps as the recursive
call when the bit is 1.  The boxed portions of \texttt{set} express that
for zero it costs 0, and for successor it costs the recursive call plus
the cost of \texttt{inc} on the potential of the output of the recursive
call.  However, because we will show that \texttt{inc} turns out to be
constant amortized time, we do not need to bound the potential of the
output of \texttt{set}.  Intuitively, to see that \texttt{inc} has
constant amortized time, observe that the \texttt{slrec} will always
supply the $\infty$ bit as the head of the list, which by definition of
the coalesced sum is both true and false, so the case is effectively the
maximum of $2$ and $\pi_1 \pi_2 \pi_1 p$.  Thus, we effectively have
recurrence where $T_{\texttt{inc}}(0) = 2$ and $T_{\texttt{inc}}(n) = 2
\vee T_{\texttt{inc}}(n-1)$, which solves to $T(n) = 2$ by induction.
Substituting this into the recurrence for \texttt{set}, we have
essentially $T_{\texttt{set}}(0) = 0$ and $T_{\texttt{set}}(n) =
T_{\texttt{set}}(n-1) + 2$, which is of course $O(n)$.  More formally,
we can show by induction that for all $n \geq 0$,
$(\scott{\norm{\texttt{inc}}_p}()(n))_c \leq 2$, and that for all $n$,
$(\scott{\norm{\texttt{set}}_p}()(n))_c \leq 2n$,
establishing bounds on these recurrences in this denotational semantics
in preorders.  

By the bounding theorem (Corollary~\ref{cor:true-cost}), we have that,
for the true operational cost $m$ of evaluating $\texttt{set}(n)
\downarrow^{(m,r)} v$, we have $m \le_\bbbc {\norm{\texttt{set}}_p}(n)_c$
in terms of the syntactic preorder judgment in $\lambda^\bbbc$.  By the
soundness of the interpretation in preorders
(Theorem~\ref{thm:soundness-inequality}), we have that $m
\le_{\mathbb{Z} \sqcup \infty} \scott{\norm{\texttt{set}}_p}()(n)_c$ in
the preorder model.  Therefore, by transitivity, we have $m \le 2n$ in
the preorder model, so our technique proves that the true operational
cost $m$ of setting the binary counter to $n$ is in fact $O(n)$,
as desired.


\section{Variable-Credit Extension}
\label{sec:ex}

The version of $\lambda^A$ described thus far supports amortized analyses
where the amount of credit stored on each element of a data structure is
fixed (e.g. $\listty{!_2 A}$ is a list with 2 credits on each element).
However, in some important amortized analyses, different amounts of credit
must be stored in different parts of a data structure---e.g. for balanced
binary search trees implemented via splay trees~\cite{sleator-tarjan-85},
the number of credits stored on each node is a function of the size of the
subtree rooted at that node.  In this section, we show that adding
existential quantification over credit amounts to $\lambda^A$ suffices to
analyze such examples, using a portion of splay trees as an example.  Using
existentials, a value of type $\exists \alpha.!_\alpha A$ is a value of type
$A$ which carries $\alpha$ credits, for some $\alpha$; for example, a tree
whose elements are of type $\exists \alpha.!_\alpha \mathbb{N}$ stores a
variable number of credits with the number on each node. In keeping with our
methodology of doing as much of an analysis as possible in the recurrence
language and its semantics, the fact that a particular piece of code uses
existentials to implement a desired credit policy will not be tracked by the
type system, but proved after recurrence extraction.  An alternative
approach would be to enrich $\lambda^A$ with some form of indexed or
dependent types to track the sizes of data structures in the type system,
but such an extension is not necessary for our approach.  
The proofs of the results in this section
\begin{icfp2020}can be found in the full version of this
paper~\citep{cutler-et-al:icfp2020-full}.\end{icfp2020}
\begin{arxiv}are in Appendix~\ref{sec:appendix}.\end{arxiv}


\subsection{Existential Types in $\lambda^A$}
To support existential quantifiers over credits, we extend the main typing judgment to be one of the form $\Delta | \Gamma \vdash_f M : A$, where $\Delta = \alpha_1,\dots,\alpha_n$ is a list of ``credit variables''. Any of the $\alpha_i$ can occur free in the types in $\Gamma$, the resource term $f$, the term $M$, or the type $A$. Credit variables $\alpha$ range over \textit{credit terms} $c$, which are (finite) sums of credit variables like $\alpha,\beta$ and credit constants $\ell$ --- i.e. $\alpha_1 + \alpha_2 + \ldots + \alpha_n + l$.  We write $\Delta \vdash c \texttt{  credit}$ to mean that a credit term is well-formed from the variables in $\Delta$.  We consider credit terms up to the usual equations for addition on natural numbers.  These credit terms can then be used as the ``bank'' in resource terms: the resource term $3x + 2y + (\alpha + 2)$ describes a context where one can use $x$ $3$ times, $y$ twice, and has access to the credit term $\alpha + 2$ credits. Most importantly, credit terms are now allowed to appear in the subscript of the $!$ modality (generalizing the natural number constants $\ell$ allowed above): a term $\alpha \mid \Gamma \vdash_f M : !_\alpha A$ with is an $A$ with $\alpha$ credits attached.
We add a new type $\exists \alpha . A$ for existentially quantifying over credit variables.
A value of type $\exists \alpha . A$ is a value of type $A[c/\alpha]$, for some credit term $c$.  Such a value does not store the ability to \emph{use} the credits $c$ --- it stores a number of credits itself.
However, combining the existential with the $!$ modality,
a value of type $\exists \alpha. !_\alpha A$ is an $A$ with $c$ credits attached, for some credit term $c$.
The operational semantics is defined for terms with no free credit variables, so its structure remains unchanged.

\begin{figure}
  \input{figs/la-ex-rules}
  \caption{Extension of $\lambda^A$ with existential types}
  \label{fig:la-ex-rules}
\end{figure}

The typing rules and operational semantics for existential types are presented in Figure~\ref{fig:la-ex-rules}.
The terms for existentials are standard $\texttt{pack}$/$\texttt{unpack}$ terms.
The operational semantics of \texttt{pack} and \texttt{unpack} are also standard; because we only evaluate closed terms, the credit term being packed/unpacked with the value will always be a (closed) natural number $\ell$.




The rest of the rules for $\lambda^A$ are mostly unchanged, so we do not repeat them: they are obtained from the rules in Figure~\ref{fig:la-ty-rules} by carrying the credit variable context $\Delta$ through all of the rules, and,
in the $!^k_c$ modality and the \texttt{save}, \texttt{transfer},
\texttt{create}, and \texttt{spend} terms, the natural number constants
$\ell$ are generalized to credit terms $c$ constructed from these variables.
Finally, since the resource terms may contain free credit variables, the ordering judgment on resource terms must be augmented with a credit variable context, and the ordering itself extended to contain the coefficient-wise ordering on credit variables.
The operational semantics for these constructs in unchanged, because closed credit terms are
precisely the credit values $\ell$ used above.

For this extension, substitution and type preservation are stated as follows:

\begin{restatable}[Substitution]{theorem}{substext}\label{thm:subst-ext}
$\;$
\begin{itemize}
  \item If $\Delta \vdash c \texttt{ credit}$ and $\Delta,\alpha \vdash c' \texttt{ credit}$, then $\Delta \vdash c'[c/\alpha] \texttt{ credit}$
  \item If $\Delta \vdash c \texttt{ credit}$ and $\Delta,\alpha|\Gamma\vdash_f M : A$, then $\Delta|\Gamma[c/\alpha] \vdash_{f[c/\alpha]} M[c/\alpha] : A[c/\alpha]$
\end{itemize}
\end{restatable}

\begin{restatable}[Preservation]{theorem}{presext}\label{thm:pres-ext}
If $\cdot | \cdot \vdash_a M : A$ and $M \downarrow^{(n,r)} v$, then $a + r \geq 0$ and $\cdot | \cdot \vdash_{a + r} v : A$.
\end{restatable}

\subsection{Extracting Recurrences for Existentials}
\begin{figure}
  \input{figs/lc-ex}
  \caption{Recurrence extraction for credit existentials}
  \label{fig:la-ex}
\end{figure}

Recall that the recurrence extraction in Figure~\ref{fig:rec-extr}
erases the $!^k_\ell A$ modalities and translates $\wait \ell M$ and $\disc \ell M$ by adding/subtracting $\ell$ to/from the amortized cost.
Since we now allow credit variables $\alpha$, such as those coming from unpacking an existential type, in the credit position of $\waitname$/$\discname$, the recurrence extraction will need to refer to the values chosen for $\alpha$ in order to know how much to add/subtract to/from the amortized cost.
Thus, we add a type $\$$ to the recurrence language, the values of which are numbers of credits, represented by natural numbers.  The credit context $\Delta$ is translated to recurrence language variables of type $\$$
(i.e. $\angles{\Delta,\alpha} = \angles{\Delta},\alpha : \$$), while existential types $\exists \alpha.A$ are translated to pairs  $\$ \times \angles{A}$.  A simple pair suffices because the $!$ modality is erased by $\angles{\cdot}$, and this is the only place where credit terms can occur in the syntax of types, so all occurrences of $\alpha$ under the binder are removed, and $\angles{A}$ is a closed type.

We show the new and changed cases of recurrence extraction in Figure~\ref{fig:la-ex}.  The introduction and elimination rules for $\exists \alpha . A$ translate to the corresponding introduction and elimination forms for $\$ \times \angles{A}$.
For \texttt{create} and \texttt{spend}, in principle, we would like the cost component of $\wait c M$ to be $c + \norm{M}_c$, but this will not type check, given that $c : \$$ but $\norm{M}_c : \bbbc$.
Recalling that costs $\bbbc$, though axiomatized as a monoid with some operations, are morally integers, we add a coerction $\texttt{to}\mathbb{C} : \$ \to \bbbc$, which is morally the inclusion of natural numbers into integers.


\begin{restatable}[Extraction Preserves Types]{theorem}{extrsoundex}\label{thm:extr-sound-ex}
If $\Delta | \Gamma \vdash_f M : A$, then $\angles{\Delta},\angles{\Gamma} \vdash \norm{M} : \norm{A}$
\end{restatable}

\subsection{Bounding Relation and Bounding Theorem}


The definition of the bounding relation for values (Definition~\ref{def:bounding}) is extended  with
\begin{itemize}
\item
  $\pack \alpha \ell v \valbd^{\exists \alpha.A,a} E$ iff $\ell \leq_{\$} \pi_1 E$ and $v \valbd^{A[\ell/\alpha],a} \pi_2 E$
\end{itemize}
Recalling that $E : \angles{\exists \alpha. A} = \$ \times \angles{A}$, this simply states that the amount of credit packed by $\alpha$ is bounded by the amount described by $\pi_1 E$, and that the value packed with the credit amount is in fact bounded by $\pi_2 E$. We remark that this definition may give the careful reader pause-- inducting on a substitution instance of an existential type where the existential variable ranges over \textit{types} leads to well-definedness issues.
But, our existential variables range over \textit{credits}, so we may simply regard a closed substitution instance of a type $\alpha \vdash A \, \mathsf{type}$ as a smaller type than $A$.

The definition of the bounding relation for open terms must also be modified to quantify over closing substitutions for the credit context, as well as the term context.
First, if $\omega$ is a substitution of credit amounts $\ell$ for credit variables, and $\Omega$ is a substitution of closed terms of type $\$$ for recurrence language variables, then $\omega \bdby^{\Delta} \Omega$ means that for all $\alpha \in \Delta$, $\omega(\alpha) \leq_\$ \Omega(\alpha)$.
Then for $\Delta | \Gamma \vdash_f M : A$ we write $M \bdby^A E$ if for all $\omega \bdby^{\Delta} \Omega$ and for all $\theta \bdby^{\Gamma[\omega],\sigma} \Theta$, we have that $M[\omega,\theta] \bdby^{A[\omega],f[\omega,\sigma]} E[\Omega,\Theta]$.
Using this notation, the bounding theorem is
\begin{restatable}[Bounding Theorem]{theorem}{boundingex}
\label{thm:bounding-ex}
If $\Delta | \Gamma \vdash_f M : A$, then $M \bdby^A \norm{M}$
\end{restatable}
\noindent and the cases which differ from the original Theorem~\ref{thm:bounding} are proved in the supplementary materials.

\subsection{Splay Tree Analysis}
We now describe somewhat informally how to use the above machinery to
analyze splay trees; the complete formalism is given 
\begin{icfp2020}in the full version of this
paper~\citep{cutler-et-al:icfp2020-full}.\end{icfp2020}
\begin{arxiv}in Appendix~\ref{sec:appendix} (Figure~\ref{fig:trec-rules}).\end{arxiv}
Following 
Okasaki's presentation~\cite{okasaki:purely-functional-data-structures},
the key operation is
a $\texttt{split} : (A \times \tree A) \to \tree A \times \tree A$ function that splits a given tree into elements larger and smaller than a given pivot. Insertion, deletion, union, intersection, difference etc. can be all implemented from \texttt{split} and a $\texttt{join}$ operation that combines two sorted trees where all the elements of the first are less than the elements of the second. Showing that $\texttt{split}$ is amortized $O(\log n)$ time, where $n$ is the size of the tree, is the most difficult part of the amortized analysis, and implies the desired time bounds for the other operations.  The key idea of splay trees is that each access rearranges the tree so that accessing the same element twice in a row is quicker the second time. In Okasaki's presentation, this rearrangement takes place in $\texttt{split}$, which performs a series of tree rotations. These rotations ensure that the amortized cost of $\texttt{split}$ (amortized over any sequence of binary search tree operations) is $O(\log n)$, even though the tree is not always balanced.
The most challenging cases of the code unpack the tree to depth two, and rotate the output if they traverses the same direction twice while searching for the pivot:
$$
\begin{array}{l}
split \; p \; (N (x,N (y,a_{11},a_{12}), N (z,a_{21},a_{22}))) | \; x \geq p \mathbin{\&\&} y \geq p = \\
\hspace{1em} (small, N (y,big,N (x,a_{12},N (z,a_{21},a_{22})))) \texttt{ where } (small,big) = \texttt{split} \; p \; a_{11}\\
\end{array}
$$
Okasaki's analysis of split maintains the invariant that there are $\varphi(t) = \lceil{\lg (|t| + 1)}\rceil$ credits associated with the root of every subtree $t$ in a splay tree, and uses the potential/physicists method to analyze the amortized cost.  

The addition of existentials to $\lambda^A$ allows us to encode this analysis, by giving \texttt{split} the type $A \otimes \tree {\exists   \alpha.!^1_\alpha A} \loli \tree {\exists \alpha.!^1_\alpha A} \otimes \tree {\exists \alpha.!^1_\alpha A}$, and using code to maintain the invariant that each of these $\alpha$'s are precisely $\varphi(t)$.


\subsubsection{Creating Variable Amounts of Credit.}

\begin{figure}
  \vspace{-0.1in}
  \input{figs/ghost-loop}
  \vspace{-0.2in}
  \caption{$\lambda^A$ term for the \texttt{spawn} function}
  \label{fig:ghost-loop}
\end{figure}

To maintain this invariant, we will sometimes need to \waitname\/ amounts of credit determined by a run-time natural number, like $\varphi(t)$ for some tree $t$---but the primitive \wait{c}{M} term allows for waiting only for a credit term $c$, which cannot depend on run-time values.
However, we can write a recursive loop that spawns a number of credits
dependent on a run-time value, and package this as a function
$\texttt{spawn} : \N \loli \exists \alpha . !^1_\alpha 1$
such that the $\alpha$ packed in the result of $\texttt{spawn}(n)$ is (the credit term representing) $n$.
The implementation of \texttt{spawn}\/ is shown in Figure~\ref{fig:ghost-loop}---at a high level,
the term loops $\texttt{create}_1$ in a $\N$-recursor, using a credit existential as a counter variable.
In this example, and throughout this section, we use pattern-matching notation as syntactic sugar for the elimination rules for positive types like $\exists,!,\otimes$, with the convention that matching on the result of a thunked recursive call implicitly forces it.

In Section~\ref{ssec:la-sem}, we argued that
the $n$ component in the operational cost semantics $M \downarrow^{n,r} v$ captures the actual operational cost of an erasure to simply-typed $\lambda$-calculus, as long as $\texttt{tick}$s in $\lambda^A$ are inserted for each STLC $\beta$-redex.  Because we do not include any \texttt{tick} terms in \texttt{spawn}, its abstract operational cost $n$ is zero.  Thus, to realize this cost semantics, $\texttt{spawn}$ must be erased before actually running the program.  Fortunately, a simple program optimization suffices to do this: translate $\lambda^A$ to simply-typed $\lambda$-calclus by dropping both the $\exists$ and $!$ types and the associated terms, at which point \texttt{spawn} has type $\N \to 1$; then replace all terms of type $1$ with the trivial value.
That is, we think of \texttt{spawn} as a \emph{ghost loop} --- code that is meant for the extracted recurrence, but not intended to actually be run.  

\subsubsection{Definition of Trees in $\lambda^A$.}
Extending $\lambda^A$ with the requisite \texttt{tree} type constructor and its rules follows both previous work~\cite{danner-et-al:icfp15}
and the pattern illustrated with lists above.
The type of trees is essentially $\tree A = \texttt{Emp} \; | \; \texttt{N of } A \otimes \N \otimes \tree A \otimes \tree A $.
The $\N$ argument caches the size of the tree, making the function $\texttt{size} : \tree A \loli \N \otimes \tree A$ --- which projects out that field and then rebuilds the tree\footnote{The tree can be rebuilt   because values of type $\N$ are duplicable--- there is a diagonal map   $\N \loli \N \otimes \N$. Also, we will often use \texttt{size} as a   function $\tree A \loli \N$, and silently contract the second   projection for re-use of the argument.} --- constant time.
To support coding the \texttt{split} function described above, we directly add a recursor that performs a two-level pattern match,
with cases for the empty tree, for a node with one child or the other empty and the other is another node, and for a node with two nodes as children; in the latter case, the recursor provides recursive calls on
all four subtrees.  


\subsubsection{Splay Tree Implementation.}
We define a \textit{splay tree} to be a binary search tree $t : \tree {\exists \alpha. !^\infty_\alpha A}$ satisfying the property that if $\texttt{size}(t) = n$, then if $t = N(\_,m,t_0,t_1)$, then $t_0$ and $t_1$ are splay trees, and for $\scott{\norm{t}_p} = N((\alpha,\_),\_,\_)$, we have $\alpha = \phi(n)$.
In other words, the credit invariant holds at each node in the tree. We note that each element of the tree not only carries $\alpha$ credits, but is also infinitely usable since we are required to compare nodes in the tree more than constantly many times. This causes no issues for the extracted recurrences, because keys in the tree are always values. We then prove a lemma which states that \texttt{split} preserves the splay tree property --- i.e. that the existentially quantified credits stored in the tree satisfy the desired invariant.  

\begin{lemma} \label{lem:splay-tree-invariant}
If $t : \tree {\exists \alpha. !^\infty_\alpha A}$ is a splay tree and $\texttt{split}(t) \downarrow (t_0,t_1)$, then $t_0$ and $t_1$ are also splay trees.
\end{lemma}

To illustrate the $\lambda^A$ term for \texttt{split}, we show one key
case of the recursor, which corresponds to the snippet given at the
beginning of this section and to \cite[Theorem
  5.2]{okasaki:purely-functional-data-structures}.  For this case, we
are in the situation where the root, labeled by $x$, has two subtrees,
$y$ with subtrees $a_{11},a_{12}$, and $z$ with subtrees
$,a_{21},a_{22}$. If the pivot is less than both $x$ and $y$, we recur
on the leftmost subtree $a_{11}$, which produces the elements of
$a_{11}$ that are smaller and bigger than the pivot.  Then $smaller$
contains all the elements of the original tree smaller than the pivot.
The elements bigger than the pivot are $bigger$ and everything else from
the original tree; we combine these together into a new tree, performing
a rotation to put $y$ at the root.

The $\lambda^A$ version of this term, presented in Figure~\ref{fig:split}, annotates the above code with some additional information about the sizes of trees, and with some code for manipulating credits.  The variables $x,y,z$ are the values of type $A$ at the root and its immediate children; these come with existentially-quantified numbers of credits $\alpha,\beta,\gamma$ ($\alpha$ credits are stored with $x$, $\beta$ with $y$, and $\gamma$ with $z$), and also with natural numbers caching the sizes of the subtrees that they are the roots of ($n_1,n_2,n_3$ respectively).
The variables $a_{ij}$ stand for the four subtrees with their (suspended) recursive call outputs; we write $\texttt{split}(p,a_{11})$ for projecting and forcing the recursive call, and write $a_{ij}$ for projecting the other subtrees.  The credit manipulation involves spending the credits $\alpha$ and $\beta$ stored with $x$ and $y$ in the input tree (we do not spend $z$, because the $z$ node is left unchanged in the output), calculating the sizes of the new nodes $t'$ and $s'$ that will be part of the output, and \texttt{spawn}ing credits corresponding to $\varphi$ of these sizes. The term presented in Figure~\ref{fig:split} is one branch of one of the step functions passed to the $\texttt{treerec}$ which forms the outermost structure of \texttt{split}.

\begin{figure}
  \vspace{-0.1in}
  \input{figs/split}
  \vspace{-0.3in}
  \caption{Part of the $\lambda^A$ term for \texttt{split}}
    \vspace{-0.1in}
  \label{fig:split}
\end{figure}

To analyze splay trees, we pass this $\lambda^A$ term through recurrence extraction and the preorder semantics and then prove the following:
\begin{theorem}
If $t : \tree {\exists \alpha . !^\infty_\alpha A}$ is a splay tree with $\texttt{size}(t) = n$, then for any $v : A$, $\scott{\norm{\texttt{split}(t,v)}_c} \leq 1 + 2\varphi(\scott{\norm{\texttt{size}}_p(t)}) \in O(\lg n)$.
\end{theorem}
\begin{proof}
As an example, we show the case for the code in Figure~\ref{fig:split}. The cost component of the extracted recurrence is
\[
1-\alpha - \beta + \scott{\norm{\texttt{split}(p,a_{11})}} + \varphi(1 + n_{12} + n_3) + \varphi(2 + n_{big} + n_{12} + n_3)
\]
The 1 comes from the $\texttt{tick}$; $\alpha$ and $\beta$ are subtracted because they are \texttt{spent}; and the $\varphi$ of the sizes of $t'$ and $s'$ are added because they are \texttt{create}d.  
By definition, $1 + n_{12} + n_3 = \scott{\norm{\texttt{size}}_p(t')}$ and $2 + n_{big} + n_{12} + n_3 = \scott{\norm{\texttt{size}}_p(s')}$. By the credit invariant, $\alpha = \varphi(\scott{\norm{\texttt{size}}_p(t)})$, and $\beta = \varphi(\scott{\norm{\texttt{size}}_p(s)})$, where $s$ is the subtree of $t$ rooted at $y$. Rewriting by these and commuting terms, the extracted recurrence is precisely
\[1 + \scott{\norm{\texttt{split}(p,a_{11})}} + \varphi(\scott{\norm{\texttt{size}}_p(s')}) + \varphi(\scott{\norm{\texttt{size}}_p(t')}) - \varphi(\scott{\norm{\texttt{size}}_p(s)}) - \varphi(\scott{\norm{\texttt{size}}_p(t)})
\]
which Okasaki~\cite[Theorem 5.2]{okasaki:purely-functional-data-structures} proves is bounded by $1 + 2\varphi(\texttt{size}(t))$, as required.
\end{proof}

\section{Related work} \label{sec:related-work}

Techniques for extracting (asymptotic)
cost information from high-level program source
code is a project that is almost as old as studying programming languages.
For non-amortized analysis of
functional languages, we have examples from the 1970s and
1980s by \citet{wegbreit:cacm75}, \citet{lematayer:toplas88}, and
Rosendahl~\cite{rosendahl:auto_complexity_analysis}.  The idea of
simultaneously extracting information about cost and size, and defining the
size of a function to be a function itself (leading to higher-order
recurrences) has its roots in \citet{danner-royer:ats-lmcs},
which in turn draws from ideas in \citet{shultis:complexity},
\citet{sands:thesis}, and Van Stone~\cite{vanstone:thesis}.  Using
bounded modal operators to describe resource usage goes back at least to
\citet{girard-et-al:tcs92:bll}, and Orchard et~al.\ have
recently incorporated these ideas into the Granule
language~\cite{orchard-et-al:icfp19:graded-modal-types}.  Perhaps the
work that is closest in spirit to ours is Benzinger's ACA system for
analyzing call-by-name \textsc{Nuprl} programs~\cite{benzinger:tcs04}.  From
a cost-annotated operational semantics, he extracts a ``symbolic semantics''
that is similar in flavor to our recurrence language and extracted
recurrences, although without amortization.  The symbolic semantics yields
higher-order recurrences, which he reduces to first-order recurrences that
can be analyzed with a computer algebra system.  

There is also extensive work on recurrence extraction from first-order
imperative languages.  The COSTA project
\citep{albert-et-al:jar11,albert-et-al:tcs12:cost-analysis,albert-et-al:tocl13:inference}
takes Java bytecode as its source language, extracts cost relations
(essentially, non-deterministic cost recurrences), and solves them for upper
bounds.  In this line of work, \citet{alonso-blas-genaim:sas12} and
\citet{flores-montoya:fm16} investigate the failure to derive tight upper
bounds in settings where amortized analysis is typically deployed.  They
trace the issue to the fact that typically cost relations do not depend on
the results of the analyzed functions.
Making this possible allows more
precise constraints which, when solved, yield tighter bounds.  The
dependency on output corresponds roughly to total accumulated savings,
and they infer an appropriate potential function (in the terminology of the
physicist's method),
modulo a choice of templates.
To analogize with our work, they delay the determination of the credit
policy until solving for upper bounds of extracted recurrences, whereas we
specify the credit policy as part of the source program, which directly
yields a recurrence for cost that takes the policy into account.


Two recent approaches that handle amortized analysis for functional programs
are Timed~ML (TiML, \cite{wang-et-al:oopsla17:timl}) and automatic amortized
resource analysis (AARA,
\cite{hoffmann-et-al:toplas12:multivariate-amortized,hoffmann-shao:esop15:parallel,hoffmann-et-al:popl17,niu-hoffmann:lpar18}).
In TiML, ML type and function definitions are annotated with indices that
convey size information.  The notion of size is left unspecified and
the indices are very flexible, and can
include constraints such as those required to define red-black trees.  Type
inference generates verification conditions.  Depending on the details of
the annotations, solving the verification conditions provides exact or
asymptotic bounds on the cost of the original program.  The focus is on
worst-case analysis, but the annotation language is sufficiently rich to
encode the physicist's method of amortized analysis.  Although it is not
part of their focus, the formalism does not appear to enable analysis of
higher-order functions whose cost depends on the complexity behavior of the
function arguments.

AARA provides a type inference system for resource bound analysis of
higher-order functional programs that incorporates amortization.  Credit
allocation is built into the type system itself.
Soundness says that the
net credit change during evaluation is bounded by the net credit change
described by the typing.  AARA focuses primarily on strict languages, but
\citet{jost-et-al:jar17} use similar ideas to analyze programs
under lazy evaluation.
In AARA, the credit allocation and usage is described
in the type judgment.
Type inference generates constraints, and the solution of these constraints
is essentially a credit allocation strategy.  
Our approach describes usage in the type judgment, but
requires the strategy to be explicit in the program (via 
$\texttt{save}$, $\waitname$, $\discname$, etc.), which
places a greater burden on the programmer.  However, reasoning about that
strategy (e.g., establishing a credit invariant) in the semantics may
provide more flexibility, though that requires more investigation.

We note that the technical differences between TiML and AARA and our
approach arise from a difference in what we might consider the
philosophical underpinnings.  TiML and AARA introduce novel type systems
with a goal of inferring cost bounds to the greatest extent possible.  Those
bounds are extracted as part of the type inference procedure.  This is not
how most programmers conceptualize a cost analysis, and our
interest is in staying as close to typical informal analyses as we can.
While $\lambda^A$ is a novel type system, the novelty exists solely in order
to make the programmer be explicit about how credits are allocated and used.
This task is part of a banker's-method analysis, though it is usually stated
informally (``put one credit on each~$1$ in the bit list'').  After that, it
is extraction of ordinary (semantic) recurrences which one hopes to be able
to bound using whatever methods are at the programmer's disposal.

\section{Future Work}
\label{sec:future-work}

%
%

We expect that the techniques used in~\citet{kavvos-et-al:popl20} to handle
general recursion in the source language can be adapted to the approach we
have taken here to handle amortization, though work remains to be done to see
whether typical non-structurally recursive amortized algorithms would satisfy
the necessary typing constraints.  
A useful project
would then be to do the analyses that 
\citet{okasaki:purely-functional-data-structures} 
describes via recurrence extraction,
which focus on amortized cost of sequences of arbitrary 
data structure operations (e.g., typical usage of a functional queue), where
the data structure is used ephemerally.  Adding a type for memoizing thunks
to the source language,
or more generally lazy evaluation and datatypes, would
permit analysis of persistent usage.


Recalling our ``big-picture'' goal of formalizing as closely as possible the
process by which programmers actually perform cost analyses, we are not
there yet.  Let us consider how an analysis of the usual splay-tree
implementation of the abstract set type actually proceeds.  First we define
the splay-tree implementation, and then reason about its operations \emph{in
isolation} to conclude that, provided the it is used ephemerally, the
amortized cost of those operations is $O(\lg n)$, where $n$ is the size of
the tree.  We then would typically analyze an algorithm that uses the
abstract set type ephemerally \emph{under the assumption that the set
operations are $O(\lg n)$-time,} where $n$ is the size of the set.  In other
words, the analysis of the data structure (which may use techniques such as
amortized analysis, and depends crucially on the structure of the tree) is
separated from the analysis of the program that uses the interface that it
implements (which uses only information about the size of the set).  In the
context of what we have presented here, while the splay-tree type would be
something like our $\tree {\exists \alpha. !^\infty_\alpha A}$, the programs
that use it would use an abstract $\set A$ type, and in particular the
abstract type would not refer to the type constructors we have introduced to
manage credits.  Codifying this would require something like abstract type
declarations (or more generally existential types as
in~\cite{mitchell-plotkin:popl85}), but where the denotation of the abstract
type (corresponding to the abstract notion of size) is not the same as that
of the concrete type of the implementation.  This is an ongoing project.

%

We have neglected any discussion of automating either the front end of this
process (annotating a program with the constructs used for amortization) or
the back end (automatic solving of recurrences). 
Fully automating the annotation of a source program may be too much to ask
(amortized analysis is hard!), but one could hope for a process that elaborates
a program with higher-level annotations (e.g.\ written as comments) into $\lambda^A$, inserting $\waitname$,
$\discname$, etc.  On the back end, the
syntactic inequality judgment from Figure~\ref{fig:syn-ord} can be used to
simplify recurrences in $\lambda^\bbbc$ as opposed to interpreting into a
model of the recurrence language and then simplifying there. Ideally, one could
add enough rules to the judgment (and perhaps enrich its structure) to be able
to simplify a large class of standard recurrences, and then apply proof
search techniques to automate the process.  We would still have higher-order
recurrences, and it would be worthwhile to see if the techniques used
by Benzinger~\cite{benzinger:tcs04} can be used to reduce them to first-order
recurrences that could be solved by a recurrence solver such as
OCRS~\cite{kincaid-et-al:popl18:ocrs}.


\begin{acks}
The authors would like to thank the anonymous ICFP reviewers for their very
thoughtful feedback. We also thank the reviewers from ESOP for suggesting an
analysis of splay trees as a way to demonstrate the expressiveness of
$\lambda^A$. Finally, the first author would like to thank Alex Kavvos for
numerous clarifying conversations during the course of this work.

This material is based upon work supported by the 
\grantsponsor{nsf}
             {National Science Foundation}
			 {https://nsf.gov}
under Grant Number \grantnum{nsf}{CCF-1618203}, the 
\grantsponsor{afosr}
             {Air Force Office of Scientific Research}
			 {}
under award number \grantnum{afosr}{FA9550-16-1-0292}, and the 
\grantsponsor{afrl}
             {United States Air Force Research Laboratory}
			 {}
under agreement number \grantnum{afrl}{FA9550-15-1-0053}.
The U.S. Government is authorized to reproduce and distribute reprints for
Governmental purposes notwithstanding any copyright notation thereon.
The views and conclusions contained herein are those of the authors and
should not be interpreted as necessarily representing the official
policies or endorsements, either expressed or implied, of the United
States Air Force Research Laboratory, the U.S. Government or Carnegie
Mellon University.
\end{acks}

\doclicenseThis

\bibliographystyle{ACM-Reference-Format}
\bibliography{refs}

\appendix

\section{Appendix}
\label{sec:appendix}
\lastructural*
\begin{proof}
All follow by straightforward induction on judgments.
\end{proof}
\fusion*
\begin{proof}
We present terms going in both directions for each caes.
\begin{enumerate}
  \item $x : !^{k_1k_2}_{\ell_1 + k_1\ell_2} \vdash_x \tsfer 1 {k_1k_2} {\ell_1 + k_1\ell_2} y x {\save {k_1} {\ell_1} {(\save {k_2} {\ell_2} {y})}} : !^{k_1}_{\ell_1} !^{k_2}_{\ell_2} A$ and $x : !^{k_1}_{\ell_1} !^{k_2}_{\ell_2} A \vdash_x \tsfer 1 {k_1} {\ell_1} y x {\tsfer {k_1} {k_2} {\ell_2} z y {\save {k_1k_2} {\ell_1 + k_1\ell_2} {z}}} : !^{k_1k_2}_{\ell_1 + k_1\ell_2}$
  
  \item $x : !^k_{\ell_1 + \ell_2}(A \otimes B) \vdash_x \tsfer 1 k {\ell_1 + \ell_2} y x {\asplit {k'} y {z_1} {z_2} {(\save k {\ell_1} {z_1}, \save k {\ell_2} {z_2})}} : !^k_{\ell_1} A \otimes !^k_{\ell_1} B$ and $x : !^k_{\ell_1} A \otimes !^k_{\ell_2} B \vdash_x \asplit 1 x {z_1} {z_2} {\tsfer 1 k {\ell_1} {y_1} {z_1} {\tsfer 1 k {\ell_2} {y_2} {z_2} {\save k {\ell_1 + \ell_2} (y_1,y_2)}}} : !^k_{\ell_1 + \ell_2}(A \otimes B)$.
  
  \item $x : !^k_\ell (A \oplus B) \vdash_x \tsfer 1 k {\ell} y x {\acase k y {z_1} {\inl {(\save k \ell {z_1})}} {z_2} {\inr {(\save k \ell {z_2})}}} : !^k_\ell A \oplus !^k_\ell B$ and $x : !^k_\ell A \oplus !^k_\ell B \vdash_x \acase 1 x {z_1} {\tsfer 1 k {\ell} y {z_1} {\save k \ell {(\inl y)}}} {z_2} {\tsfer 1 k {\ell} y {z_2} {\save k \ell {(\inr y)}}} : !^k_\ell (A \oplus B)$
\end{enumerate}
\end{proof}
\pres*
\begin{proof}
By induction on $M \downarrow v$.
\begin{itemize}
\item (Values): Suppose $\cdot \vdash_a v : A$ and $v \downarrow^{(0,0)} v$. Then, $a + 0 \geq 0$ (because $a \geq 0$), and $\cdot \vdash_{a+0 = a} v : A$.
\item (Tick): Immediate by IH.

\item ($!$-I):  Suppose $\cdot \vdash_b \save k l M : !^k_l A$, and $\save k l M \downarrow^{(\_,kr)} \save k l v$. We must show that $b + kr \geq 0$ and that $\cdot \vdash_{b+kr} \save k l v : !^k_l A$. Inverting the rules, we have that $\cdot \vdash_a M : A$ with $ka+l \leq b$, and that $M \downarrow^{(n,r)} v$. By IH, $\cdot \vdash_{a+r} v : A$ with $a + r \geq 0$. Since $k,l \geq 0$, $0 \leq k(a+r) + l = ka+l+kr$, which, since $ka+l \leq b$, is less than or equal to $b+kr$. So, $b+kr \geq 0$ and $\cdot \vdash_{b+kr} \save k l v : !^k_l A$, as required.

\item ($!$-E): For this case, suppose $\cdot \vdash_{k'a+b} \tsfer {k'} k l x M N : C$, and  $\tsfer{k'} k l x M N \downarrow^{(\_,k'r_1+r_2)} v$. We want to show that: $k'a + b + k'r_1 + r_2 \geq 0$ and $\cdot \vdash_{k'a + b + k'r_1 + r_2} v : C$. By inversion, $\cdot \vdash_a M : !^k_l A$, and $x : A \vdash_{b+k'(kx+l)} N :C$, as well as $M \downarrow^{(\_,r_1)} \save k l {v_1}$ and $N[v_1/x] \downarrow^{(\_,r_2)} v$. By IH, we know that $\cdot \vdash_{a+r_1} \save k l {v_1} : !^k_l A$ and $a+r_1 \geq 0$, so by inversion, there is a $d$ such that $kd+l \leq a+r_1$, and $\cdot \vdash_d v_1 : A$, and so substitution gives that $\cdot \vdash_{b+k'(kd+l)} N[v_1/x] : C$. But $kd+l \leq a+r_1$, so by structural weakening we have $\cdot \vdash_{b+k'a+k'r_1} N[v_1/x] : C$. Again by IH, $\cdot \vdash_{b+r_2 + k'a+k'r_1} v : C$ and $b+r_2 + k'a+k'r_1 \geq 0$, as required.

\item (\waitname): Suppose $\cdot \vdash_a \wait l M : A$ and $\wait l M \downarrow^{(n,r+l)} v$. Inverting, we have $\cdot \vdash_{a+l} M : A$, and $M \downarrow^{(n,r)} v$. By IH, we have that $a + l + r \geq 0$, and $\cdot \vdash_{a+r+l} v$. But, we wanted to show that $a + r + l \geq 0$ and that $\cdot \vdash_{a + r + l} v : A$, and so we are done.
\item (\discname): Suppose $\cdot \vdash_{a+l} \disc l M : A$, and $\disc l M \downarrow^{(\_,r-l)} v$. We want to show that $a + l + r - l = a + r \geq 0$, and that $\cdot \vdash_{a+l+r-l=a+r} v$ Inverting, we have that $\cdot \vdash_a M : A$ and $M \downarrow^{(n,r)} v$. By IH, $a + r \geq 0$ and $\cdot \vdash_{a+r} v : A$, as required.

\item($\otimes$-I): For this case, let $\cdot \vdash_{a+b}(M,N) : A \otimes B$, and $(M,N) \downarrow^{(\_,k_1+k_2)} (v_1,v_2)$. We must show that $a + b + k_1 + k_2 \geq 0$, and that $\cdot \vdash_{a + b + k_1 + k_2 }(v_1,v_2)$
Inverting, we get the four premises $\cdot \vdash_a M : A$, $\cdot \vdash_b N : B$, and $M \downarrow^{(\_,k_1)}v_1$ and $N \downarrow^{(\_,k_2)} v_2$. Using the IH on these two pairs, we get that $a + k_1 \geq 0$, $b + k_2 \geq 0$, $\cdot \vdash_{a+k_1} v_1 : A$, and $\cdot \vdash_{b + k_2} v_2 : B$. Adding the two inequalities and applying $\otimes$-I to the judgments gives the desired result.

\item($\oplus$-E): Suppose $\cdot \vdash_{a+b_1+b_2} \acase {k'} M x {N_1} y {N_2} : C$ and $\acase {k'} M x {N_1} y {N_2} \downarrow^{(\_,k'r_1 + r_2)} v$. 
Inverting the typing judgment, $\cdot \vdash_a M : A \oplus B$, $x : A \vdash_{b_1 + k'x} N_1 : C$ and $y : B \vdash_{b_2 + k'} N_2 : C$. Inverting the evaluation judgment gives two symmetric cases, so suppose that $M \downarrow^{(\_,r_1)} \inl {v_1}$ and $N_1[v_1/x] \downarrow^{(\_,r_2)} v$. By IH, $\cdot \vdash_{a+r_1} \inl {v_1} : A \oplus B$ and $a + r_1 \geq 0$. So, $\cdot \vdash_{a+r_1} v_1 : A$. By substitution, $\cdot \vdash_{b_1 + k'(a+r_1)} N_1[v_1/x] : C$. By IH, $\cdot \vdash_{k'a + b_1 + k'r_1 +  r_2} v : C$ and $k'a + b_1 + k'r_1 +  r_2 \geq 0$. Since $b_2 \geq 0$, by structural weakening, $\cdot \vdash_{k'a + b_1 + b_2 + k'r_1 +  r_2} v : C$ and $k'a + b_1 + b_2 + k'r_1 +  r_2 \geq 0$, as required.

\item ($\loli$-E):  Let $\cdot \vdash_{a+b} M \, N : B$, and $M \, N \downarrow^{(\_,k_1+k_2+k_3)} v$. We want to show that $a + b + k_1 + k_2 + k_3 \geq 0$, and that $\cdot \vdash_{a + b + k_1 + k_2 + k_3} v : B$. We invert both judgments to get $\cdot \vdash_a M : A \loli B$, and $\cdot \vdash_b N : A$, and $M\downarrow^{(\_,k_1)} \lambda x.M'$, and $N \downarrow^{(\_,k_2)} v_1$, and that $M'[v_1/x] \downarrow^{(\_,k_3)} v$. Applying the IH to the first evaluation, we have that $\cdot \vdash_{a+k_1} \lambda x.M' : A \loli B$. Inverting the proof of that judgment, we get that $x : A \vdash_{a+k_1+x} M' : B$. By IH again, $\cdot \vdash_{b+k_2} v_1 : A$, and by substitution, $\cdot \vdash_{a+b+k_1+k_2} M'[v_1/x]$. By IH once more, $a + b + k_1 + k_2 + k_3 \geq 0$, and $\cdot \vdash_{a + b + k_1 + k_2 + k_3} v : B$, as required.

\item ($\N$-E) Suppose $\cdot \vdash_{a+b_1+b_2} \nrec {M} {N_1} {N_2} : C$. By inversion, $\cdot \vdash_a M : \N$, $\cdot \vdash_{b_1} N_1 : 1 \loli C$, and $\cdot \vdash_{b_2} N_2 : !^\infty_0(\N \tensor (1 \loli C) \loli C)$. We have two evaluation cases to consider.
\begin{itemize}
  \item Suppose $\nrec M {N_1} {N_2} \downarrow^{(\_,r_1+r_2+r_3+r_3)} v$ by way of $M \downarrow^{(\_,r_1)} 0 : \N$, $N_1 \downarrow^{(\_,r_2)} \lambda x.N_1'$, $N_2 \downarrow^{(\_,r_3)} \_$, and $N_1'[()/x] \downarrow^{(\_,r_4)} v$. Then, by IH, we have the following:
  \begin{itemize}
    \item $\cdot \vdash_{a+r_1} 0 : \N$, and $a + r_1 \geq 0$
    \item $\cdot \vdash_{b_1+r_2} \lambda x.N_1' : 1 \loli C$, $b_1 + r_2 \geq 0$.
    \item $b_2 + r_3 \geq 0$
  \end{itemize}
  Since $\cdot \vdash_0 () : 1$, $\cdot \vdash_{b_1 + r_1} N_1'[()/x] : C$. By IH, $\cdot \vdash_{b_1 + r_2 + r_4} v : C$. By structural weakening, $\cdot \vdash_{a+b_1+b_2 + r_1 + r_2 + r_3 + r_4} v : C$, as required.
  \item Suppose $\nrec M {N_1} {N_2} \downarrow^{(\_,r_1+r_2+r_3+r_3)} v$ by way of $M \downarrow^{(\_,r_1)} S(v_1) : \N$, $N_1 \downarrow^{(\_,r_2)} \lambda x.N_1'$, $N_2 \downarrow^{(\_,r_3)} \save \infty 0 {(\lambda x.N_2')}$, and $N_2'[(v_1,\lambda z.\nrec {v_1} {\lambda x.N_1'} {\save \infty 0 {(\lambda x.N_2')}})/x] \downarrow^{(\_,r_4)} v$. Then, by IH we have:
  \begin{itemize}
    \item $\cdot \vdash_{a+r_1} S(v_1) : \N$, $a + r_1 \geq 0$
    \item $\cdot \vdash_{b_1 + r_2} \lambda x.N_1' : 1 \loli C$, $b_1 + r_2 \geq 0$
    \item $\cdot \vdash_{b_2 + r_3} \save \infty 0 {(\lambda x.N_2')} : !^\infty_0(\N \otimes (1 \loli C) \loli C)$, and $b_2 + r_3 \geq 0$.
  \end{itemize}
  By $\N$-strengthening, $\cdot \vdash_0 S(v_1) : \N$, and so $\cdot \vdash_0 v_1 : \N$. Since $\cdot \vdash_{b_2 + r_3} \save \infty 0 {(\lambda x.N_2')} : !^\infty_0(\N \otimes (1 \loli C) \loli C)$, there is a $c \geq 0$ so that $\infty \cdot c \leq b_2 + r_3$ with $\cdot \vdash_c \lambda x.N_2' : \N \otimes (1 \loli C) \loli C$. Then, $\cdot \vdash_{b_1 + r_3 + \infty \cdot c} \nrec {v_1} {\lambda x.N_1'} {\save \infty 0 {(\lambda x.N_2')}} : C$, and so $\cdot \vdash_{a+b_1+r_1+r_2 + \infty \cdot c} (v_1,\lambda z.\nrec {v_1} {\lambda x.N_1'} {\save \infty 0 {(\lambda x.N_2')}}) : \N \otimes (1 \loli C)$. Thus, since $x : \N \otimes (1 \loli C) \vdash_{x + c} N_2' : C$,
  $$
  \cdot \vdash_{a+b_1+r_1+r_2 + \infty \cdot c} N_2'[(v_1,\lambda z.\nrec {v_1} {\lambda x.N_1'} {\save \infty 0 {(\lambda x.N_2')}})/x] : C  
  $$
  since $\infty \cdot c + c = \infty \cdot c$. So, by IH, $\cdot \vdash_{a+b_1 + r_1 + r_2 + \infty \cdot c + r_4} v : C$, and because $\infty \cdot c \leq b_2 + r_3$, we have by weakening that $\cdot \vdash_{a+b_1+b_2+r_1+r_2+r_3+r_4} v : C$ as required.
\end{itemize}
  
\item ($[A]$-E) Suppose $\cdot \vdash_{a + b_1 + b_2} \lrec M {N_1} {N_2} : C$. Then, $\cdot \vdash_a M : A$,  $\cdot \vdash_{b_1} N_1 : 1 \to C$, and $\cdot \vdash_{b_2} : N_2 : !^\infty_0(A \otimes (\listty A \amp C) \loli C)$. We have two evaluation cases to consider. 

Firstly, suppose $\lrec M {N_1} {N_2} \downarrow^{(\_,r_1+r_2+r_3+r_4)} v$ by way of $M \downarrow^{(\_,r_1)} v$, $N_1 \downarrow^{(\_,r_2)} \lambda x.N_1'$, $N_2 \downarrow^{(\_,r_3)} v'$, and $N_1'[()/x] \downarrow^{(\_,r_4)} v$. Then, by IH, $a + r_1 \geq 0$, which means that $\cdot \vdash_{a+r_1} () : 1$. By IH, $\cdot \vdash_{b_1 + r_2} \lambda x.N_1'$ and $b_1 + r_2 \geq 0$. So, by inversion and then substitution, $\cdot \vdash_{a+b_1+r_1+r_2} N_1'[()/x] : C$. By IH, $b_2 + r_3 \geq 0$, so by weakening, $\cdot \vdash_{a+b_1+b_2+r_1+r_2+r_3} N_1'[()/x]$. Finally, by IH, $a+b_1+b_2+r_1+r_2+r_3 + r_4 \geq 0$ and $\cdot \vdash_{a+b_1+b_2+r_1+r_2+r_3+r_4} v : C$.

\sloppypar Now, suppose $\lrec M {N_1} {N_2} \downarrow^{(\_,r_1+r_2+r_3+r_4)} v$ by way of $M \downarrow^{(\_,r_1)} \cons{v_1} {v_2}$, $N_1 \downarrow^{(\_,r_2)} \lambda x.N_1'$, $N_2 \downarrow^{(\_,r_3)} \save \infty 0 {(\lambda x.N_2')}$, and $N_2'[(v_1,\amppair {v_2} {\lrec {v_2} {\lambda x.N_1'} {\save \infty 0 {(\lambda x.N_2')}}})/x] \downarrow^{(\_,r_4)} v$. By IH, $\cdot \vdash_{a + r_1} \cons {v_1} {v_2} : \listty A$ and $a + r_1 \geq 0$. By inversion, there are $d_1,d_2 \geq 0$ so that $a+r_1 = d_1 + d_2$ and $\cdot \vdash_{d_1} v_1 : A$ and $\cdot \vdash_{d_2} v_2 : \listty A$. By two more applications of the IH, $\cdot \vdash_{b_1 + r_2} \lambda x.N_1' : 1 \loli C$, $\cdot \vdash_{b_2 + r_3} \save \infty 0 {(\lambda x.N_2')} : !^\infty_0(A\otimes(\listty A \amp C) \loli C)$, with $b_1 + r_2 \geq 0$ and $b_2 + r_3 \geq 0$. By inversion, there is some $c \geq 0$ with $\infty \cdot c \leq b_2 + r_3$ such that $\cdot \vdash_c \lambda x.N_2' : A \otimes (\listty A \amp C) \loli C$. Next,
$$
   \infer{
     \cdot \vdash_{d_2 + b_1 + r_2 + \infty \cdot c} \lrec {v_2} {\lambda x.N_1'} {\save \infty 0 {(\lambda x.N_2')}} : C
   }{
     \cdot \vdash_{d_2} v_2 : \listty A
     &
     \cdot \vdash_{b_1 + r_2} \lambda x.N_1' : 1 \loli C
     &
     \cdot \vdash_{\infty \cdot c} \save \infty 0 {(\lambda x.N_2')} : !^\infty_0 (A \otimes (\listty A \amp C) \loli C)
   }
$$
then, with $\cdot \vdash_{d_2 + b_1 + r_2 + \infty \cdot c} v_2 : \listty A$,
we have that
$$
\cdot \vdash_{d_2 + b_1 + r_2 + \infty \cdot c} \amppair {v_2} {\lrec {v_2} {\lambda x.N_1'} {\save \infty 0 {(\lambda x.N_2')}}} : \listty A \amp C
$$
and since $\cdot \vdash_{d_1} v_1 : A$,
$$
\cdot \vdash_{a + r_1 + b_1 + r_2 + \infty \cdot c} (v_1,\amppair {v_2} {\lrec {v_2} {\lambda x.N_1'} {\save \infty 0 {(\lambda x.N_2')}}}) : A \otimes (\listty A \amp C)
$$
and so by substitution, and using the fact that $c + \infty \cdot c = \infty \cdot c$, $\cdot \vdash_{a + b_1 + r_1 + r_2 + \infty \cdot c} N_2'[(v_1,\amppair {v_2} {\lrec {v_2} {\lambda x.N_1'} {\save \infty 0 {(\lambda x.N_2')}}})/x] : C$. By weakening, since $\infty \cdot c \leq b_2 + r_3$, 
$\cdot \vdash_{a + b_1 + b_2 + r_1 + r_2 + r_3} N_2'[(v_1,\amppair {v_2} {\lrec {v_2} {\lambda x.N_1'} {\save \infty 0 {(\lambda x.N_2')}}})/x] : C$. Finally, by IH, $\cdot \vdash_{a+b_1+b_2+r_1+r_2+r_3+r_4} v : C$, and $a+b_1+b_2+r_1+r_2+r_3+r_4 \geq 0$, as required.

\item ($\amp$-I): Immediate.
\item ($\amp$-E): By symmetry, it suffices to only consider the $\pi_1$ case. Let $\cdot \vdash_a \pi_1 M : A$ and $\pi_1 M \downarrow^{(\_,r_1+r_2)} v$. By inversion, we have that $\cdot \vdash_a M : A \amp B$, and that $M \downarrow^{(\_,r_1)} \amppair {N_1} {N_2}$ and $N_1 \downarrow^{(\_,r_2)}$. We must show that $\cdot \vdash_{a+r_1+r_2} v : A$, and that $a+r_1+r_2 \geq 0$. By IH, $\cdot \vdash_{a+r_1} \amppair {N_1} {N_2} : A \amp B$. Inverting this, we get that $\cdot \vdash_{a+r_1} N_1 : A$, and so again by IH, $\cdot_{a+r_1+r_2} v : A$, and $a+r_1+r_2 \geq 0$, as required.
\end{itemize}
\end{proof}

\valevalzero*
\begin{proof}
By inspection of cases.
\end{proof}
\natstren*
\begin{proof}
By canonical forms, $v = \overline{n}$, proceed by induction on $n$.
\end{proof}
\extrsound*
\begin{proof}
By induction on $\Gamma \vdash_f M : A$
\end{proof}
\weakening*
\begin{proof}
We prove 1 and 2 simultaneously by induction on $A$.
\begin{enumerate}
  \item Suppose $M \bdby^{A,a} E$, and $E \leq_{\mathbb{C} \times \angles{A}} E'$. We need to show that $M \bdby^{A,a} E'$. Suppose $M \downarrow^{(n,r)} v$.
  We need to show:
  \begin{itemize}
    \item $n \leq E'_c - r$
    \item $v \valbd^{A,a+r} E'_p$
  \end{itemize}
  But, since $M \bdby^{A,a} E$
  \begin{itemize}
    \item $n \leq E_c - r$
    \item $v \valbd^{A,a+r} E_p$
  \end{itemize} 
  so, it suffices to show that $E_c \leq_{\mathbb{C}} E'_c$ and $E_p \leq_{\angles A} E'_p$, which is true by the $\pi_1(-)$ and $\pi_2(-)$ congruences, recalling that $(-)_c$ and $(-)_p$ are simply $\pi_1$ and $\pi_2$.
  \item Let $E \leq_{\angles A} E'$. We have a few cases to consider.
  \begin{itemize}
    \item[($!$)] Suppose $\save k l v \valbd^{!^k_l A,a} E$. We must show that $\save k l v \valbd^{!^k_l A,a} E'$. We know that there is a $d \geq 0$ such that $ka+l \leq d$, and $v \valbd^{A,d} E$. So, by IH, $v \valbd^{A,d} E'$, and hence $\save k l v \valbd^{!^k_l A,a} E'$, as required.
    \item[($\loli$)] Suppose $\lambda x.M \valbd^{A\loli B,a} E$. We need to show that $\lambda x.M \valbd^{A\loli B,a} E'$. Let $v \valbd^{A,b} E_v$. Then, $M[v/x] \bdby^{B,a+b} E \; E_v$. Using the application congruence and 1, $M[v/x] \bdby^{B,a+b} E' \; E_v$. Since $v,b,E_v$ were chosen arbitrarily, $\lambda x.M \valbd^{A \loli B,a} E'$ as required.
    \item[($\tensor$)] Suppose $(v_1,v_2) \valbd^{A_1\otimes A_2,a} E$. Then, there are $a_1,a_2$ such that $a_1 + a_2 = a$, and $v_i \valbd^{A_i,a_i} \pi_i E$, and so by $\pi_i$-congruence and the IH, $v_i \valbd^{A_i,a_i} \pi_i E'$, so $(v_1,v_2) \valbd^{A_1\otimes A_2,a} E'$, as required.
    \item[($\listty{A}$)] Both cases are immediate by transitivity.
    \item[($\N$)] Both cases are immediate by transitivity.
    \item[($\oplus$)] Both cases are immediate by transitivity.
    \item[($A \amp B$)] Suppose $\amppair {M_1} {M_2} \valbd^{A_1 \amp A_2,a} E$. Then, for $i \in \{1,2\}$, $M_i \bdby^{A_i,a} \pi_i E$. By $\pi_i$-congruence, $\pi_i E \leq_{\norm{A}} \pi_i E'$, and so by IH from 1, we know that $M_i \bdby^{A_i,a} \pi_i E'$, and are done.
  \end{itemize}
\end{enumerate}
\end{proof}

\credwkn*
\begin{proof}
  We prove the two claims simultaneously.
  \begin{enumerate}
   \item[(1)] Suppose $M \bdby^{A,a_1} E$. To show $M \bdby^{A,a_2} E$, suppose $M \downarrow^{(n,r)} v$. We must show that
   such that
   \begin{itemize}
     \item $n \leq E_c - r$
     \item $v \valbd^{A,a_2+r} E_p$
   \end{itemize}
   But, since $M \bdby^{A,a_1} E$, we have
   \begin{itemize}
     \item $n \leq E_c - r$
     \item $v \valbd^{A,a_1 + r} E_p$
   \end{itemize}
  Since $a_1 \leq a_2$, $a_1 + r \leq a_2+ r$, so we are done by (2).
  \item[(2)] By lexicographic induction on first $A$ and then the size of $v$.
  \begin{itemize}
       \item[($!$)] Let $\save k l v \valbd^{!^k_l A,a_1} E$. Then, there is a $d \geq 0$ such that $kd + l \leq a_1$ and $v \valbd^{A,d} E$. But $kd+l\leq a_1 \leq a_2$, and so $\save k l v \valbd^{!^k_l A,a_2} E$
       \item[($\loli$)] Let $\lambda x.M \valbd^{A \loli B,a_1} E$. Suppose $v \valbd^{A,b} E'$. Then, $M[v/x] \bdby^{B,a_1 + b} E \; E'$, and so by (1), $M[v/x] \bdby^{B,a_2 + b} E \; E'$. Since $v$ was chosen arbitrarily, $\lambda x.M \valbd^{A \loli B,a_2} E$, as required.
    \item[($\tensor$)] Let $(v_1,v_2) \valbd^{A_1 \otimes A_2,a_1} E$. Then, there are $b_1,b_2$ with $b_1 + b_2$ such that $b_1 + b_2 = a_1$, and $v_i \valbd^{A_i,a_1} \pi_i E$ for $i \in \{1,2\}$. By the IH on $v_1$, we have that $v_1 \valbd^{A_1,b_1 + a_2 - a_1} \pi_1 E$, and so $(v_1,v_2) \valbd^{A_1 \otimes A_2,a_2} E$, as required.
    \item[($\listty{A}$)] The empty case is immediate. Suppose $\cons {v_1} {v_2} \valbd^{\listty A, a_1} E$. Then, there are $E_1, E_2,b_1,b_2$ such that $b_1 + b_2 = a_1$, $\cons {E_1} {E_2} \leq E$, $v_1 \valbd^{A,b_1} E_1$, and $v_1 \valbd^{\listty A,b_2} E_2$. By IH, $v_1 \valbd^{A,b_1 + a_2 - a_1} E_1$, and so $\cons {v_1} {v_2} \valbd^{\listty A, a_2} E$.
    \item[($\N$)] The zero case is immediate. Suppose $S(v) \valbd^{\N,a_1} E$. Then, there is $E'$ such that $S(E') \leq E$, and $v \valbd^{\N,a_1} E'$. Since $v$ is a smaller term than $S(v)$, we can apply the IH to see that $v \valbd^{\N,a_2} E'$, and so $S(v) \valbd^{\N,a_2} E$, as desired.
    \item[($\oplus$)] The two cases are symmetric, so we present only one. Suppose $\inl v \valbd^{A \oplus B,a_1} E$. Then we have $E'$ such that $\inl E' \leq E$, and $v \valbd^{A,a_1} E'$, which, by IH, means that $v \valbd^{A,a_2} E'$, and so $\inl v \valbd^{A \oplus B,a_2} E$.
    \item[($A \amp B$)] Immediate by IH.
  \end{itemize}
  \end{enumerate}
\end{proof}

\nreclemma*
\begin{proof}
Proceed by induction on $n$.

For notational simplicity, let $E_2^* = \lambda p. E_2 (\pi_1 p,(\lambda z. \pi_2 p))$

\begin{itemize}
  \item ($n = 0$): To show $\nrec 0 {\lambda x.N_1'} {\save \infty 0 {(\lambda x.N_2')}} \bdby^{C,c_3 + \infty \cdot d} \nrec {E} {E_1} {E_2^*}$, suppose that $\nrec 0 {\lambda x.N_1'} {\save \infty 0 {(\lambda x.N_2')}} \downarrow^{(n,r)} v$ by way of $N_1'[()/x] \downarrow^{(n,r)}$.
  
  We must show that :
  \begin{itemize}
    \item $n \leq \nrec E {E_1} {E_2^*}_c - r$
    \item $v \valbd^{C,c_3 + \infty \cdot d} \nrec E {E_1} {E_2^*}_p$
  \end{itemize}
  We know $N_1[()/x] \bdby^{C,c_3} E_1 \; ()$, since $() \valbd^{1,0} ()$, and so:
  \begin{itemize}
    \item $n \leq (E_1 \; ())_c - r$
    \item $v \valbd^{C,c_3} (E_1 \; ())_p$
  \end{itemize}
  
  Since $0 \valbd^{\N,0} E$, $0 \leq_\N E$, and so $E_1 \; () \leq \nrec 0 {E_1} {E_2^*} \leq \nrec E {E_1} {E_2^*}$.
  
  \item \sloppypar ($n > 0$): Suppose that $\nrec {S(\overline{n})} {\lambda x.N_1'} {\save \infty 0 {(\lambda x.N_2')}} \downarrow^{(n',r)} v$ by way of 
$$
  N_2'[(\overline{n},\lambda z.\nrec {\overline{n}} {\lambda x.N_1'} {\save \infty 0 {(\lambda x.N_2')}})] \downarrow^{(n',r)} v.
$$ 
We must show that:
  \begin{itemize}
    \item $n' \leq \nrec {E} {E_1} {E_2^*}_c - r$;
    \item $v \valbd^{C,c_3+\infty \cdot d + r} \nrec E {E_1} {E_2^*}_p$.
  \end{itemize}
  Since $S(\overline{n}) \valbd^{\N,0} E$, there is an $E'$ so that $S(E') \leq_\N E$, and $\overline{n} \valbd^{\N,0} E'$. For notational convenience, let $E^* = (E',\nrec {E'} {E_1} {E_2^*})$. Note that $E' \leq \pi_1 E^*$, and that $\nrec {E'} {E_1} {E_2^*} \leq \pi_2 E^*$. By IH, $\nrec {\overline{n}} {\lambda x.N_1'} {\save \infty 0 {(\lambda x.N_2')}} \bdby^{C,c_3 + \infty \cdot d} \nrec {E'} {E_1} {E_2^*}$, and thus by weakening $\nrec {\overline{n}} {\lambda x.N_1'} {\save \infty 0 {(\lambda x.N_2')}} \bdby^{C,c_3 + \infty \cdot d} \pi_2 E^*$. For some variable $z$ not free in the term on the left, $\lambda z. \nrec {\overline{n}} {\lambda x.N_1'} {\save \infty 0 {(\lambda x.N_2')}} \valbd^{1 \loli C,c_3 + \infty \cdot d} \lambda z. \pi_2 E^*$, and so $(\overline{n},\lambda z. \nrec {\overline{n}} {\lambda x.N_1'} {\save \infty 0 {(\lambda x.N_2')}}) \valbd^{\N \otimes (1 \loli C),c_3 + \infty \cdot d} (\pi_1 E^*,\lambda z. \pi_2 E^*s)$, and since $\lambda x.N_2' \valbd^{\N \times (1 \loli C) \loli C,d} E_2$, using the fact that $\infty \cdot d + d = \infty \cdot d$
  $$N_2'[(\overline{n},\lambda z. \nrec {\overline{n}} {\lambda x.N_1'} {\save \infty 0 {(\lambda x.N_2')}})/x] \bdby^{C,c_3 + \infty \cdot d} E_2 \; (\pi_1 E^*,\lambda z.\pi_2 E^*)$$
  but,
  \begin{align*}
   E_2 \; (\pi_1 E^*,\lambda z.\pi_2 E^*) &\leq (\lambda p. E_2 \; (\pi_1 p, \lambda z.\pi_2 p)) E^*\\
   &= E_2^* (E',\nrec {E'} {E_1} {E_2^*})\\
   &\leq \nrec {S(E')} {E_1} {E_2^*}\\
   &\leq \nrec {E} {E_1} {E_2^*}
  \end{align*}
  and so we are done by weakening.
\end{itemize}
\end{proof}

\lreclemma*
\begin{proof}
We proceed by induction on the derivation of $\cdot \vdash_d v : \listty A$.
First, suppose $v = []$.  To show that $\lrec {[]} {\lambda x.N_1'} {\save \infty 0 {(\lambda x.N_2')}} \bdby^{C,c_1+d+\infty \cdot c_2} \lrec E {E_1} {\lambda x. E_2 (\pi_1 x,((0,\pi_1 \pi_2 x),\pi_2\pi_2 x))}$, assume that  $\lrec {[]} {\lambda x.N_1'} {\save \infty 0 {(\lambda x.N_2')}} \downarrow^{(n,r)} v$. By inversion, it was by way of $N_1'[()/x] \downarrow^{(n,r)} v$. It suffices to show 
\begin{itemize}
   \item $n \leq \lrec E {E_1} {\lambda x. E_2 (\pi_1 x,((0,\pi_1 \pi_2 x),\pi_2\pi_2 x))}_c - r$
   \item $v\ \valbd^{C,c_1+d+\infty \cdot c_2 + r} \lrec E {E_1} {\lambda x. E_2 (\pi_1 x,((0,\pi_1 \pi_2 x),\pi_2\pi_2 x)) }_p$
\end{itemize}
Since $() \leq_1 ()$, $() \valbd^{1,d} ()$, so $N_1'[()/x] \bdby^{C,c_1+d} E_1 \; ()$, and so
\begin{itemize}
  \item $n \leq (E_1 \; ())_c - r$
  \item $v \valbd^{C,c_1+d+r} (E_1 \; ())_p$
\end{itemize}
But, $\infty \cdot c_2 > 0$ since $c_2 > 0$, and so by credit weakening, $v \valbd^{C,c_1+d+\infty \cdot c_2 +r} (E_1 \; ())_p$. Note that, by assumption, $[] \valbd^{1,d} E$, which means that $[] \leq_{\listty {\angles A}} E$. So,
$$
E_1 \; () \leq \lrec {[]} {E_1} {\lambda x. E_2 (\pi_1 x,((0,\pi_1 \pi_2 x),\pi_2\pi_2 x)) } \leq \lrec E {E_1} {\lambda x. E_2 (\pi_1 x,((0,\pi_1 \pi_2 x),\pi_2\pi_2 x)) }
$$
and so we are done by weakening.

Otherwise, suppose $v = \cons {v_1} {v_2}$. To show that $\lrec {\cons {v_1} {v_2}} {\lambda x.N_1'} {\save \infty 0 {(\lambda x.N_2')}} \bdby^{C,c_1+d+\infty \cdot c_2} \lrec E {E_1} {\lambda x. E_2 (\pi_1 x,((0,\pi_1 \pi_2 x),\pi_2\pi_2 x)) }$, suppose $\lrec {\cons {v_1} {v_2}} {\lambda x.N_1'} {\save \infty 0 {(\lambda x.N_2')}} \downarrow^{(n,r)} v$. By inversion, it was by $N_2'[(v_1,\amppair {v_2} {\lrec {v_2} {\lambda x.N_1'} {\save \infty 0 {(\lambda x.N_2')}}})/x] \downarrow^{(n,r)} v$. It suffices to show:
\begin{itemize}
  \item $n \leq \lrec E {E_1} {\lambda x. E_2 (\pi_1 x,((0,\pi_1 \pi_2 x),\pi_2\pi_2 x)) }_c - r$
  \item $v \valbd^{C,c_1+d+\infty \cdot c_2 + r} \lrec E {E_1} {\lambda x. E_2 (\pi_1 x,((0,\pi_1 \pi_2 x),\pi_2\pi_2 x)) }_p$
\end{itemize}

Since $\cons {v_1} {v_2} \valbd^{\listty A,d} E$, there are $d_1,d_2 \geq 0$ such that $d_1 + d_2 = d$, along with $E'$,$E''$ such that $v_1 \valbd^{A,d_1} E'$ and $v_2 \valbd^{\listty A,d_2} E''$, and $\cons {E'} {E''} \leq E$

By IH, $\lrec {v_2} {\lambda x.N_1'} {\save \infty 0 {(\lambda x.N_2')}} \bdby^{C,c_1 + d_2 + \infty \cdot c_2} \lrec {E''} {E_1} {\ldots}$. 
Since $v_2 \valbd^{\listty A,d_2} E''$, $v_2 \bdby^{\listty A,d_2} (0,E'')$, 
and since $c_1 + \infty \cdot c_2 \geq 0$, we have by credit weakening that $v_2 \bdby^{\listty A,c_1 + d_2 + \infty \cdot c_2} (0,E'')$. 
So, 
$\amppair {v_2} {\lrec {v_2} {\lambda x.N_1'} {\save \infty 0 {(\lambda x.N_2')}}} \valbd^{\listty A \amp C,c_1 + d_2 + \infty \cdot c_2} ((0,E''),\lrec {E''} {E_1} {\ldots})$. 
Further, using the fact that $d_1 + d_2 = d$, 

$$
\begin{array}{l}
(v_1,\amppair {v_2} {\lrec {v_2} {\lambda x.N_1'} {\save \infty 0 {(\lambda x.N_2')}}}) \valbd^{A \otimes (\listty A \amp C), c_1 + d + \infty \cdot c_2}\\ (E',((0,E''),\lrec {E''} {E_1} {\ldots}))
\end{array}
$$
Thus, since $\lambda x.N_2' \valbd^{A \otimes (\listty A \amp C) \loli C,c_2} E_2$, we have (using the fact that $c_2 + \infty \cdot c_2 = \infty \cdot c_2$)
$$
N_2'[(v_1,\amppair {v_2} {\lrec {v_2} {\lambda x.N_1'} {\save \infty 0 {(\lambda x.N_2')}}})/x] \bdby^{C,c_1+d+\infty \cdot c_2} E_2 \; (E',((0,E''),\lrec {E''} {E_1} {\ldots}))
$$
By definition, this means that
\begin{itemize}
  \item $n \leq (E_2 \; (E',((0,E''),\lrec {E''} {E_1} {\ldots})))_c - r$
  \item $v \valbd^{c_1+d+\infty \cdot c_2 + r} (E_2 \; (E',((0,E''),\lrec {E''} {E_1} {\ldots})))_p$
\end{itemize}
We then compute:
\begin{align*}
&E_2 \; (E',((0,E''),\lrec {E''} {E_1} {\ldots}))\\
&\leq (\lambda x. E_2 (\pi_1 x,((0,\pi_1 \pi_2 x),\pi_2\pi_2 x))) \; (E',(E'',\lrec {E''} {E_1} {\lambda x. E_2 (\pi_1 x,((0,\pi_1 \pi_2 x),\pi_2\pi_2 x))}))\\
&\leq \lrec {\cons {E'} {E''}} {E_1} {\lambda x. E_2 (\pi_1 x,((0,\pi_1 \pi_2 x),\pi_2\pi_2 x))s}\\
&\leq \lrec {E} {E_1} {\lambda x. E_2 (\pi_1 x,((0,\pi_1 \pi_2 x),\pi_2\pi_2 x)) }
\end{align*}

and hence we are done by weakening.
\end{proof}

\bounding*
\begin{proof}
By induction on $\Gamma \vdash_f M : A$.
\begin{itemize}

\item[($!$-I)] 
Let $\Gamma \vdash_g \save k l M : !^k_l A$. By inversion, we have $\Gamma \vdash_f M : A$ with $kf + l \leq g$. Let $\theta \subbd^{\Gamma,\sigma} \Theta$. To show $\save k l {M}[\theta] \bdby^{!^k_l A,g[\sigma]} (k\norm{M}[\Theta]_c,\norm{M}[\Theta]_p)$, it suffices to show $\save k l {M}[\theta] \bdby^{!^k_l A,kf[\sigma] + l} (k\norm{M}[\Theta]_c,\norm{M}[\Theta]_p)$ by credit weakening. So, let $\save k l M \downarrow^{(n,kr)} \save k l v$ by way of $M \downarrow^{(n,r)} v$. It suffices to show, using the fact that $kf[\sigma] + l + kr = k(f[\sigma] + r) + l$
\begin{itemize}
  \item $n \leq k\norm{M}[\Theta]_c - kr$
  \item $\save k l v \valbd^{!^k_l A,k(f[\sigma] + r) + l} \norm{M}[\Theta]_p$
\end{itemize}

To show $\save k l v \valbd^{!^k_l A,k(f[\sigma] + r) + l} \norm{M}[\Theta]_p$, it suffices to provide $d \geq 0$ such that $kd+l \leq k(f[\sigma] + r) + l$, and $v \valbd^{A,d} \norm{M}[\Theta]_p$.
By IH, $M[\theta] \bdby^{A,f[\sigma]} \norm{M}[\Theta]$, which means that
\begin{itemize}
 \item $n \leq \norm{M}[\Theta]_c - r$
 \item $v \valbd^{A,f[\sigma] + r} \norm{M}[\Theta]_p$
\end{itemize}
So, $d = f[\sigma] + r$, and the inequality $n \leq kn \leq k\norm{M}[\Theta]_c - kr$ follows by multiplying the above one by $k$ (since $k \geq 1$).

\item[($!$-E)]  Let $\Gamma \vdash_{k'f + g} \tsfer {k'} k l x M N : C$. By inversion, we have that $\Gamma \vdash_f M : !^k_l A$, as well as $\Gamma,x : A\vdash_{g+k'(kx+l)} N : C$. Suppose $\theta \subbd^{\Gamma,\sigma} \Theta$. 
We need to show that $\tsfer {k'} k l x {M[\theta]} {N[\theta]} \bdby^{C,k'f[\sigma] + g[\sigma]} \norm{M}[\Theta]_c +_c \norm{N}[\Theta,\norm{M}[\Theta]_p/x]$. 
Suppose $\tsfer {k'} k l x {M[\theta]} {N[\theta]} \downarrow^{(n_1+n_2,k'r_1+r_2)} v$. By inversion, it was by $M[\theta] \downarrow^{(n_1,r_1)} \save k l {v_1}$ and $N[\theta,v_1/x] \downarrow^{(n_2,r_2)} v$.
It suffices to show that
\begin{itemize}
  \item $n_1 + n_2 \leq k'\norm{M}[\Theta]_c + \norm{N}[\Theta,\norm{M}[\Theta]_p/x]_c - (k'r_1 + r_2)$
  \item $v \valbd^{C,k'f[\sigma] + g[\sigma] + k'r_1 + r_2} \norm{N}[\Theta,\norm{M}[\Theta]_p/x]_p$
\end{itemize}
By IH, we have that $M[\theta] \bdby^{!^k_l A, f[\sigma]} \norm{M}[\Theta]$, which means that there are $b_1,c_1$ with $b_1 + c_1 = f[\sigma]$ and
\begin{itemize}
  \item $n_1 \leq \norm{M}[\Theta]_c - r_1$
  \item $\save k l {v_1} \valbd^{!^k_l A,f[\sigma] + r_1} \norm{M}[\Theta]_p$
\end{itemize}
Since $\save k l {v_1} \valbd^{!^k_l A,f[\sigma] + r_1} \norm{M}[\Theta]_p$, there is a $d \geq 0$ such that $kd+l \leq f[\sigma] + r_1$, and $v_1 \valbd^{A,d} \norm{M}[\Theta]_p$. Thus, $(\theta,v_1/x) \subbd^{(\Gamma,x:A),(\sigma,x\mapsto d)} (\Theta,\norm{M}[\Theta]_p/x)$, and so by IH, $N[\theta,v_1/x] \bdby^{C,g[\sigma] + k'(kd+l)} \norm{N}[\Theta,\norm{M}[\Theta]_p/x]$.
By credit weakening, since $kd+l \leq f[\sigma] + r_1$,  $N[\theta,v_1/x] \bdby^{C,g[\sigma] + k'(f[\sigma] + r_1)} \norm{N}[\Theta,\norm{M}[\Theta]_p/x]$. This gives us that
\begin{itemize}
  \item $n_2 \leq \norm{N}[\Theta,\norm{M}[\Theta]_p/x]_c - r_2$
  \item $v \valbd^{C,g[\sigma] + k'(f[\sigma] + r_1) + r_2} \norm{N}[\Theta,\norm{M}[\Theta]_p/x]_p$
\end{itemize}
To establish the desired inequality, we multiply the first inequality by $k'$, to find that $k'n_1 \leq k'\norm{M}[\Theta]_c - k'r_1$. But $k' \geq 1$,
so $n_1 \leq k'n_1 \leq k'\norm{M}[\Theta]_c - k'r_1$. Therefore, $n_1 + n_2 \leq k'\norm{M}[\Theta]_c + \norm{N}[\Theta,\norm{M}[\Theta]_p/x]_c - (k'r_1 + r_2)$ as required. For value bounding, we note that $g[\sigma] + k'(f[\sigma] + r_1) + r_2 = k'f[\sigma] + g[\sigma] + k'r_1 +r_2$, and are done.

\item[(\discname)] Let $\Gamma \vdash_{f + l} \disc l M : A$.  By inversion, $\Gamma \vdash_f M : A$.
 To show $\disc l M \bdby^{A} (-l) +_c \norm{M}$, suppose $\theta \subbd^{\Gamma,\sigma} \Theta$. 
 To show $\disc l M[\theta] \bdby^{A,f[\sigma] + l} (-l) +_c \norm{M}[\Theta]$, suppose $\disc l M[\theta] \downarrow^{(n,r-l)} v$. 
 By inversion we also have that $M[\theta] \downarrow^{(n,r)} v$. 
 It suffices to show
 \begin{itemize}
   \item $n \leq -l + \norm{M}[\Theta]_c - (r - l)$
   \item $v \valbd^{A,f[\sigma] + l + r - l} \norm{M}[\Theta]_p$
 \end{itemize}
 or, canceling, it suffices to show $n \leq \norm{M}[\Theta]_c  - r$ and $v \valbd^{A,f[\sigma] + r} \norm{M}[\Theta]_p$, which is precisely what we get from the IH.

\item[(\waitname)] Let $\Gamma \vdash_f \wait l M : A$. By inversion, $\Gamma \vdash_{f+l} M : A$. To show $\wait l M \bdby^{A} l +_c \norm{M}$, suppose $\theta \subbd^{\Gamma,\sigma} \Theta$, 
to show $\wait l M[\theta] \bdby^{A,f[\sigma]} l +_c \norm{M}[\Theta]$, 
suppose $\wait l M[\theta] \downarrow^{(n,r+l)} v$. 
By inversion, $M[\theta] \downarrow^{(n,r)} v$.
It suffices to show
\begin{itemize}
  \item $n \leq l + \norm{M}[\theta]_c - (r + l)$
  \item $v \valbd^{f[\sigma] + r + l}$
\end{itemize}
By IH, we have that $M[\theta] \bdby^{A,f[\sigma] + l} \norm{M}[\Theta]$, so
\begin{itemize}
  \item $n \leq \norm{M}[\Theta]_c - r$
  \item $v \valbd^{A,f[\sigma] + l + r}$
\end{itemize}
and so we are done, canceling the $l$s in the first inequality.

\item[(tick)] Immediate from IH, canceling $1$s.

\item[($\loli$-I)] Let $\Gamma \vdash_f \lambda x.M : A \loli B$. By inversion, $\Gamma, x : A \vdash_{f + x} M : B$. 
Let $\theta \subbd^{\Gamma,\sigma} \Theta$. 
To show $\lambda x.M[\theta] \bdby^{A\loli B,f[\sigma]} (0,\lambda x.\norm{M}[\Theta])$, let $\lambda x.M[\theta] \downarrow^{(0,0)} \lambda x.M[\theta]$.
The first condition is trivial ($0 \leq 0$). We need to show that $\lambda x.M \valbd^{f[\sigma]} \lambda x.\norm{M}[\Theta]$. Let $v \valbd^{A,d} E$. We must show that $M[\theta,v/x] \bdby^{B,f[\sigma] + d} (\lambda x.\norm{M}[\Theta]) E$, or by weakening, that $M[\theta,v/x] \bdby^{B,f[\sigma]+d} \norm{M}[\Theta,E/x]$. But, since $v \valbd^{A,d} E$, we have $(\theta,v/x) \subbd^{(\Gamma,x:A),(\sigma, x \mapsto d)} (\Theta,E/x)$, and so by IH, $M[\theta,v/x] \bdby^{B,f[\sigma] + d} \norm{M}[\Theta,E/x]$, as required.

\item[($\loli$-E)] Let $\Gamma \vdash_{f + g} M \, N : B$.  Inversion gives $\Gamma \vdash_f M : A \loli B$ and $\Gamma \vdash_g N : A$. Let $\theta \subbd^{\Gamma,\sigma} \Theta$. We must show $M[\theta] \, N[\theta] \bdby^{B,f[\sigma] + g[\sigma]} (\norm{M}[\Theta]_c + \norm{N}[\Theta]_c) +_c \norm{M}[\Theta]_p\norm{N}[\Theta]_p$. Suppose $M[\theta] \, N[\theta] \downarrow^{(n_1+n_2+n_3,r_1+r_2+r_3)} v$. Inversion gives us that $M[\theta] \downarrow^{(n_1,r_1)} \lambda x.M'$, $N[\theta] \downarrow^{(n_2,r_2)} v_1$, and $M'[v_1/x] \downarrow^{(n_3,r_3)} v$. It remains to show that 
\begin{itemize}
  \item $n_1 + n_2 + n_3 \leq \norm{M}[\Theta]_c + \norm{N}[\Theta]_c + (\norm{M}[\Theta]_p\norm{N}[\Theta]_p)_c - (r_1 + r_2 + r_3)$
  \item $v \valbd^{B,f[\sigma] + g[\sigma] + r_1 + r_2 + r_3} (\norm{M}[\Theta]_p\norm{N}[\Theta]_p)_p$
\end{itemize}
 By the IH applied to $\Gamma \vdash_f M : A \loli B$, we know that $M[\theta] \bdby^{A\loli B,f[\sigma]} \norm{M}[\Theta]$, so
 \begin{itemize}
  \item $n_1 \leq \norm{M}[\Theta]_c - r_1$
  \item $\lambda x.M' \valbd^{A \loli B,f[\sigma] + r_1} \norm{M}[\Theta]_p$.
 \end{itemize}  
 Again applying the IH to $\Gamma \vdash_g N : A$, we know $N[\theta] \bdby^{A,g[\sigma]} \norm{N}[\Theta]$, so
\begin{itemize}
  \item $n_2 \leq \norm{N}[\Theta]_c - r_2$
  \item $v_1 \valbd^{A,g_[\sigma] + r_2} \norm{N}[\Theta]_p$
\end{itemize}
  
But since $\lambda x.M' \valbd^{A \loli B,f[\sigma] + r_1} \norm{M}[\Theta]_p$ and $v_1 \valbd^{A,g[\sigma] + r_2} \norm{N}[\Theta]_p$, we have $M'[v_1/x] \bdby^{B,f[\sigma] + g[\sigma] + r_1 + r_2} \norm{M}[\Theta]_p\norm{N}[\Theta]_p$, which means that:
\begin{itemize}
  \item $n_3 \leq (\norm{M}[\Theta]_p\norm{N}[\Theta]_p)_c - r_3$
  \item $v \valbd^{B,f[\sigma] + g[\sigma] + r_1 + r_2 + r_3} (\norm{M}[\Theta]_p\norm{N}[\Theta]_p)_p$
\end{itemize}
We add the inequalities together, and are done.

\item[($\tensor$-I)] Let $\Gamma \vdash_{f_1 + g_1} (M_1,M_2) : A_1 \tensor A_2$. By inversion, we have that $\Gamma \vdash_{f_i} M_i : A_i$ for $i = 1,2$. Let $\theta \subbd^{\Gamma,\sigma} \Theta$. Towards proving $(M_1[\theta],M_2[\theta]) \bdby^{A_1\tensor A_2,f_1[\sigma] + f_2[\sigma]} (\norm{M_1}[\Theta]_c + \norm{M_2}[\Theta]_c,(\norm{M_1}[\Theta]_p,\norm{M_2}[\Theta]_p))$, assume $(M_1[\theta], M_2[\theta]) \downarrow^{(n_1+n_2,r_1+r_2)} (v_1,v_2)$. By inversion, it must also be that $M_i[\theta] \downarrow^{(n_i,r_i)} v_i$ for $i = 1,2$.
If suffices to show:
\begin{itemize}
  \item $n_1 + n_2 \leq \norm{M_1}[\Theta]_c + \norm{M_2}[\Theta] - (r_1 + r_2)$
  \item $(v_1,v_2) \valbd^{A_1 \tensor A_2,f_1[\sigma] + f_2[\sigma] + r_1 + r_2} (\norm{M_1}[\Theta]_p,\norm{M_2}[\Theta]_p)$
\end{itemize}
By IH, we have that, for $i \in \{1,2\}$
\begin{itemize}
  \item $n_i \leq \norm{M_i}[\Theta]_c - r_i$
  \item $v_i \valbd^{A_i,f[\sigma]_i + r_i} \norm{M_i}[\Theta]_p$
\end{itemize}
Adding the two inequalities and applying the definition of value bounding at $\otimes$, we are done.

\item[($\tensor$-E)]
Let $\Gamma \vdash_{k'f + g} \asplit {k'} M x y N : C$. Inversion gives $\Gamma \vdash_f M : A \tensor B$, and $\Gamma,x : A, y : B \vdash_{g + k'(x + y)} N : C$. Let $\theta \subbd^{\Gamma,\sigma} \Theta$. We must show that
$$
\asplit {k'} {M[\theta]} x y {N[\theta]} \bdby^{C,k'f[\sigma] + g[\sigma]} k'\norm{M}[\Theta]_c +_c \norm{N}[\Theta,\pi_1\norm{M}[\Theta]_p/x,\pi_2\norm{M}[\Theta]_p/y]
$$
Suppose that $\asplit {k'} {M[\theta]} x y {N[\theta]} \downarrow^{(n_1 + n_2,k'r_1 + r_2)} v$ by way of $M[\theta] \downarrow^{(n_1,r_1)} (v_1,v_2)$ and $N[\theta,v_1/x,v_2/y] \downarrow^{(n_2,r_2)} v$. It remains to show that
\begin{itemize}
  \item $n_1 + n_2 \leq k'\norm{M}[\Theta]_c + \norm{N}[\Theta,\pi_1\norm{M}[\Theta]_p/x,\pi_2\norm{M}[\Theta]_p/y]_c - (k'r_1 + r_2)$
  \item $v \valbd^{C,k'f[\sigma] + g[\sigma] + k'r_1 + r_2} \norm{N}[\Theta,\pi_1\norm{M}[\Theta]_p/x,\pi_2\norm{M}[\Theta]_p/y]_p$
\end{itemize}
By IH, $M[\theta] \bdby^{A \otimes B,f[\sigma]} \norm{M}[\Theta]$, so
\begin{itemize}
  \item $n_1 \leq \norm{M}[\Theta]_c - r_1$
  \item $(v_1,v_2) \valbd^{A \otimes B,f[\sigma] + r_1} \norm{M}[\Theta]_p$
\end{itemize}
and so there are $c_1,c_2 \geq 0$ so that $c_1 + c_2 = f[\sigma] + r_1$ and $v_1 \valbd^{A,c_1} \pi_1\norm{M}[\Theta]_p$ and $v_2 \valbd^{B,c_2} \pi_2\norm{M}[\Theta]_p$. So, $(\theta,v_1/x,v_2/y) \subbd^{(\Gamma,x:A,y:B),(\sigma,x\mapsto c_1,y\mapsto c_2)} (\Theta,\pi_1\norm{M}[\Theta]_p/x,\pi_2\norm{M}[\Theta]_p/y)$. So, by IH,
$N[\theta,v_1/x,v_2/y] \bdby^{C,g[\sigma] + k'(f[\sigma] + r_1)} \norm{N}[\Theta,\pi_1\norm{M}[\Theta]_p/x,\pi_2\norm{M}[\Theta]_p/y]$, so
\begin{itemize}
  \item $n_2  \leq \norm{N}[\Theta,\pi_1\norm{M}[\Theta]_p/x,\pi_2\norm{M}[\Theta]_p/y]_c - r_2$
  \item $v \valbd^{C,k'f[\sigma] + g[\sigma] + k'r_1 + r_2} \norm{N}[\Theta,\pi_1\norm{M}[\Theta]_p/x,\pi_2\norm{M}[\Theta]_p/y]_p$
\end{itemize}
Then,
\begin{align*}
 n_1 + n_2 &\leq k'n_1 + n_2\\
 &\leq k'\norm{M}[\Theta]_c + \norm{N}[\Theta,\pi_1\norm{M}[\Theta]_p/x,\pi_2\norm{M}[\Theta]_p/y]_c - (k'r_1 + r_2)
\end{align*}
as required.

\item[($\oplus$-E)] Let $\Gamma \vdash_{k'f + g_1 + g_2} \acase {k'} M x {N_1} y {N_2} : C$. By inversion, $\Gamma \vdash_f M : A \oplus B$, $\Gamma,x:A \vdash_{g_1 + k'x} N_1 : C$, and $\Gamma, y : B \vdash_{g_2 + k'y} N_2 : C$. Let $\theta \subbd^{\Gamma,\sigma} \Theta$. We must show that
$$
\acase {k'} {M[\theta]} x {N_1[\theta]} y {N_2[\theta]} \bdby^{C,k'f[\sigma] + g_1[\sigma] + g_2[\sigma]} k'\norm{M}[\Theta]_c +_c \ccase {\norm{M}[\Theta]_p} x {\norm{N_1}} y {\norm{N_2}}
$$
\sloppypar Because the two cases are symmetric, we consider only the following evaluation: $\acase {k'} {M[\theta]} x {N_1[\theta]} y {N_2[\theta]} \downarrow^{(n_1 + n_2,k'r_1 + r_2)} v$ by way of $M[\theta] \downarrow^{(n_1,r_1)} \inl {v_1}$ and $N_1[\theta,v_1/x] \downarrow^{(n_2,r_2)}v$. We must show that
\begin{itemize}
  \item $n_1 + n_2 \leq k'\norm{M}[\Theta]_c + \ccase {\norm{M}[\Theta]_p} x {\norm{N_1}} y {\norm{N_2}}_c - (k'r_1 + r_2)$
  \item $v \valbd^{C,k'f[\sigma]+g_1[\sigma] + g_2[\sigma] + k'r_1 + r_2} \ccase {\norm{M}[\Theta]_p} x {\norm{N_1}} y {\norm{N_2}}_p$
\end{itemize}
By IH, $M[\theta] \bdby^{A \oplus B,f[\sigma]} \norm{M}[\Theta]$, so
\begin{itemize}
  \item $n_1 \leq \norm{M}[\Theta]_c - r_1$
  \item $\inl {v_1} \valbd^{A \oplus B,f[\sigma] + r_1}\norm{M}[\Theta]_p$
\end{itemize}
so there is an $E$ such that $\inl E \leq_{\angles A + \angles B} \norm{M}[\Theta]_p$ and $v_1 \valbd^{A,f[\sigma] + r_1} E$. So, $(\theta,v_1/x) \subbd^{(\Gamma,x:A),(\sigma,x\mapsto f[\sigma]+r_1)} (\Theta,E/x)$ and hence by IH, $N_1[\theta,v_1/x] \bdby^{g_1[\sigma]  + k'(f[\sigma] + r_!)} \norm{N_1}[\Theta,E/x]$, so
\begin{itemize}
  \item $n_2 \leq \norm{N_1}[\Theta,E/x]_c - r_2$
  \item $v \valbd^{C,k'f[\sigma] + g_1[\sigma] + k'r_1 + r_2} \norm{N_1}[\Theta,E/x]_p$
\end{itemize}
Since $g_2[\sigma] \geq 0$, we have by credit weakening that $v \valbd^{C,k'f[\sigma] + g_1[\sigma] + g_2[\sigma] + k'r_1 + r_2} \norm{N_1}[\Theta,E/x]_p$.
Then, we compute:
\begin{align*}
\norm{N_1}[\Theta,E/x] &\leq_{\norm C} \textsc{\ccase {\inl E} x {\norm{N_1}[\Theta]} y {\norm{N_2}[\Theta]}}\\
&\leq \ccase {\norm{M}[\Theta]_p} x {\norm{N_1}[\Theta]} y {\norm{N_2}[\Theta]}
\end{align*}
which gives us the value bounding condition, and again compute:
\begin{align*}
n_1 + n_2 &\leq k'n_1 + n_2\\
&\leq k'\norm{M}[\Theta]_c + \ccase {\norm{M}[\Theta]_p} x {\norm{N_1}} y {\norm{N_2}}_c - (k'r_1 + r_2)
\end{align*}
which gives us the cost bounding condition.

\item[($\oplus$-I)] The cases for $\inl M$ and $\inr M$ are symmetric, so we let $\Gamma \vdash_f \inl M : A \oplus B$. Inversion gives $\Gamma \vdash_f M : A$. Let $\theta \subbd^{\Gamma,\sigma} \Theta$. 
To show that $\inl {M[\theta]} \bdby^{A \oplus B,f[\sigma]} (\norm{M}[\Theta]_c,\inl{\norm{M}[\Theta]_p})$,
we let $\inl{M[\theta]} \downarrow^{(n,r)} \inl v$. 
Inverting, we have $M[\theta] \downarrow^{(n,r)} v$. 
It suffices to show that $n \leq \norm{M}[\Theta]_c - r$, and that $\inl v \valbd^{A \oplus B,f[\sigma] + r} \inl {\norm{M}[\Theta]_p}$. By IH we have $n \leq \norm{M}[\Theta]_c - r$, and $v \valbd^{A,f[\sigma] + r} \norm{M}[\Theta]_p$. So we are done by the definition of value bounding at $\oplus$ for $\texttt{inl}$.

\item[($\listty A$-I, cons)] Let $\Gamma \vdash_{f+g} \cons M N : \listty A$. 
By inversion, $\Gamma \vdash_f M : A$ and $\Gamma \vdash_g N : \listty A$.
Let $\theta \subbd^{\Gamma,\sigma} \Theta$. 
To show that $\cons M N \bdby^{\listty A, f[\sigma] + g[\sigma]} (\norm{M}[\Theta]_c + \norm{N}[\Theta]_c, \cons {\norm{M}[\Theta]_p} {\norm{N}[\Theta]_p})$, 
let $\cons M N \downarrow^{(n_1 + n_2,r_1 + r_2)} \cons {v_1} {v_2}$. 
By inversion, $M \downarrow^{(n_1,r_1)} v_1$ and $N \downarrow^{(n_2,r_2)} v_2$. It suffices to provide $b,c$ where $c \geq 0$ and $b + c = f[\sigma] + g[\sigma]$ and that
\begin{itemize}
  \item $n_1 + n_2 \leq\norm{M}[\Theta]_c + \norm{N}[\Theta]_c - (r_1 + r_2)$
  \item $\cons {v_1} {v_2} \valbd^{\listty A, f[\sigma] + g[\sigma] + r_1 + r_2} \cons {\norm{M}[\Theta]_p} {\norm{N}[\Theta]_p}$.
\end{itemize}
By IH,
\begin{itemize}
  \item $n_1 \leq \norm{M}[\Theta]_c - r_1$
  \item $v_1 \valbd^{A,f[\sigma] + r_1} \norm{M}[\Theta]_p$
\end{itemize}
and by IH on $N$, there are $b_2,c_2$ with $c_2 \geq 0$ and $b_2 + c_2 = g[\sigma]$ such that
\begin{itemize}
  \item $n_2 \leq \norm{N}[\Theta]_c - r_2$
  \item $v_2 \valbd^{\listty A,g[\sigma] + r_2} \norm{N}[\Theta]_p$
\end{itemize}
Thus, the desired inequality follows from adding the two inductively computed ones, and the value bounding relation for $\cons {v_1} {v_2}$ is immediate by the definition.

\item[($\listty A$-E)] Suppose $\Gamma \vdash_{f+g_1+g_2} \lrec M {N_1} {N_2} : C$. By inversion, $\Gamma \vdash_f M : \listty A$, $\Gamma \vdash_{g_1} N_1 : 1 \loli C$, $\Gamma \vdash_{g_2} N_2 : !^\infty_0 (A \otimes (\listty A \amp C) \loli C)$. Let $\theta \subbd^{\Gamma,\sigma} \Theta$. To show that
\begin{align*}
\lrec {M[\theta]} {N_1[\theta]} {N_2[\theta]} \bdby^{C,f[\sigma] + g_1[\sigma] + g_2[\sigma]} &(\norm{M}[\Theta]_c + \norm{N_1}[\Theta]_c + \norm{N_2}[\Theta]_c) +_c\\ 
&\lrec {\norm{M}[\Theta]_p} {\norm{N_1}[\Theta]_p} {\lambda (a,(as,r)). \norm{N_2}[\Theta]_p\;  (a,((0,as),r))}
\end{align*}
we break into the two evaluation cases. Firstly, suppose that $\lrec {M[\theta]} {N_1[\theta]} {N_2[\theta]} \downarrow^{(n_1+n_2+n_3+n_4,r_1+r_2+r_3+r_4)} v$ by way of $M[\theta] \downarrow^{(n_1,r_1)} \elist$, $N_1[\theta] \downarrow^{(n_2,r_2)} \lambda x.N_1'$, $N_2[\theta] \downarrow^{(n_3,r_3)} \save \infty 0 {(\lambda x.N_2')}$, and $N_1'[()/x] \downarrow^{(n_4,r_4)} v$. From here, denote we denote $\lambda x. \norm{N_2}[\Theta]_p (\pi_1 x,((0,\pi_1 \pi_2 x),\pi_2\pi_2 x))$ as $\norm{N_2}^*$.

 It suffices to show that:
\begin{itemize}
  \item $n_1 + n_2 + n_3 + n_4 \leq \norm{M}[\Theta]_c + \norm{N_1}[\Theta]_c + \norm{N_2}[\Theta]_c + \lrec {\norm{M}[\Theta]_p} {\norm{N_1}[\Theta]_p} {\norm{N_2}^*}_c - (r_1 + r_2 + r_3 + r_4)$
  \item $v \valbd^{C,f[\sigma] + g_1[\sigma] + g_2[\sigma] + r_1 + r_2 + r_3 + r_4} \lrec {\norm{M}[\Theta]_p} {\norm{N_1}[\Theta]_p} {\norm{N_2}^*}_p$
\end{itemize}
By IH, $M[\theta] \bdby^{\listty A,f[\sigma]} \norm{M}[\Theta]$, so
\begin{itemize}
  \item $n_1 \leq \norm{M}[\Theta]_c - r_1$
  \item $\elist \valbd^{f[\sigma] + r_1} \norm{M}[\Theta]_p$
\end{itemize}
The second condition tells us that $\elist \leq \norm{M}[\Theta]_p$.
Again by IH, $N_2[\theta] \bdby^{!^\infty_0(A \otimes (\listty A \amp C) \loli C),g_2[\sigma]} \norm{N_2}[\Theta]$, so
\begin{itemize}
  \item $n_3 \leq \norm{N_2}[\Theta]_p - r_3$
  \item $\save \infty 0 {(\lambda x.N_2')} \valbd^{!^\infty_0(A \otimes (\listty A \amp C) \loli C),g_2[\sigma] + r_3} \norm{N_2}[\Theta]_p$
\end{itemize}
In particular, by preservation, $g_2[\sigma] + r_3 \geq 0$.
Thirdly by IH, $N_1[\theta] \bdby^{1 \loli C,g_1[\sigma]} \norm{N_1}[\Theta]$, which means
\begin{itemize}
  \item $n_2 \leq \norm{N_1}[\Theta]_c - r_2$
  \item $\lambda x.N_1' \valbd^{1 \loli C,g_1[\sigma] + r_2} \norm{N_1}[\Theta]_p$
\end{itemize}
Since $() \leq ()$, $() \valbd^{1,f[\sigma] + r_1} ()$. Hence, $N_2'[()/x] \bdby^{C,g_1[\sigma] + f[\sigma] + r_1 + r_2} \norm{N_1}[\Theta]_p \; ()$. This means that
\begin{itemize}
  \item $n_4 \leq (\norm{N_1}[\Theta]_p \; ())_c - r_4$
  \item $v \valbd^{C,f[\sigma] + g_1[\sigma] + r_1 + r_2 + r_4} (\norm{N_1}[\Theta]_p \; ())_p$
\end{itemize}
But by credit weakening, since $g_2[\sigma] + r_3 \geq 0$, we have 
$v \valbd^{C,f[\sigma] + g_1[\sigma] + g_2[\sigma] + r_1 + r_2 + r_3 + r_4} (\norm{N_1}[\Theta]_p \; ())_p$. But, we can compute:
$$
\norm{N_1}[\Theta]_p \; () \leq \lrec {\elist} {\norm{N_1}[\Theta]_p} {\norm{N_2}[\Theta]_p} \leq \lrec {\norm{M}[\Theta]_p} {\norm{N_1}[\Theta]_p} {\norm{N_2}^*}
$$
and we are done by weakening.\\

Otherwise, assume $M[\theta] \downarrow^{(n_1,r_1)} \cons {v_1} {v_2}$, $N_2[\theta] \downarrow^{(n_2,r_2)} \save \infty 0 {(\lambda x.N_2')}$, $N_1[\theta] \downarrow^{(n_3,r_3)} \lambda x.N_1'$, and $$N_2'[(v_1,\amppair {v_2} {\lrec {v_2} {\lambda x.N_1'} {\save \infty 0 {(\lambda x.N_2')}}})] \downarrow^{(n_4,r_4)} v$$.
Just like the previous case, it suffices to show
\begin{itemize}
  \item $n_1 + n_2 + n_3 + n_4 \leq \norm{M}[\Theta]_c + \norm{N_1}[\Theta]_c + \norm{N_2}[\Theta]_c + \lrec {\norm{M}[\Theta]_p} {\norm{N_1}[\Theta]_p} {\norm{N_2}^*}_c - (r_1 + r_2 + r_3 + r_4)$
  \item $v \valbd^{C,f[\sigma] + g_1[\sigma] + g_2[\sigma] + r_1 + r_2 + r_3 + r_4} \lrec {\norm{M}[\Theta]_p} {\norm{N_1}[\Theta]_p} {\norm{N_2}[\Theta]_p}_p$
\end{itemize}
By IH, $M[\theta] \bdby^{\listty A,f[\sigma]} \norm{M}[\Theta]$, so
\begin{itemize}
  \item $n_1 \leq \norm{M}[\Theta]_c - r_1$
  \item $\cons {v_1} {v_2} \valbd^{\listty A,f[\sigma] + r_1} \norm{M}[\Theta]_p$
\end{itemize}
By the second condition, we know that there are $d_1,d_2 \geq 0$ with $d_1 + d_2 = f[\sigma] + r_1$, and $E_1,E_2$ with $\cons {E_1} {E_2} \leq \norm{M}[\Theta]_p$ such that $v_1 \valbd^{A,d_1} E_1$, and $v_2 \valbd^{\listty A,d_2} E_2$.
By IH, $N_1[\theta] \bdby^{1 \loli C,g_1[\sigma]} \norm{N_1}[\Theta]$, so
\begin{itemize}
  \item $n_3 \leq \norm{N_1}[\Theta]_c - r_3$
  \item $\lambda x.N_1' \valbd^{1 \loli C,g_1[\sigma] + r_3} \norm{N_1}[\Theta]_p$
\end{itemize}
Again by IH, $N_2[\theta] \bdby^{!^\infty_0(A \otimes (\listty A \amp C) \loli C),g_2[\sigma]} \norm{N_2}[\Theta]$, which means that
\begin{itemize}
  \item $n_2 \leq \norm{N_2}[\theta]_c - r_2$
  \item $\save \infty 0 {(\lambda x.N_2')} \valbd^{!^\infty 0 (A \otimes (\listty A \amp C) \loli C),g_2[\sigma] + r_2} \norm{N_2}[\Theta]_p$
\end{itemize}
The second condition means by definition that there is a $c \geq 0$ such that $\infty \cdot c \leq g_2[\sigma] + r_2$, and $\lambda x.N_2' \valbd^{A \otimes (\listty A \amp C) \loli C,c} \norm{N_2}[\Theta]_p$. We claim that 

$$
\begin{array}{l}
N_2'[(v_1,\amppair {v_2} {\lrec {v_2} {\lambda x.N_2'} {\save \infty 0 {(\lambda x.N_2')}}})] \bdby^{f[\sigma] + g_1[\sigma] + g_2[\sigma] + r_1 + r_2 + r_3}\\ \norm{N_2} \; (E_1,((0,E_2),\lrec {E_2} {\norm{N_1}[\Theta]_p} {\norm{N_2}^*}))
\end{array}
$$

To prove this claim, we split into cases on the finitude of $g_2[\sigma] + r_2$.
\begin{itemize}
  \item Suppose $g_2[\sigma] + r_2$ is finite. Then $c = 0$, and $g_2[\sigma] + r_2 = 0$, and so by the list recursor lemma with $E_1 = \norm{N_1}[\Theta]_p$, $E_2 = \norm{N_2}[\Theta]_p$, and $E = E_2$, we have that $\lrec {v_2} {\lambda x.N_1'} {\save \infty 0 {(\lambda x.N_2')}} \bdby^{C,d_2 + g_1[\sigma] + r_3} \lrec {E_2} {\norm{N_1}[\Theta]_p} {\norm{N_2}^*}$. Since $v_2 \valbd^{\listty A,d_2} E_2$, we have that $v_2 \bdby^{\listty A,d_2} (0,E_2)$, and by credit weakening, since $g_1[\sigma] + r_3 \geq 0$, $v_2 \bdby^{\listty A,d_2 + g_1[\sigma] + r_3} (0,E_2)$. Thus:
   $$
   \begin{array}{l}
   \amppair {v_2} {\lrec {v_2} {\lambda x.N_1'} {\save \infty 0 {(\lambda x.N_2')}}} \bdby^{\listty A \amp C,d_2 + g_1[\sigma] + r_3}\\ ((0,E_2),\lrec {E_2} {\norm{N_1}[\Theta]_p} {\norm{N_2}^*})   
   \end{array}
   $$
    Next, since $v_1 \valbd^{A,d_1} E_1$, $d_1 + d_2 = f[\sigma] + r_1$, and $\lambda x.N_2' \valbd^{A \tensor (\listty A \amp C),0} \norm{N_2}[\Theta]_p$,
  $$
   \begin{array}{l}
    N_2'[(v_1,\amppair {v_2} {\lrec {v_2} {\lambda x.N_1'} {\save \infty 0 {(\lambda x.N_2')}}})/x] \bdby^{f[\sigma] + g_1[\sigma] + r_1 + r_3} \\\norm{N_2}[\Theta]_p \; (E_1,((0,E_2),\lrec {E_2} {\norm{N_1}[\Theta]_p} {\norm{N_2}^*}))     
   \end{array}
  $$
  Which is exactly what we wanted to show, since $g_2[\sigma] + r_2 = 0$.
  \item Suppose $g_2[\sigma] + r_2 = \infty$. Then, there is a $c \geq 0$ such that $\infty \cdot c \leq \infty$, and $\lambda x.N_2' \bdby^{A \otimes (\listty A \amp C) \loli C,c} \norm{N_2}[\Theta]_p$ By credit weakening, we may assume $c > 0$. By the list recursor lemma, $\lrec {v_2} {\lambda x.N_1'} {\save \infty 0 {(\lambda x.N_2')}} \bdby^{C,d_2 + g_1[\sigma] + r_3 + \infty \cdot c} \lrec {E_2} {\norm{N_1}[\Theta]_p} {\norm{N_2}^*}$. By the same reasoning as in the previous case, 
  $$
  \begin{array}{l}
  (v_1,\amppair {v_2} {\lrec {v_2} {\lambda x.N_1'} {\save \infty 0 {(\lambda x.N_2')}}}) \bdby^{A \tensor (\listty A \amp C),f[\sigma] + g_1[\sigma] + r_1 + r_3}\\ (E_1,((0,E_2),\lrec {E_2} {\norm{N_1}[\Theta]_p} {\norm{N_2}^*}))  
  \end{array}
  $$ 
  Then, since $\infty \cdot c + c = \infty \cdot c$, 
  $$
  \begin{array}{l}
  N_2'[(v_1,\amppair {v_2} {\lrec {v_2} {\lambda x.N_1'} {\save \infty 0 {(\lambda x.N_2')}}})/x] \bdby^{C,f[\sigma] + g_1[\sigma] + r_1 + r_3 + \infty \cdot c}\\ \norm{N_2}[\Theta]_p \; (E_1,((0,E_2),\lrec {E_2} {\norm{N_1}[\Theta]_p} {\norm{N_2}^*}))  
  \end{array}
  $$
  which, $\infty \cdot c \leq g_2[\sigma] + r_2$, gives us our goal by credit weakening.
\end{itemize}

From this result, we have by definition that
\begin{itemize}
  \item $n_4 \leq (\norm{N_2}[\Theta]_p \; (E_1,((0,E_2),\lrec {E_2} {\norm{N_1}[\Theta]_p} {\norm{N_2}^*})))_c - r_4$
  \item $v \valbd^{C,f[\sigma] + g_1[\sigma] + g_2[\sigma] + r_1 + r_2 + r_3 + r_4} (\norm{N_2}[\Theta]_p \; (E_1,((0,E_2),\lrec {E_2} {\norm{N_1}[\Theta]_p} {\norm{N_2}^*})))_p$
\end{itemize}
Then we can compute:
\begin{align*}
\norm{N_2}[\Theta]_p \; (E_1,((0,E_2),\lrec {E_2} {\norm{N_1}[\Theta]_p} {\norm{N_2}^*})) &\leq \norm{N_2}^* \; (E_1,(E_2,\lrec {E_2} {\norm{N_1}[\Theta]_p} {\norm{N_2}^*}))\\
&\leq \lrec {\cons {E_1} {E_2}} {\norm{N_1}[\Theta]_p} {\norm{N_2}^*}\\
&\leq \lrec {\norm{M}[\Theta]_p} {\norm{N_1}[\Theta]_p} {\norm{N_2}^*}
\end{align*}

and so we are done by weakening.

\item[($\N$-E)] 

Suppose $\Gamma \vdash_{f+g_1+g_2} \nrec M {N_1} {N_2} : C$. 
By inversion, we have that $\Gamma \vdash_f M : \N$, $\Gamma \vdash_{g_1} N_1 : 1 \loli C$, and $\Gamma \vdash_{g_2} N_2 : !^\infty_0 (\N \otimes (1 \loli C) \loli C)$. 
Let $\theta \subbd^{\Gamma,\sigma} \Theta$. For convenience, let $\norm{N_2}[\Theta]_p^* = \lambda p.\norm{N_2}[\Theta]_p(\pi_1 p,\lambda z. \pi_2 p)$.
We must show: 
$$
\begin{array}{l}
\nrec {M[\theta]} {N_1[\theta]} {N_2[\theta]} \bdby^{C,f[\sigma] + g_1[\sigma] + g_2[\sigma]}\\
(\norm{M}[\Theta]_c + \norm{N_1}[\Theta]_c + \norm{N_2}[\Theta]_c) +_c \nrec {\norm{M}[\Theta]_p} {\norm{N_1}[\Theta]_p} {\norm{N_2} [\Theta]_p^*}
\end{array}
$$
In order to show this, we have two evaluation cases to consider.
Suppose 
$$\nrec {M[\theta]} {N_1[\theta]} {N_2[\theta]} \downarrow^{(n_1+n_2+n_3+n_4,r_1+r_2+r_3+r_4)} v$$
by way of $M[\theta] \downarrow^{(n_1,r_1)} 0$, $N_1[\theta] \downarrow^{(n_2,r_2)} \lambda x.N_1'$, $N_2[\theta] \downarrow^{(n_3,r_3)} v'$, and $N_1'[()/x] \downarrow^{(n_4,r_4)} v$. 
It suffices to show that:
\begin{itemize}
  \item $n_1+n_2+n_3+n_4 \leq b + \norm{M}[\Theta]_c + \norm{N_1}[\Theta]_c + \norm{N_2}[\Theta]_c + \nrec {\norm{M}[\Theta]_p} {\norm{N_1}[\Theta]_p} {\norm{N_2} [\Theta]_p^*}_c - (r_1 + r_2 + r_3 + r_4)$
  \item $v \valbd^{C,f[\sigma] + g_1[\sigma] + g_2[\sigma] + r_1 + r_2 + r_3 + r_4} \nrec {\norm{M}[\Theta]_p} {\norm{N_1}[\Theta]_p} {\norm{N_2} [\Theta]_p^*}_p$
\end{itemize}

By IH, $M[\theta] \bdby^{\N,f[\sigma]} \norm{M}[\Theta]$, so
\begin{itemize}
  \item $n_1 \leq \norm{M}[\Theta]_c - r_1$
  \item $0 \valbd^{\N,f[\sigma] + r_1} \norm{M}[\Theta]_p$
\end{itemize}
since $0 \valbd^{\N,f[\sigma] + r_1} \norm{M}[\Theta]_p$, $0 \leq_\N \norm{M}[\Theta]_p$. 
By IH, $N_1 [\theta] \bdby^{1 \loli C,g_1[\sigma]} \norm{N_1}[\Theta]$, so
\begin{itemize}
  \item $n_2 \leq \norm{N_1}[\Theta]_c - r_2$
  \item $\lambda x.N_1' \valbd^{1 \loli C,g_1[\sigma] + r_2} \norm{N_1}[\Theta]_p$
\end{itemize}
By IH, $N_2[\theta] \bdby^{!^\infty_0(\cdots),g_2[\sigma]} \norm{N_2}[\Theta]$, and so
\begin{itemize}
  \item $n_3 \leq \norm{N_2}[\Theta]_c - r_3$
\end{itemize}
We omit the value bounding condition since it does not factor into the rest of the proof. Since $() \leq_1 ()$, $() \valbd^{1,f[\sigma] + r_1} ()$. So: $N_1'[()/x] \bdby^{C,f[\sigma] + g_1[\sigma] + r_1 + r_2} \norm{N_1}[\Theta] \; ()$. Thus,
\begin{itemize}
  \item $n_4 \leq (\norm{N_1}[\Theta] \; ())_c - r_4$
  \item $v \valbd^{C,f[\sigma] + g_1[\sigma] + r_1 + r_2 + r_4} (\norm{N_1}[\Theta] \; ())_p$
\end{itemize}
Since $g_2[\sigma] + r_3 \geq 0$, we know by credit weakening, $v \valbd^{C,f[\sigma] + g_1[\sigma] + g_2[\sigma] + r_1 + r_2 + r_3 + r_4} (\norm{N_1}[\Theta] \; ())_p$.
Since $0 \leq \norm{M}[\Theta]_p$, we compute: 
\begin{align*}
\norm{N_1}[\Theta] \; () &\leq_C \nrec 0 {\norm{N_1}[\Theta]_p} {\norm{N_2}[\Theta]_p^*}\\
&\leq \nrec {\norm{M}[\Theta]_p} {\norm{N_1}[\Theta]_p} {\norm{N_2}[\Theta]_p^*}
\end{align*}

So we are done by weakening.\\

Suppose $\nrec {M[\theta]} {N_1[\theta]} {N_2[\theta]} \downarrow^{(n_1+n_2+n_3+n_4,r_1+r_2+r_3+r_4)} v$ by way of 
$M[\theta] \downarrow^{(n_1,r_1)} S(v_1)$, 
$N_2[\theta] \downarrow^{(n_2,r_2)} \save \infty 0 {\lambda x.N_2'}$, 
$N_1[\theta] \downarrow^{(n_3,r_3)} \lambda x.N_1'$, and 
$$N_2'[(v_1,\nrec {v_1} {\lambda z. {(\nrec {v_1} {\lambda x.N_1'} {\save \infty 0 {\lambda x.N_2'}})}})/x] \downarrow^{(n_4,r_4)} v$$
\begin{itemize}
  \item $n_1 + n_2 + n_3 + n_4 \leq b + \norm{M}[\Theta]_c + \norm{N_1}[\Theta]_c + \norm{N_2}[\Theta] + \nrec {\norm{M}[\Theta]_p} {\norm{N_1}[\Theta]_p} {\norm{N_2}[\Theta]_p^*}_c - (r_1+r_2+r_3+r_4)$
  \item $v \valbd^{C,f[\sigma] + g_1[\sigma] + g_2[\sigma] + r_1+r_2+r_3+r_4} \nrec {\norm{M}[\Theta]_p} {\norm{N_1}[\Theta]_p} {\norm{N_2}[\Theta]_p^*}_p$
\end{itemize}

By IH, $M[\theta] \bdby^{\N,f[\sigma]} \norm{M}[\Theta]$, so
\begin{itemize}
  \item $n_1 \leq \norm{M}[\Theta]_c - r_1$
  \item $S(v_1) \valbd^{\N,f[\sigma] + r_1} \norm{M}[\Theta]_p$
\end{itemize}
Since $S(v_1) \valbd^{\N,f[\sigma] + r_1} \norm{M}[\Theta]_p$, there is an $E$ such that $v_1 \valbd^{\N,f[\sigma] + r_1} E$, and $S(E) \leq \norm{M}[\Theta]_p$.

By IH, $N_1[\theta] \bdby^{1 \loli C,g_1[\sigma]} \norm{N_1}[\Theta]$, and so
\begin{itemize}
  \item $n_3 \leq \norm{N_1}[\Theta]_c - r_3$
  \item $\lambda x.N_1' \valbd^{1 \loli C,g_1[\sigma] + r_3} \norm{N_1[\Theta]}_p$
\end{itemize}

By IH, $N_2[\theta] \bdby^{!^\infty_0(\N \otimes (1 \loli C) \loli C),g_2[\sigma]}\norm{N_2}[\Theta]$, so by definition,
\begin{itemize}
  \item $n_2 \leq \norm{N_2}[\Theta]_c - r_2$
  \item $\save \infty 0 {\lambda x.N_2'} \valbd^{!^\infty_0(\N\otimes (1 \loli C) \loli C),g_2[\sigma] + r_2}\norm{N_2}[\Theta]_p$
\end{itemize}
and so there is a $d \geq 0$ so that $\lambda x.N_2' \valbd^{\N \otimes (1 \loli C) \loli C,d} \norm{N_2}[\Theta]_p$, and $\infty \cdot d \leq g_2[\sigma] + r_3$. By the $\N$-recursor lemma, $\nrec {v_1} {\lambda x.N_1'} {\save \infty 0 {(\lambda x.N_2')}} \bdby^{C,g_1[\sigma] + r_2 + \infty \cdot d} \nrec {E} {\norm{N_1}[\Theta]_p} {\norm{N_2}[\Theta]_p^*}$.
Let $E^* = (E,\nrec {E} {\norm{N_1}[\Theta]_p} {\norm{N_2}[\Theta]_p^*})$. Note that $v_1 \valbd^{\N,f[\sigma] + r_1} \pi_1 E^*$ and $$\lambda z. \nrec {v_1} {\lambda x.N_1'} {\save \infty 0 {(\lambda x.N_2')}} \valbd^{1 \loli C,g_1[\sigma] + r_2 + \infty \cdot d} \lambda z. \pi_2 E^*$$, and so:
$$
(v_1,\lambda z. \nrec {v_1} {\lambda x.N_1'} {\save \infty 0 {(\lambda x.N_2')}}) \valbd^{\N \otimes (1 \loli C),f[\sigma] + g_1[\sigma] + \infty \cdot d + r_1 + r_2} (\pi_1  E^*,\lambda z.\pi_2 E^*)
$$.

Thus, because $\lambda z.N_2' \valbd^{\N \otimes (1 \loli C) \loli C,d} \norm{N_2}[\Theta]_p$, and $\infty \cdot d + d = \infty \cdot d$

$$
N_2'[(v_1,\lambda z. \nrec {v_1} {\lambda x.N_1'} {\save \infty 0 {(\lambda x.N_2')}})] \bdby^{C,f[\sigma] + g_1[\sigma] + \infty \cdot d + r_1 + r_2} \norm{N_2}[\Theta]_p(\pi_1  E^*,\lambda z.\pi_2 E^*)
$$
and so:
\begin{itemize}
  \item $n_4 \leq (\norm{N_2}[\Theta]_p \; (\pi_1  E^*,\lambda z.\pi_2 E^*))_c - r_4$
  \item $v \valbd^{C,f[\sigma] + g_1[\sigma] + \infty \cdot d + r_1 + r_2 + r_4}  (\norm{N_2}[\Theta]_p \; (\pi_1  E^*,\lambda z.\pi_2 E^*))_p$
\end{itemize}
but, $\infty \cdot d \leq g_2[\sigma] + r_3$, and
\begin{align*}
  \norm{N_2}[\Theta]_p \; (\pi_1  E^*,\lambda z.\pi_2 E^*) &\leq (\lambda p. \norm{N_2}[\Theta]_p (\pi_1 p,\lambda z.\pi_2 p))E^*\\
  &\leq \norm{N_2}[\Theta]_p^* \; (E,\nrec {E} {\norm{N_1}[\Theta]_p} {\norm{N_2}[\Theta]_p^*})\\
  &\leq \nrec {S(E)} {\norm{N_1}[\Theta]_p} {\norm{N_2}[\Theta]_p}\\
  &\leq \nrec {\norm{N_1}[\Theta]_p} {\norm{N_1}[\Theta]_p} {\norm{N_2}[\Theta]_p}
\end{align*}
and so we are done by weakening and credit weakening.

\item[($\amp$-I)] Suppose $\Gamma \vdash_f \amppair M N : A \amp B$, and let $\theta \subbd^{\Gamma,\sigma} \Theta$. We can invert to find that $\Gamma \vdash_f M : A$ and $\Gamma \vdash_f N : B$. Since $\amppair {M[\theta]} {N[\theta]} \downarrow^{(0,0)} \amppair  {M[\theta]} {N[\theta]} $, to show that $\amppair  {M[\theta]} {N[\theta]}  \bdby^{A \amp B,f[\sigma]} (0,(\norm{M}[\Theta],\norm{N}[\Theta]))$ we must show that $0\leq 0$ (done!) and that $\amppair {M[\theta]} {N[\theta]}  \valbd^{A \amp B,f[\sigma]} (\norm{M}[\Theta],\norm{N}[\Theta])$. For this, it suffices by weakening to show that $M[\theta] \bdby^{A,f[\sigma]} \norm{M}[\Theta]$ and $N[\theta] \bdby^{B,f[\sigma]} \norm{N}[\Theta]$, which are precisely the inductive hypotheses.

\item[($\amp$-E)] By symmetry, it suffices to present the $pi_1$ case. Suppose $\Gamma \vdash_f \pi_1 M : A$. By inversion, $\Gamma \vdash_f M : A \amp B$. Let $\theta \subbd^{\Gamma,\sigma} \Theta$. To show that $\pi_1 M[\theta] \bdby^{A,f[\sigma]} \norm{M}[\Theta]_c +_c \pi_1(\norm{M}[\Theta]_p)$, assume $\pi_1 M[\theta] \downarrow^{(n_1+n_2,r_1+r_2)} v$. By inversion, it was by way of $M[\theta] \downarrow^{(n_1,r_1)} \amppair {N_1} {N_2}$ and $N_1 \downarrow^{(n_2,r_2)} v$. We must show that:
\begin{itemize}
   \item $n_1 + n_2 \leq \norm{M}[\Theta]_c + (\pi_1\norm{M}[\Theta]_p)_c - (r_1 + r_2)$
   \item $v \valbd^{A,f[\sigma] + r_1 + r_2} (\pi_1\norm{M}[\Theta]_p)_p$
\end{itemize}
By IH, $M[\theta] \bdby^{A \amp B,f[\sigma]} \norm{M}[\Theta]$, so
\begin{itemize}
  \item $n_1 \leq \norm{M}[\Theta]_c - r_1$
  \item $\amppair {N_1} {N_2} \valbd^{A \amp B,f[\sigma] + r_1} \norm{M}[\Theta]_p$
\end{itemize}
where the second condition means, in particular, that $N_1 \bdby^{A,f[\sigma] + r_1} \pi_1 \norm{M}[\Theta]_p$. So, since $N_1 \downarrow^{(n_2,r_2)} v$,
\begin{itemize}
  \item $n_2 \leq (\pi_1 \norm{M}[\Theta]_p)_c - r_2$
  \item $v \valbd^{A,f[\sigma] + r_1 + r_2} (\pi_1 \norm{M}[\Theta]_p)_p$
\end{itemize}
as required.

\item[(var)] Suppose $\Gamma, x : A \vdash_{x+f} x : A$. Let $(\theta,v/x) \subbd^{(\Gamma,x:A),(\sigma,x\mapsto a)}(\Theta,E/x)$. We know that $v \valbd^{A,a} E$.
We must show that $v \bdby^{A,a+f[\sigma]} (0,E)$. We know that $v \downarrow^{(0,0)} v$. Of course, $0 \leq 0$. Since $f[\sigma] \geq 0$, we are done by credit weakening.

\end{itemize}
\end{proof}

\auxsemlemma*
\begin{proof}
Let $A,B,C,G$ be posets. Note that these are not required to be in the image of $\scott{\cdot}$. For each case we must show two statements: the function is in fact montonic, and that the functions in its image (an exponential poset) are themselves monotonic.
\begin{enumerate}
  \item Suppose $(f,g) \leq (f',g')$ as elements of ${\left(C^1\right)}^G\times {\left(C^{\N\times C}\right)}^G$. To show that $\texttt{snrec}(f,g) \leq \texttt{snrec}(f',g')$, it suffices to show that for all $\gamma,n$, that $\snrec(f,g)(\gamma,n) \leq \snrec(f',g')(\gamma,n)$. Proceed by induction on $n$.
  \begin{itemize}
    \item $n = 0$. By the definition of $\snrec$, it suffices to show that $f(\gamma)() \leq f'(\gamma)()$, which is true since $f \leq f'$.
    \item $n+1$ By definition of $\snrec$, it suffces to show $g(\gamma)(n,\snrec(f,g)(\gamma,n)) \vee f(\gamma)() \leq g'(\gamma)(n,\snrec(f',g')(\gamma,n)) \vee f(\gamma)()$. We have already shown that $f(\gamma)() \leq f'(\gamma)()$, so it remains to show $g(\gamma)(n,\snrec(f,g)(\gamma,n)) \leq g'(\gamma)(n,\snrec(f',g')(\gamma,n))$. Since $g \leq g'$, $g(\gamma) \leq g'(\gamma)$. By reflexivity, $n \leq n$. By IH, $\snrec(f,g)(\gamma,n) \leq \snrec(f',g')(\gamma)(n)$, and so we are done.
    
    Now, let $(f,g) \in {\left(C^1\right)}^G\times {\left(C^{\N\times C}\right)}^G$. We must show that if $(\gamma,n) \leq (\gamma',n')$ in $G \times \N$, then $\snrec(f,g)(\gamma,n) \leq \snrec(f,g)(\gamma',n')$. Proceed by induction on $n$. We have three cases to consider.
    \begin{itemize}
      \item $n = n' = 0$. By definition of $\snrec$, it suffices to show that $f(\gamma)() \leq f(\gamma')()$, which is true since $\gamma \leq \gamma'$.
      \item $n = 0$, $n' + 1$: By the definition of $\snrec$, we must show that $f(\gamma)() \leq g(\gamma')(n',\snrec(f,g)(\gamma',n')) \vee f(\gamma')()$, for which it suffices to show $f(\gamma)() \leq f(\gamma')()$, which we already argued was true.
      \item $n+1$, $n'+1$. Expanding definitions again and simplifying, it suffices to show that $g(\gamma)(n,\snrec(f,g)(\gamma,n)) \leq g(\gamma')(n',\snrec(f,g)(\gamma',n'))$. Since $g$ is monotonic, $g(\gamma) \leq g(\gamma')$. Since $n + 1 \leq n' + 1$, $n \leq n'$. By IH, $\snrec(f,g)(\gamma,n) \leq \snrec(f,g)(\gamma',n')$, and so $g(\gamma)(n,\snrec(f,g)(\gamma,n)) \leq g(\gamma')(n',\snrec(f,g)(\gamma',n'))$, as required.
    \end{itemize}
  \end{itemize}
  \item Let $(f,g) \leq (f',g) \in {\left(C^1\right)^G} \times {\left(C^{A \times (\N \times C)}\right)^G}$. We want to show that $\slrec(f,g) \leq \slrec(f',g')$. Fix $\gamma \in G$, we prove by induction on $n$ that for all $n \in \N$, $\slrec(f,g)(\gamma,n) \leq \slrec(f',g')(\gamma,n)$.
    \begin{itemize}
      \item $n = 0$: expanding the definition of $\slrec$, we must show that $f(\gamma)() \leq f'(\gamma)()$, which is true because $f \leq f'$.
      \item $n > 0$. It suffices to show that $g(\gamma)(\infty,(n,\slrec(f,g)(\gamma,n))) \leq g'(\gamma)(\infty,(n,\slrec(f',g')(\gamma,n)))$. Since $g \leq g'$, $g(\gamma) \leq g'(\gamma)$. Further, $\infty \leq \infty$, $n \leq n$, and by IH, $\slrec(f,g)(\gamma,n) \leq \slrec(f',g')(\gamma,n)$, as required.
    \end{itemize}
    
    Now, let $(f,g) \in {\left(C^1\right)^G} \times {\left(C^{A \times (\N \times C)}\right)^G}$. We must show that if $(\gamma,n) \leq (\gamma',n')$, $\slrec(f,g)(\gamma,n) \leq \slrec(f,g)(\gamma,n')$. We again prove this by induction on $n$. There are three cases we must consider.
    \begin{itemize}
      \item $n = n' = 0$. Immediate.
      \item $n = 0, n' > 0$. Identical to the similar case for \texttt{snrec}.
      \item $n,n' > 0$. To show that 
      $$g(\gamma)(\infty,(n,\snrec(f,g)(\gamma,n))) \vee f(\gamma)() \leq g(\gamma')(\infty,(n',\snrec(f,g)(\gamma',n))) \vee f(\gamma')()$$
      it suffices to show that $f(\gamma) \leq f(\gamma')$ (which is true because $\gamma \leq \gamma'$ and $f$ is monotonic) and $g(\gamma)(\infty,(n,\snrec(f,g)(\gamma,n))) \leq g(\gamma')(\infty,(n',\snrec(f,g)(\gamma',n')))$. Since $n + 1 \leq n'+1$, $n \leq n'$, and so the desired result follows from IH and the fact that $g(\gamma)$
    \end{itemize}
    
  \item Let $(f,g) \leq (f',g') \in C^{G \times A}\times C^{G \times B}$. We must show that $\scase(f,g) \leq \scase(f',g')$ in $C^{G \times (A + B)}$ Let $(\gamma,x) \in G \times (A+B)$. The two cases for $x$ are symmetrical, so we consider when $x = \inl a$. Then,
    \begin{align*}
      \scase(f,g)(\gamma,\inl a) &= f(\gamma,a) \vee g(\gamma,\infty)\\
                                 &\leq f'(\gamma,a) \vee g(\gamma,\infty)\\
                                 &= \scase(f',g')(\gamma,\inl a)
    \end{align*}
    as required.
    
    Now, fix $(f,g) \in C^{G \times A}\times C^{G \times B}$. We must show that for all $(\gamma,x) \leq (\gamma',y)$, $\scase(f,g)(\gamma,x) \leq \scase(f,g)(\gamma',y)$. We have two symmetric cases to consider, so we present the case where $x = \inl a$ and $y = \inl a'$. Then,
    \begin{align*}
      \scase(f,g)(\gamma,\inl a) &= f(\gamma,a) \vee g(\gamma,\infty)\\
                                 &\leq f(\gamma',a') \vee g(\gamma',\infty)\\
                                 &= \scase(f,g)(\gamma',\inl a')
    \end{align*}
    as required.
\end{enumerate}
\end{proof}
\semsubst*
\begin{proof}
By induction on $\Gamma, x : T_1 \vdash E : T_2$.
\begin{itemize}
  \item (\texttt{nrec}): Suppose $\Gamma, x : T_1 \vdash \nrec E {E_1} {E_2} : T_2$. By inversion, $\Gamma x : T_1 \vdash E : \N $, $\Gamma, x : T_1 \vdash E_1 : 1 \to T_2$, and $\Gamma, x: T_1 \vdash E_2 : \N \times T_2 \to T_2$. By IH,
  \begin{itemize}
   \item  $\scott{\Gamma \vdash E[E'/x] : \N} = (1_{\scott{\Gamma}},\scott{\Gamma \vdash E' : T_1}) ; \scott{\Gamma, x : T_1 \vdash E : \N}$  
   \item $\scott{\Gamma \vdash E_1[E'/x] : 1 \to T_2} = (1_{\scott{\Gamma}},\scott{\Gamma \vdash E' : T_1}) ; \scott{\Gamma, x : T_1 \vdash E_1 : 1 \to T_2}$
   \item $\scott{\Gamma \vdash E_2[E'/x] : \N \times T_2 \to T_2} = (1_{\scott{\Gamma}},\scott{\Gamma \vdash E' : T_1}) ; \scott{\Gamma, x: T_1 \vdash E_2 : \N \times T_2 \to T_2}$
  \end{itemize}. For ease of notation, we let $f = \scott{\Gamma \vdash E' : T_1}$, $g = \scott{\Gamma, x : T_1 \vdash E : \N}$, $h_1 =  \scott{\Gamma, x : T_1 \vdash E_1 : 1 \to T_2}$, and $h_2 = \scott{\Gamma, x: T_1 \vdash E_2 : \N \times T_2 \to T_2}$.
  Then, we compute:
  \begin{align*}
  &\scott{\Gamma \vdash \left(\nrec E {E_1} {E_2}\right)[E'/x] : T_2} \\
  &= \scott{\Gamma \vdash \nrec {E[E'/x]} {E_1[E'/x]} {E_2[E'/x]}}\\
  &= (1_{\scott{\Gamma}}, \scott{\Gamma \vdash E[E'/x] : \N}) ; \snrec(\scott{\Gamma \vdash E_1[E'/x] : 1 \to T_2},\scott{\Gamma \vdash E_2[E'/x] : \N \times T_2 \to T_2})\\
  &= (1_{\scott{\Gamma}},(1_{\scott{\Gamma}},f) ; g) ; \snrec((1_{\scott{\Gamma}},f) ; h_1,(1_{\scott{\Gamma}},f) ; h_2)
  \end{align*}
  It remains to show that
  $$
(1_{\scott{\Gamma}},(1_{\scott{\Gamma}},f) ; g) ; \snrec((1_{\scott{\Gamma}},f) ; h_1,(1_{\scott{\Gamma}},f) ; h_2) = (1_{\scott{\Gamma}},f);(1_{\scott{\Gamma,x:T_1}},g);\snrec(h_1,h_2)  
  $$
  Let $\gamma \in \scott{\Gamma}$.
  Applying the left hand side to $\gamma$, we get
  $$
  \snrec((1_{\scott{\Gamma}},f);h_1,(1_{\scott{\Gamma}},f);h_1)(\gamma,g(\gamma,f(\gamma)))
  $$
  and on the right:
  $$
  \snrec(h_1,h_2)((\gamma,f(\gamma)),g(\gamma,f(\gamma)))  
  $$
  Letting $\gamma' = (\gamma,f(\gamma))$, we must show that $\snrec((1_{\scott{\Gamma}},f);h_1,(1_{\scott{\Gamma}},f);h_1)(\gamma,g(\gamma')) = \snrec(h_1,h_2)(\gamma',g(\gamma'))$. We proceed by induction on $n = g(\gamma')$.
  \begin{itemize}
    \item $n = 0$. 
    \begin{align*}
      \snrec((1_{\scott{\Gamma}},f);h_1,(1_{\scott{\Gamma}},f);h_1)(\gamma,0) &= ((1_{\scott{\Gamma}},f);h_1)(\gamma)()\\
      &= h_1(\gamma,f(\gamma))()\\
      &= h_1(\gamma')()\\
      \snrec(h_1,h_2)(\gamma',0) &= h_1(\gamma')()
    \end{align*}
    as required.
    \item $n+1$:
    \begin{align*}
    &\snrec((1_{\scott{\Gamma}},f);h_1,(1_{\scott{\Gamma}},f);h_1)(\gamma,n+1)\\
    &= ((1_{\scott{\Gamma}},f);h_1)(\gamma)(n,\snrec((1_{\scott{\Gamma}},f);h_1,(1_{\scott{\Gamma}},f);h_1)(\gamma,n)) \vee h_1(\gamma')()\\
    &= h_1(\gamma')(n,\snrec(h_1,h_2)(\gamma',n)) \vee h_1(\gamma')()\\
    &= \snrec(h_1,h_2)(\gamma',n+1)
    \end{align*}
  \end{itemize}
  
  \item (\texttt{lrec}): Suppose $\Gamma, x : T_1 \vdash \lrec {E'} {E_1} {E_2} : T_2$, and $\Gamma \vdash E : T_1$. By inversion, $\Gamma,x:T_1 \vdash E : \listty T$, $\Gamma, x : T_1 \vdash E_1 : 1 \to T_2$, and $\Gamma,x:T_1 \vdash E_2 : T \times (\listty T \times T_2) \to T_2$.
  By IH, we have that:
  \begin{itemize}
    \item $\scott{\Gamma \vdash E'[E/x] : \listty T} = (1_{\scott{\Gamma}},\scott{\Gamma \vdash E : T_1}) ; \scott{\Gamma,x:T_1 \vdash E : \listty T}$
    \item $\scott{\Gamma \vdash E_1[E/x] : 1 \to T_2} = (1_{\scott{\Gamma}},\scott{\Gamma \vdash E : T_1}) ; \scott{\Gamma, x : T_1 \vdash E_1 : 1 \to T_}$
    \item $\scott{\Gamma \vdash E_2[E/x] : T \times (\listty T \times T_2) \to T_2} = (1_{\scott{\Gamma}},\scott{\Gamma \vdash E : T_1}) ; \scott{\Gamma,x:T_1 \vdash E_2 : T \times (\listty T \times T_2) \to T_2}$
  \end{itemize}
  Let $f = \scott{\Gamma \vdash E : T_1}$, $g = \scott{\Gamma,x : T_1 \vdash E' : \listty T}$, $h_1 = \scott{\Gamma,x:T_1 \vdash E_1 : 1 \to T_2}$, and $h_2 = \scott{\Gamma,x:T_1 \vdash E : T \times (\listty T \times T_2) \to T_2}$

  We must show that 
  $$
   (1_{\scott{\Gamma}},f) ; (1_{\scott{\Gamma} \times \scott{T_1}},g) ; \slrec(h_1,h_2) = (1_{\scott{\Gamma}},(1_{\scott{\Gamma}},f);g) ; \slrec((1_{\scott{\gamma}},f) ; h_1,(1_{\scott{\gamma}},f) ; h_2)
  $$
  
  Let $\gamma \in \scott{\Gamma}$, and let $\gamma' = (\gamma,f(\gamma))$. We must then show that
  
  $$\slrec(h_1,h_2)(\gamma',g(\gamma')) = \slrec((1_{\scott{\gamma}},f) ; h_1,(1_{\scott{\gamma}},f) ; h_2)(\gamma,g(\gamma'))$$
  
  We proceed by induction on $n = g(\gamma')$.
  
  \begin{itemize}
    \item ($n = 0$): The LHS is $\slrec(h_1,h_2)(\gamma',0) = h_1(\gamma')()$, and the RHS is 
    $$\slrec((1_{\scott{\gamma}},f) ; h_1,(1_{\scott{\gamma}},f) ; h_2)(\gamma,0) = h_1(\gamma,f(\gamma))() = h_1(\gamma')()$$.
    \item ($n > 0$): The LHS is:
    \begin{align*}
      &\slrec(h_1,h_2)(\gamma',n+1)\\
      &= h_2(\gamma')(\infty,(n,\slrec(h_1,h_2)(\gamma',n))) \vee h_1(\gamma')()
    \end{align*}
    and the RHS (applying the IH in the 2nd step) is
    \begin{align*}
     &\slrec((1_{\scott{\Gamma}},f) ; h_1,(1_{\scott{\Gamma}},f) ; h_2)(\gamma,n+1)\\
     &= h_2(\gamma')(\infty,(n,\slrec((1_{\scott{\Gamma}},f) ; h_1,(1_{\scott{\Gamma}},f) ; h_2)(\gamma,n))) \vee  h_1(\gamma')()\\
     &= h_2(\gamma')(\gamma,(n,\slrec(h_1,h_2)(\gamma',n))) \vee h_1(\gamma')()
    \end{align*}
    as required.
  \end{itemize}
\end{itemize}
\end{proof}
\begin{proof}
By induction on $\Gamma \vdash E : T$.
\begin{itemize}
  \item ($\texttt{nrec}$): Let $\Gamma \vdash \nrec E {E_1} {E_2} : T$. By inversion, $\Gamma \vdash E : \N$, $\Gamma \vdash E_1 : 1 \to T$, and $\Gamma \vdash E_2 : \N \times C \to C$. By IH, $\scott{\Gamma \vdash E : \N} \in \Hom(\scott{\Gamma},\N)$. Then, $(1_{\Gamma},\scott{\Gamma \vdash E : \N}) \in \Hom(\scott{\Gamma},\scott{\Gamma}\times \N)$. By IH, $\scott{\Gamma \vdash E_1 : 1 \to T} \in \Hom(\scott{\Gamma},\scott{T}^1)$ and $\scott{\Gamma \vdash E_2 : \N \times T \to T} \in \Hom(\scott{\Gamma},\scott{T}^{\N \times \scott{T}})$. Then, by Theorem~\ref{thm:aux-sem-lemma} and composition, $(1_{\Gamma},\scott{\Gamma \vdash E : \N}) ; \snrec(\scott{\Gamma \vdash E_1 : 1 \to T},\scott{\Gamma \vdash E_2 : \N \times T \to T}) \in \Hom(\scott{\Gamma},\scott{T})$, as required.
  
  \item ($\texttt{lrec}$): Let $\Gamma \vdash \lrec E {E_1} {E_2} : T$. By inversion, $\Gamma \vdash E : \listty {T'}$, $\Gamma \vdash E_2 : 1 \to T$, and $\Gamma \vdash E_2 : T' \times (\listty {T'} \times T) \to T$. Applying the IH to all of these premises, we have that $\scott{\Gamma \vdash E : \listty {T'}} \in \Hom(\scott{\Gamma},\N)$, $\scott{\Gamma \vdash E_2 : 1 \to T} \in \Hom(\scott{\Gamma},\scott{T}^1)$, and $\scott{\Gamma \vdash E_2 : T' \times (\listty {T'} \times T) \to T} \in \Hom(\scott{\Gamma},\scott{T}^{\scott{T'}\times(\N\times\scott{T})})$. By Theorem~\ref{thm:aux-sem-lemma} and composition, $(1_{\scott{\Gamma}},\scott{\Gamma \vdash E : \listty {T'}}) ; \slrec(\scott{\Gamma \vdash E_2 : 1 \to T},\scott{\Gamma \vdash E_2 : T' \times (\listty {T'} \times T) \to T}) \in \Hom(\scott{\Gamma},\scott{T})$ as required.
\end{itemize}
\end{proof}

\preordsound*
\begin{proof}
By induction on $\Gamma \vdash E \leq E'$. The new cases ($\snrec$ and $\slrec$) follow easily from the definitions.
\end{proof}
\substext*
\begin{proof}
By induction on $\Delta,\alpha \vdash c' \texttt{ credit}$ and $\Delta,\alpha|\Gamma\vdash_f M : A$, respectively.
\end{proof}
\presext*
\begin{proof}
The cases for all pre-existing rules are identical-- the only new cases are for \texttt{pack}, \texttt{unpack}, and \texttt{trec}. We present only the final case of \texttt{trec}, as it is the most illustritive.
\begin{itemize}
  \item (\texttt{pack}): Suppose that $\cdot | \cdot \vdash_a \pack \alpha \ell M : \exists \alpha$ and $\pack \alpha \ell M \downarrow^{(n,r)} \pack \alpha \ell v$ by way of $\cdot | \cdot \vdash_a M : A[\ell/\alpha]$ and $M \downarrow^{(n,r)} v$. By IH, $\cdot | \cdot \vdash_{a+r} v : A[\ell/\alpha]$ and $a + r \geq 0$. By the rule for $\texttt{pack}$, $\cdot | \cdot \vdash_{a + r} \pack \alpha \ell v : \exists \alpha. A$, as required.
  \item (\texttt{unpack}): Suppose that $\cdot | \cdot \vdash_{a + b} \unpack \alpha x M N : C$ by way of $\cdot | \cdot \vdash_a M : \exists \alpha. A$ and $\alpha | x : A \vdash_{b+x} N : C$ with $\Delta \vdash C$, and that $\unpack \alpha x M N \downarrow^{(n_1+n_2,r_1+r_2)} v$ by way of $M \downarrow^{(n_1,r_1)} \pack \alpha \ell {v_1}$ and $N[\ell/\alpha,v_1/x] \downarrow^{(n_2,r_2)} v$. By IH, $\cdot | \cdot \vdash_{a+r_1} v_1 : A[\ell/\alpha]$. By credit variable substituion, $\cdot | x : A[\ell/\alpha] \vdash_{b+x} N[\ell/\alpha] :C$. By substitution, $\cdot | \cdot \vdash_{b + a + r_1} N[\ell/\alpha,v_1/x] L C$ By IH, $\cdot | \cdot \vdash_{a+b+r_1+r_2} v : C$ and $a + b + r_1 + r_2 \geq 0$ as required.
  \item (\texttt{trec}): Suppose:
  $$
  \infer{\cdot | \cdot \vdash_{a + \sum b_i} \trec M {N_1} {N_2} {N_3} {N_4} {N_4}}
  {
   \begin{array}{l}
  \cdot | \cdot \vdash_f M : \tree A\\
  \cdot | \cdot \vdash_{b_1} N_1 : !^\infty_0(1 \loli C)\\
  \cdot | \cdot \vdash_{b_2} N_2 : !^\infty_0(A \loli C)\\
  \cdot | \cdot \vdash_{b_3} N_3 : !^\infty_0(A \otimes \N \otimes A \otimes N \otimes (\tree A \amp C)^2 \loli C)\\
  \cdot | \cdot \vdash_{b_4} N_4 : !^\infty_0(A \otimes \N \otimes A \otimes N \otimes (\tree A \amp C)^2 \loli C)\\
  \cdot | \cdot \vdash_{b_5} N_5 : !^\infty_0(A \otimes \N \otimes A \otimes N \otimes A \otimes N (\tree A \amp C)^4 \loli C)\\
  \end{array}
  }
  $$
  and
  $$
  \infer{\trec M {N_1} {N_2} {N_3} {N_4} {N_5} \downarrow^{(\sum n_i,\sum r_i)} v}
  {
  \begin{array}{l}
  M \downarrow^{(n_0,r_0)} N(v_1,n_1,N(v_2,n_2,t_{00},t_{01}),N(v_3,n_7,t_{10},t_{11}))\\
  N_i \downarrow^{(n_i,r_i)} \save \infty 0 {v_i'} \qquad (1 \leq i \leq 4)\\
  N_5 \downarrow^{(n_5,r_5)} \save \infty 0 {(\lambda x.N_5')}\\
  N_5'[(v_1,n_1,v_2,n_2,v_3,n_3,\amppair{t_{00},\texttt{trec}({t_{00}},{\save \infty 0 {v_1'}},{\dots},{\save \infty 0 {(\lambda x.N_5')}})},\dots)/x]
  \end{array}
  }  
  $$
  By IH, $\cdot | \cdot \vdash_{a+r_0} N(\dots)$. Hence, there are $d_1,\dots,d_n$, all non-negative, so that $\sum d_i = a + r_0$, and $\cdot | \cdot \vdash_{d_i} w_i: A_i$ where $w_i$ is the $i$th value in the value which $M$ evaluates to (in particular, $\cdot | \cdot \vdash_{d_1} v_1 : A$, and $\cdot | \cdot \vdash_{d_6} t_{00} : \tree A$).  Again by IH, there are $c_1,\dots,c_5$ so that $\infty c_i \leq b_i + r_i$, with $\cdot | \cdot \vdash_{c_i} v_i'$. Thus, $\cdot | \cdot \vdash_{d_6 + \sum c_i} \amppair{t_{00}} {\texttt{trec}(t_00,\save \infty 0 {v_1'},\dots)}$, and similarly for the rest of the subtrees. This immediately implies $$
  \cdot | \cdot \vdash_{\sum {d_i} + 4\sum c_i} (v_1,n_1,v_2,n_2,v_3,n_3,\amppair{t_{00}} {\texttt{trec}(t_00,\save \infty 0 {v_1'},\dots)},\dots) : (A\otimes \N)^3 \otimes (\tree A \amp C)^4  
  $$
  then by substitution
  $$
  \cdot | \cdot \vdash_{\sum d_i + c_5 + 4\sum c_i} N_5'[ (v_1,n_1,v_2,n_2,v_3,n_3,\amppair{t_{00}} {\texttt{trec}(t_00,\save \infty 0 {v_1'},\dots)},\dots)/x] : C
  $$
  The result follows immediately by weakening ($\infty c_i \leq b_i + r_i$) and IH.
\end{itemize}
\end{proof}
\extrsoundex*
\begin{proof}
By induction on $\Delta | \Gamma \vdash_f M : A$.
\end{proof}
\boundingex*
\begin{proof}
$\;$
\begin{itemize}
  \item (\texttt{pack}): Suppose $\Delta | \Gamma \vdash_f \pack \alpha c M : \exists \alpha. A$ by way of $\Delta | \Gamma \vdash_f M : A[c/\alpha]$. Let $\omega \bdby_{\texttt{credit}}^\Delta \Omega$ and $\theta \subbd^{\Gamma[\omega],\sigma} \Theta$. We must show that $\pack \alpha {c[\omega]} {M[\omega,\theta]} \bdby^{\exists \alpha.A[\omega],f[\omega,\sigma]} (\norm{M}_c[\Omega,\Theta],(c[\Omega],\norm{M}_p[\Omega,\Theta]))$. Suppose $\pack \alpha {c[\omega]} {M[\omega,\theta]} \downarrow^{(n,r)} \pack \alpha {c[\omega]} v$ by way of $M[\omega,\theta] \downarrow^{(n,r)} v$. It suffices to show
  \begin{itemize}
    \item $n + r \leq \norm{M}_c[\Omega,\Theta]$
    \item $\pack \alpha {c[\omega]} v \valbd^{\exists \alpha.A[\omega],f[\omega,\sigma] + r} (c[\Omega],\norm{M}_p[\Omega,\Theta])$
  \end{itemize}
  The second item is equivalent to proving that $c[\omega] \leq c[\Omega]$ (which is true because credit terms are monotone), and that $v \valbd^{A[c/\alpha,\omega],f[\omega,\sigma] + r} \norm{M}_p[\Omega,\Theta]$, which follows immediately by IH.
  
  \item (\texttt{unpack}): Suppose that $\Delta | \Gamma \vdash_{f+g} \unpack \alpha x M N : C$ by way of $\Delta | \Gamma \vdash_f M : \exists \alpha.A$ and $\Delta,\alpha | \Gamma,x:A \vdash_{g+x} N : C$ with $\alpha$ not free in $C$. Let $\omega \bdby_\texttt{credit}^\Delta \Omega$ and $\theta \subbd^{\Gamma[\omega],\sigma} \Theta$. Suppose $ \unpack \alpha x {M[\omega,\theta]} {N[\omega,\theta]} \downarrow^{(n_1+n_2,r_1+r_2)} v$ by way of $M[\omega,\theta] \downarrow^{(n_1,r_1)} \pack \alpha \ell {v_1}$ and $N[\omega,\theta,\ell/\alpha,v_1/x]  \downarrow^{(n_2,r_2)} v$. It suffices to show that
  \begin{itemize}
    \item $n_1 + n_2 + r_1 + r_2 \leq \norm{M}_c[\Omega,\Theta] + \norm{N}_c[\Omega,\Theta,\pi_1\norm{M}_p[\Omega,\Theta]\alpha,\pi_2\norm{M}_p[\Omega,\Theta]/x]$
    \item $v \valbd^{C[\omega],f[\omega,\sigma] + g[\omega,\sigma] + r_1 + r_2} \norm{N}_p[\Omega,\Theta,\pi_1\norm{M}_p[\Omega,\Theta]\alpha,\pi_2\norm{M}_p[\Omega,\Theta]/x]$
  \end{itemize}
  By IH, $M[\omega,\theta] \bdby^{\exists \alpha.A[\omega],f[\omega,\sigma]} \norm{M}[\Omega,\Theta]$, and so
  \begin{itemize}
    \item $n_1 + r_1 \leq \norm{M}_c[\Omega,\Theta]$
    \item $\pack \alpha \ell {v_1} \valbd^{\exists \alpha.A[omega],f[\omega,\sigma] + r_1}\norm{M}_p[\Omega,Theta]$
  \end{itemize}
  which means that $\ell \leq \pi_1 \norm{M}_p[\Omega,Theta]$ and that $v_1 \valbd^{A[\omega,\ell/\alpha],f[\omega,\sigma] + r_1} \pi_2 \norm{M}_p[\Omega,Theta]$.
  Hence, $(\omega,\ell/\alpha) \bdby_{\texttt{credit}}^{\Delta,\alpha}(\Omega,\pi_1 \norm{M}_p[\Omega,\Theta]/\alpha)$, and $(\theta,v_1/x) \subbd^{\Gamma[\omega],x:A[\ell/\alpha],\sigma,x\mapsto f[\omega,\sigma] + r_1} (\Theta,\pi_2 \norm{M}_p[\Omega,\Theta]/x)$. Thus, by IH, $$N[\omega,\theta,\ell/\alpha,v_1/x] \bdby^{C[\omega],g[\omega,\sigma] + f[\omega,\sigma] + r_1} \norm{N}[\Omega,\Theta,\pi_1 \norm{M}_p[\Omega,\Theta]/\alpha,\pi_2 \norm{M}_p[\Omega,\Theta]/x]$$ By definition,
  \begin{itemize}
    \item $n_2 + r_2 \leq \norm{N}_c[\Omega,\Theta,\pi_1 \norm{M}_p[\Omega,\Theta]/\alpha,\pi_2 \norm{M}_p[\Omega,\Theta]/x]$
    \item $v \valbd^{C[\omega],f[\omega,\sigma] + g[\omega,\sigma] + r_1 + r_2} \norm{N}_p[\Omega,\Theta,\pi_1 \norm{M}_p[\Omega,\Theta]/\alpha,\pi_2 \norm{M}_p[\Omega,\Theta]/x]$
  \end{itemize}
  as required.
\end{itemize}
\end{proof}

\begin{figure}
  \input{figs/trec-rules}
  \caption{$\lambda^A$ and $\lambda^\bbbc$ \texttt{tree} extension, and recurrence extraction}
  \label{fig:trec-rules}
\end{figure}

\end{document}